%% file: main.tex
\documentclass[prx,onecolumn,superscriptaddress,longbibliography]{revtex4-2}

\usepackage[]{fontenc}

\usepackage{graphicx}
\usepackage{bm}
\usepackage{mathtools} 
\usepackage{leftindex}
\usepackage[caption=false]{subfig}
\usepackage{chngcntr} 
\usepackage{multirow}
\usepackage{subeqnarray}
\usepackage{wasysym}

\usepackage[british]{babel}
\usepackage{amsmath}
\usepackage{amssymb}
\usepackage{framed}
\usepackage{soul}
\usepackage{color}
\usepackage{alltt}
\usepackage{pstricks,pst-node,pst-tree,pst-grad,pst-text,graphics}
\usepackage{dcolumn}
\usepackage{url}
\usepackage{cases}
\usepackage{natbib}

\usepackage{ifthen}
\usepackage{afterpage}
\usepackage{comment}
\usepackage{supertabular}
\usepackage{xspace}
\usepackage{setspace}
\usepackage{textcomp}
\usepackage{tikz}
\usepackage{tabularx}
\usepackage{tikz-feynman}
\tikzfeynmanset{compat=1.1.0}
\usepackage{dsfont}
\usepackage{enumitem}

\usepackage[hidelinks,colorlinks=true,linkcolor=blue,filecolor=blue,urlcolor=blue,citecolor=blue]{hyperref}
\usetikzlibrary{decorations.pathmorphing}
\usetikzlibrary{decorations.markings}
\usetikzlibrary{arrows,shapes,decorations,automata,backgrounds,petri}

\newcommand{\defequal}{\vcentcolon=}
\newcommand{\plaind}{\mathrm{d}}

\newcommand{\ddint}[1]{\ddintx{#1}{d}}
\newcommand{\ddintx}[2]{\mathchoice{\!\plaind^{#2}#1\,}{\!\plaind^{#2}#1\,}{\!\plaind^{#2}#1\,}{\!\plaind^{#2}#1\,}}

\newcommand{\dbar}{\text{\dj}}
\newcommand{\deltabar}{\delta\mkern-6mu\mathchar'26}
\newcommand{\dintbar}[1]{\mathchoice{\!\dbar#1\,}{\!\dbar#1\,}{\!\dbar#1\,}{\!\dbar#1\,}}
\newcommand{\ddintbar}[1]{\mathchoice{\!\dbar^d#1\,}{\!\dbar^d#1\,}{\!\dbar^d#1\,}{\!\dbar^d#1\,}}

\newcommand{\Rset}{\mathbb{R}}
\newcommand{\Nset}{\mathbb{N}}

\newcommand{\gpvec}[1]{\mathbf{#1}}

\newcommand{\zerovec}{\gpvec{0}}
\newcommand{\xvec}{\gpvec{x}}

\newcommand{\qvec}{\gpvec{q}}
\newcommand{\kvec}{\gpvec{k}}

\newcommand{\phivec}{\bm{\phi}}
\newcommand{\Phivec}{\bm{\Phi}}
\newcommand{\varphivec}{\bm{\varphi}}
\newcommand{\xivec}{\bm{\xi}}

\newcommand{\varphitilde}{\tilde{\varphi}}

\newcommand{\phitilde}{\tilde{\phi}}

\renewcommand{\exp}[1]{\mathchoice%
{\mathrm{e}^{#1}}%
{\operatorname{exp}(#1)}
{\operatorname{exp}\left(#1\right)}%
{\operatorname{exp}\left(#1\right)}}

\newcommand{\elabel}[1]{\label{eq:#1}}
\newcommand{\eref}[1]{\eqref{eq:#1}}
\newcommand{\Eref}[1]{\mbox{Eq.~\eqref{eq:#1}}}
\newcommand{\Erefs}[1]{\mbox{Eqs.~\eqref{eq:#1}}}

\newcommand{\seclabel}[1]{\label{sec:#1}}

\newcommand{\Sref}[1]{Section~\ref{sec:#1}}

\newcommand{\APref}[1]{Appendix~\ref{app:#1}}

\newcommand{\Fref}[1]{Figure~\ref{fig:#1}}

\renewcommand{\exp}[1]{\mathchoice%
{e^{#1}}%
{\operatorname{exp}(#1)}%
{\operatorname{exp}(#1)}%
{\operatorname{exp}(#1)}}

\newcommand{\action}{\mathcal{A}}

\newcommand{\symgroup}{\mathcal{O}(n) \otimes \mathcal{O}(n)}

\newcommand{\kvecsquared}{\kvec^2}
\newcommand{\nablasquared}{\nabla^2}
\makeatletter
\newcommand{\Hamiltonian}[1][]{\mathcal{H}{\@ifempty{#1}{}{_#1}}}
\makeatother

\newcommand{\SuperHamiltonian}{\tilde{\Hamiltonian}}

\newcommand{\sym}{\text{sym}}

\newcommand{\duc}{d_{\rm c}}

\newcommand{\dimx}{[x]}

\newcommand{\dvec}{\bm{d}}

\newcommand{\phitildevec}{\bm{\tilde{\phi}}}

\newcommand{\phiveci}{\bm{\phi}_i}

\newcommand{\tautilde}{\tilde{\tau}}

\newcommand{\phiany}[1][]{\phi^{#1}}

\newcommand{\phione}[1][]{\phiany[#1]_1}
\newcommand{\phitwo}[1][]{\phiany[#1]_2}

\newcommand{\phiiR}[1][]{\phi_{i,R}^{#1}}

\newcommand{\phitildeiR}[1][]{\phitilde_{i,R}^{#1}}

\newcommand{\phitildeany}[1][]{\phitilde^{#1}}

\newcommand{\phitildeone}[1][]{\phitildeany[#1]_1}
\newcommand{\phitildetwo}[1][]{\phitildeany[#1]_2}
\newcommand{\phitildeveci}{\bm{\tilde{\phi}}_i}
\newcommand{\phitildevecone}{\bm{\tilde{\phi}}_1}
\newcommand{\phitildevectwo}{\bm{\tilde{\phi}}_2}

\makeatletter
\newcommand{\Qtensor}[1]{Q^{\@ifempty{#1}{abcd}{#1}}}
\newcommand{\kroneckerdelta}[1]{\delta^{\@ifempty{#1}{ab}{#1}}}
\makeatother

\newcommand{\OC}{\mathcal{O}}

\newcommand{\diagramcontribution}[1]{\mathcal{I}_{#1}}
\newcommand{\diagram}{\mathcal{J}}

\newcommand{\muepspos}{\mu^{\epsilon}}
\newcommand{\muepsneg}{\mu^{-\epsilon}}
\newcommand{\Gammafactor}{\Gamma\left( \epsilon/2 \right)}

\newcommand{\loopfactor}{\frac{\Gammafactor}{(4\pi)^{d/2}}}

\newcommand{\numcomponent}{n}

\makeatletter
\newcommand{\invproponefunc}[1]{\Gamma_{\phitilde_1 \phi_1}(\@ifempty{#1}{\kvec, \omega}{#1})}
\newcommand{\invproptwofunc}[1]{\Gamma_{\phitilde_2 \phi_2}(\@ifempty{#1}{\kvec, \omega}{#1})}
\newcommand{\invpropifunc}[1]{\Gamma_{\phitilde_i \phi_i}(\@ifempty{#1}{\kvec, \omega}{#1})}

\newcommand{\invpropiRfunc}[1]{\Gamma_{R, \phitilde_i \phi_i}\left(\@ifempty{#1}{\kvec, \omega}{#1}\right)}
\newcommand{\invpropiRhat}[1]{\hat{\Gamma}_{R, \phitilde_i \phi_i}\left(\@ifempty{#1}{\kvec, \omega}{#1}\right)}

\newcommand{\vertexnoiseonefunc}[1]{\Gamma_{\phitilde_1 \phitilde_1}\left(\@ifempty{#1}{\kvec, \omega}{#1}\right)}
\newcommand{\vertexnoisetwofunc}[1]{\Gamma_{\phitilde_2 \phitilde_2}(\@ifempty{#1}{\kvec, \omega}{#1})}
\newcommand{\vertexnoiseifunc}[1]{\Gamma_{\phitilde_i \phitilde_i}(\@ifempty{#1}{\kvec, \omega}{#1})}
\newcommand{\vertexnoiseiRfunc}[1]{\Gamma_{R, \phitilde_i \phitilde_i}(\@ifempty{#1}{\kvec, \omega}{#1})}

\newcommand{\invpropone}[2]{\Gamma_{\phitilde_1 \phi_1}^{\@ifempty{#1}{ab}{#1}}(\@ifempty{#2}{\kvec, \omega}{#2})}
\newcommand{\invproptwo}[2]{\Gamma_{\phitilde_2 \phi_2}^{\@ifempty{#1}{ab}{#1}}(\@ifempty{#2}{\kvec, \omega}{#2})}
\newcommand{\vertexnoiseone}[2]{\Gamma_{\phitilde_1 \phitilde_1}^{\@ifempty{#1}{ab}{#1}}(\@ifempty{#2}{\kvec, \omega}{#2})}
\newcommand{\vertexnoisetwo}[2]{\Gamma_{\phitilde_2 \phitilde_2}^{\@ifempty{#1}{ab}{#1}}(\@ifempty{#2}{\kvec, \omega}{#2})}
\newcommand{\vertexuone}[2]{\Gamma_{\phitildeone \phione \phione \phione  }^{\@ifempty{#1}{abcd}{#1}}(\@ifempty{#2}{\kvec}{#2})}
\newcommand{\vertexutwo}[2]{\Gamma_{\phitildetwo \phitwo \phitwo \phitwo  }^{\@ifempty{#1}{abcd}{#1}}(\@ifempty{#2}{\kvec}{#2})}
\newcommand{\vertexgone}[2]{\Gamma_{\phitildeone \phione \phitwo \phitwo  }^{\@ifempty{#1}{abcd}{#1}}(\@ifempty{#2}{\kvec}{#2})}
\newcommand{\vertexgtwo}[2]{\Gamma_{\phitildetwo \phitwo \phione \phione  }^{\@ifempty{#1}{abcd}{#1}}(\@ifempty{#2}{\kvec}{#2})}
\newcommand{\vertexquartic}[2]{\Gamma_{\phitilde_i \phi_i \phi_j \phi_j  }^{\@ifempty{#1}{abcd}{#1}}(\@ifempty{#2}{\kvec}{#2})}

\newcommand{\invpropi}[2]{\Gamma_{\phitilde_i \phi_i}^{\@ifempty{#1}{ab}{#1}}(\@ifempty{#2}{\kvec, \omega}{#2})}
\newcommand{\vertexnoisei}[2]{\Gamma_{\phitilde_i \phitilde_i}^{\@ifempty{#1}{ab}{#1}}(\@ifempty{#2}{\kvec, \omega}{#2})}
\newcommand{\vertexui}[2]{\Gamma_{\phitilde_i \phi_i \phi_i \phi_i  }^{\@ifempty{#1}{abcd}{#1}}(\@ifempty{#2}{\kvec}{#2})}

\newcommand{\invproprenone}[2]{\Gamma_{R, \phitildeone \phione}^{\@ifempty{#1}{ab}{#1}}(\@ifempty{#2}{\kvec, \omega}{#2})}
\newcommand{\invproprentwo}[2]{\Gamma_{R, \phitildetwo \phitwo}^{\@ifempty{#1}{ab}{#1}}(\@ifempty{#2}{\kvec, \omega}{#2})}
\newcommand{\vertexnoiserenone}[2]{\Gamma_{R, \phitildeone \phitildeone}^{\@ifempty{#1}{ab}{#1}}(\@ifempty{#2}{\kvec, \omega}{#2})}
\newcommand{\vertexnoiserentwo}[2]{\Gamma_{R, \phitildetwo \phitildetwo}^{\@ifempty{#1}{ab}{#1}}(\@ifempty{#2}{\kvec, \omega}{#2})}
\newcommand{\vertexurenone}[2]{\Gamma_{R, \phitildeone \phione \phione \phione}^{\@ifempty{#1}{abcd}{#1}}(\@ifempty{#2}{\kvec}{#2})}
\newcommand{\vertexurentwo}[2]{\Gamma_{R, \phitildetwo \phitwo \phitwo \phitwo}^{\@ifempty{#1}{abcd}{#1}}(\@ifempty{#2}{\kvec}{#2})}
\newcommand{\vertexrengone}[2]{\Gamma_{R, \phitildeone \phione \phitwo \phitwo}^{\@ifempty{#1}{abcd}{#1}}(\@ifempty{#2}{\kvec}{#2})}
\newcommand{\vertexrengtwo}[2]{\Gamma_{R,\phitildetwo \phitwo \phione \phione}^{\@ifempty{#1}{ab}{#1}}(\@ifempty{#2}{\kvec, \omega}{#2})}
\newcommand{\vertexquarticren}[2]{\Gamma_{R, \phitilde_i \phi_i \phi_j \phi_j }^{\@ifempty{#1}{abcd}{#1}}(\@ifempty{#2}{\kvec, \omega}{#2})}

\newcommand{\etatilde}{\tilde{\eta}}

\newcommand{\invpropiren}[2]{\Gamma_{R, \phitilde_i \phi_i}^{\@ifempty{#1}{ab}{#1}}\left(\@ifempty{#2}{\kvec, \omega}{#2}\right)}
\newcommand{\vertexnoiseiren}[2]{\Gamma_{R, \phitilde_i \phitilde_i}^{\@ifempty{#1}{ab}{#1}}(\@ifempty{#2}{\kvec, \omega}{#2})}
\newcommand{\vertexuiren}[2]{\Gamma_{R, \phitilde_i \phi_i \phi_i \phi_i  }^{\@ifempty{#1}{abcd}{#1}}(\@ifempty{#2}{\kvec}{#2})}

\makeatother

\newcommand{\bra}[1]{\left \langle #1 \right|}
\newcommand{\ket}[1]{\left | #1 \right \rangle}

\newcommand{\transpose}{\intercal}
\newcommand{\ave}[1]{\left \langle #1 \right \rangle}
\newcommand{\ident}{\mathds{1}}

\newcommand{\Done}{D_1}
\newcommand{\Dtwo}{D_2}
\newcommand{\rone}{r_1}
\newcommand{\rtwo}{r_2}
\newcommand{\tauone}{\tau_1}
\newcommand{\tautwo}{\tau_2}
\newcommand{\uone}{u_1}
\newcommand{\utwo}{u_2}
\newcommand{\gonetwo}{g_{12}}
\newcommand{\gtwoone}{g_{21}}

\newcommand{\tauRat}{t}

\newcommand{\utildei}{\tilde{u}_i}
\newcommand{\utildeiR}{\tilde{u}_{i,R}}
\newcommand{\utildeone}{\tilde{u}_1}
\newcommand{\utildetwo}{\tilde{u}_2}
\newcommand{\utilde}{\tilde{u}}
\newcommand{\gtildeonetwo}{\tilde{g}_{12}}
\newcommand{\gtildetwoone}{\tilde{g}_{21}}

\newcommand{\utildeoneR}{\tilde{u}_{1,R}}
\newcommand{\utildetwoR}{\tilde{u}_{2,R}}
\newcommand{\gtildeonetwoR}{\tilde{g}_{12,R}}
\newcommand{\gtildetwooneR}{\tilde{g}_{21,R}}

\newcommand{\utildeonestar}{\tilde{u}^*_{1,R}}
\newcommand{\utildetwostar}{\tilde{u}^*_{2,R}}
\newcommand{\utildestar}{\utilde^{*}}
\newcommand{\gtildeonetwostar}{\tilde{g}^*_{12,R}}

\newcommand{\wstar}{w^*_R}
\newcommand{\sigmastar}{\sigma^{*}_{R}}

\newcommand{\Zphione}{Z_{\phivec_1}}
\newcommand{\Zphitwo}{Z_{\phivec_2}}
\newcommand{\Zphitildeone}{Z_{\phitildevecone}}
\newcommand{\Zphitildetwo}{Z_{\phitildevectwo}}
\newcommand{\ZDone}{Z_{D_1}}
\newcommand{\ZDtwo}{Z_{D_2}}
\newcommand{\Zrone}{Z_{\tau_1}}
\newcommand{\Zrtwo}{Z_{\tau_2}}

\newcommand{\Zuone}{Z_{u_1}}
\newcommand{\Zutwo}{Z_{u_2}}
\newcommand{\Zgonetwo}{Z_{g_{12}}}
\newcommand{\Zgtwoone}{Z_{g_{21}}}

\newcommand{\logzutildei}{\ln Z_{\tilde{u}_i}}
\newcommand{\logzgtildeonetwo}{\ln Z_{\tilde{g}_{12}}}
\newcommand{\logzgtildetwoone}{\ln Z_{\tilde{g}_{21}}}

\newcommand{\gammaphiistar}{\gamma^{*}_{\phi_i}}

\newcommand{\gammaphitildeistar}{\gamma^{*}_{\tilde{\phi}_i}}
\newcommand{\gammaDonestar}{\gamma^*_{D_1}}
\newcommand{\gammaDtwostar}{\gamma^*_{D_2}}

\newcommand{\gammaGammaonestar}{\gamma^*_{\Gamma_1}}
\newcommand{\gammaGammatwostar}{\gamma^*_{\Gamma_2}}

\newcommand{\gammatautildeonestar}{\gamma^{*}_{\tautilde_1}}
\newcommand{\gammatautildetwostar}{\gamma^{*}_{\tautilde_2}}

\newcommand{\betautildei}{\beta_{\tilde{u}_i}}
\newcommand{\betautildeone}{\beta_{\tilde{u}_1}}
\newcommand{\betautildetwo}{\beta_{\tilde{u}_2}}
\newcommand{\betaw}{\beta_w}
\newcommand{\betagtildeonetwo}{\beta_{\tilde{g}_{12}}}
\newcommand{\betagtildetwoone}{\beta_{\tilde{g}_{21}}}

\newcommand{\kint}{\int \dbar^d k}

\newcommand{\omegaint}{\int \dbar \omega}

\newcommand{\xint}{\int d^d x}
\newcommand{\tint}{\int dt}

\newcommand{\deltaik}{\Delta_i(\kvec)}

\newcommand{\deltaonek}{\Delta_1(\kvec)}

\newcommand{\deltatwok}{\Delta_2(\kvec)}

\newcommand{\imag}{\mathring{\imath}}
\newcommand{\etal}{\textit{et al.}~}
\newcommand{\eg}{\textit{e.g.}~}
\newcommand{\ie}{\textit{i.e.}~}

\usepackage{color}
\definecolor{darkgreen}{rgb}{0,0.6,0}
\definecolor{darkblue}{rgb}{0,0,0.6}
\definecolor{darkred}{rgb}{0.6,0,0}
\definecolor{darkpurple}{rgb}{0.5,0,0.5}
\hypersetup{
bookmarksopen=true,
pdftitle="",
pdfauthor="", 
pdftoolbar=false, 
pdfstartview={FitH},		
pdfmenubar=true,			
pdfhighlight=/O,			
colorlinks=true,			
urlcolor=darkblue,
citecolor=darkblue,		
linkcolor=darkblue}

\makeatletter
\newcommand{\customlabel}[2]{%
   \protected@write \@auxout {}{\string \newlabel {#1}{{#2}{\thepage}{#2}{#1}{}} }%
   \hypertarget{#1}{\hspace{0pt}}
}
\makeatother

\input{diagrams_small}

\begin{document}


\newcommand{\titleText}{Critical Dynamics of Non-Reciprocally Coupled Conserved Systems}
\title{\titleText}

\author{Emir Sezik}%
 \email{emir.sezik19@imperial.ac.uk}
\affiliation{%
Department of Mathematics
and Centre of Complexity Science, 
Imperial College London, London SW7 2AZ, United Kingdom}%

\author{Gunnar Pruessner}
\email{g.pruessner@imperial.ac.uk}
\affiliation{%
Department of Mathematics
and Centre of Complexity Science, 
Imperial College London, London SW7 2AZ, United Kingdom}%

\date{\today}

\begin{abstract}
Non-reciprocal systems have been shown to sustain time-dependent patterns, most prominently travelling waves. The transition into these time-dependent states generally breaks time-translational invariance, representing a clear deviation from equilibrium dynamics. Though common implementations of non-reciprocity lead to such phenomenology, these spatio-temporal patterns are absent in other models. In the same vein, the ensuing scaling behaviour also depends on the precise way non-reciprocity is implemented. To better understand the effects of different non-reciprocal interactions, we study the critical conserved dynamics of non-reciprocally coupled spin systems. Specifically, we consider the dynamics of two $n$-component order parameter fields $\phiveci$ with $i \in\{1,2\}$. Unlike the common implementations of non-reciprocal interactions, we introduce the non-reciprocity solely through the non-linear interaction between the distinct species. Using the field-theoretic renormalisation group (RG) procedure, we perform a one-loop analysis and show that at one-loop level, the critical behaviour depends on the microscopic value of certain quantities. Using the flow functions, we elucidate the behaviour of the fixed points for different bare microscopic values. We also show that for $n \geq 4$, there is a fixed point where the ensuing critical dynamics asymptotically obey detailed-balance, implying the emergent dynamics are agnostic to the microscopic non-reciprocity on large scales. Finally, we show that the conserved dynamics reduces the number of independent scaling exponents, mimicking the effect of a standard fluctuation-dissipation relation.
\end{abstract}

                              
\maketitle

\newcommand{\longtodo}[1]{\todo[inline,size=\tiny]{#1}}

\section{Introduction}
Recently, non-reciprocal interactions have attracted much interest from the active matter community for its relevance in out-of-equilibrium systems \cite{fruchart_non-reciprocal_2021, doi:10.1073/pnas.2010318117, saha_pairing_2019, yllanes_how_2017,shankar_topological_2022}. They have been shown to possess an intriguing phenomenology ranging from synchronisation \cite{risler_universal_2004, risler_universal_2005} to the formation of "time crystals" \cite{daviet_kardar-parisi-zhang_2024}. Broadly, non-reciprocity refers to the notion that two entities interact in a way that breaks detailed balance (or Newton's Third Law). Such interactions can even arise microscopically \cite{PhysRevX.12.010501,marchetti_nr_pattern_formation,doi:10.1073/pnas.2010318117} in the context of social interactions (i.e. predator-prey systems) \cite{reichenbach_mobility_2007} or solids with odd elasticities \cite{scheibner_odd_2020}. Even though there are a myriad ways of implementing non-reciprocity in different contexts \cite{lorenzana_non-reciprocal_2024, liu_dynamic_2025,PhysRevE.101.052601, avni_non-reciprocal_2024,loos_long-range_2023, rouzaire_non-reciprocal_2024,avni_dynamical_2024,saha_scalar_2020}, most implementations lead to the same physics: dynamical steady states with travelling patterns, most prominently waves \cite{PhysRevX.12.010501, marchetti_nr_pattern_formation, doi:10.1073/pnas.2010318117, saha_scalar_2020, avni_dynamical_2024, garcia-morales_complex_2012,doi:10.1073/pnas.2407705121, johnsrud_phase_2025} that break time-translational invariance. However, this is not the only phenomenology contained in non-reciprocal interactions. Other implementations \cite{young_nonequilibrium_2020, young_nonequilibrium_2024} do not lead to spatio-temporal patterns, but to generic condensation behaviour in late times. 

In a similar vein, the ensuing scaling behaviour also depends on the precise way the non-reciprocity is implemented. For systems where the non-reciprocity induces ``run-and-chase" dynamics between species (where the species are coupled at least at the linear level), the transition into these states have been shown to be \emph{effectively thermal} \cite{risler_universal_2004, risler_universal_2005, tauber_perturbative_2014, sieberer_nonequilibrium_2014} with most of the scaling exponents, except a novel one, being that of equilibrium Model A (or Model~B if the order parameters are conserved) of Halperin's classification \cite{halperin_calculation_1972}. This is no longer true, however, when the non-reciprocity enters only through the non-linear coupling between the species \cite{young_nonequilibrium_2020, young_nonequilibrium_2024}. Here, it was argued that a true non-equilibrium fixed point emerges with strong dynamical scaling, where both fields display critical long-ranged fluctuations, with a distinct universality class. These varied results, seemingly in contradiction with universality, stem from the different underlying symmetries of the respective models. Different non-reciprocal interactions encompass different symmetries which in turn cause different critical behaviour. The systems in Refs.~\cite{tauber_perturbative_2014,saha_scalar_2020,johnsrud_phase_2025,pisegna2025nonreciprocalmixturessuspensionrole}, which include the most common implementations of non-reciprocal interactions, exhibit rotational symmetry characterized by the $U(1)\sim SO(2)$ group, whereas the system in Ref. \cite{young_nonequilibrium_2020} possesses only $\mathbb{Z}_2 \otimes \mathbb{Z}_2$, a far more restrictive symmetry. Therefore, exploring how different types of non-reciprocal interactions and symmetries affect the scaling behaviour of the system during the transition remains an intriguing research avenue. 

Motivated by this, we investigate the critical dynamics of two conserved order parameter fields $\phiveci$ where $i \in \{1,2\}$ with $\symgroup$ symmetry. We couple the spins non-reciprocally only at the non-linear level, in order to understand the behaviour of such couplings. By performing a perturbative renormalisation group (RG) calculation about the upper critical dimension, we show that to one loop, the fixed points depend on bare parameters of the model. This implies that to unveil the full universality class of the model, one needs to perform a \emph{full} $2-$loop analysis where every parameter in the model receives a non-trivial correction. A similar issue persists in the models considered in Refs. \cite{young_nonequilibrium_2020, young_nonequilibrium_2024}. However, the authors of Refs. \cite{young_nonequilibrium_2020, young_nonequilibrium_2024} only perform a partial $2-$loop analysis and we argue below that their results depend on bare parameters as well. We compute the scaling exponents of this model at this new critical point and show that for $n<4$, the universality class is distinct from equilibrium Model~B. Furthermore, we show that even though the system is out of equilibrium, the conserved dynamics ensures the anomalous exponents of correlation and response function to be the same, mimicking the effect of standard fluctuation-dissipation relations.

The rest of the paper is organised as follows. In Sec. \ref{sec:setup} we introduce the model, identify the critical points at the mean-field level and derive the field theory as well as the elements of the perturbative expansion. In Sec. \ref{sec:RG}, we introduce the central objects of field-theoretic renormalisation, perform the renormalisation group calculation and use these results to calculate the flow functions. These are then used to explore the stable fixed points of the RG flow in Section \ref{sec:criticalpoints}. We calculate the ensuing scaling exponents at these fixed points in Sec. \ref{sec:scalingexp}. Finally, we summarise and conclude in Sec. \ref{sec:conclusion}.

\section{Setup} \label{sec:setup}
In this section, we introduce our model, its associated field theory, and the elements of the perturbative expansion. 

\subsection{Model} \label{eq:model}
We are interested in the conserved relaxational dynamics of two local order parameter vector fields $\phivec_1,\phivec_2  \in \mathbb{R}^{n}$ in $d$-dimensional space $\xvec\in\Rset^d$, given by the Langevin equations
\begin{subequations} \elabel{Model}
\begin{align}
    \dot{\phivec}_1(\xvec,t) &= \nabla^2 \frac{\delta \Hamiltonian_1}{\delta \phivec_1(\xvec,t)} + \frac{\gonetwo}{2}\nabla^2 \Big( \phivec_1(\xvec,t) \phivec_2(\xvec,t) \cdot \phivec_2(\xvec,t)\Big) + \sqrt{2 \Gamma_1} \xivec_1(\xvec,t) \elabel{eom_one} \\
    \dot{\phivec}_2(\xvec,t) &= \nabla^2 \frac{\delta \Hamiltonian_2}{\delta \phivec_2(\xvec,t)} + \frac{\gtwoone}{2}\nablasquared \Big( \phivec_2(\xvec,t) \phivec_1(\xvec,t) \cdot \phivec_1(\xvec,t)\Big) + \sqrt{2 \Gamma_2} \xivec_2(\xvec,t) \elabel{eom_two}
\end{align}
\end{subequations}
where the noises $\xivec_1,\xivec_2\in\Rset^n$ are identically and independently distributed Gaussian random variables satisfying 
\begin{subequations} \elabel{noises}
\begin{align}
\ave{\xivec_i(\xvec,t)}&=0\\
\ave{\xivec_i(\xvec,t)\xivec^{\transpose}_j(\xvec',t') } &=- \ident_n \delta_{ij}  \nablasquared \delta(\xvec - \xvec') \delta(t-t') \ ,
\end{align}
\end{subequations}
with $\ident_n$ the $n\times n$ identity.
In the following, we will use subscript $i,j\in\{1,2\}$ to index species and, if needed,
superscript latin indices $a,b,c,d\in\{1,\ldots,n\}$ to index components of vectors in $\Rset^n$, such as $\phivec_1=(\phi_1^1,\phi_1^2,\ldots,\phi_1^n)^\transpose$. In this notation, the noise correlator is written $\ave{\xi^a_i(\xvec,t)\xi^b_j(\xvec',t') } = - \delta_{ij} \delta^{ab} \nablasquared \delta(\xvec - \xvec') \delta(t-t')$. Throughout, we will make use of the Einstein summation convention, where repeated indices are implicitly being summed over, in summations involving field components $a,b$. 

\Erefs{Model} corresponds to the conserved dynamics of the non-reciprocally coupled spins studied in \cite{young_nonequilibrium_2020, young_nonequilibrium_2024}. The right-hand sides of \Erefs{Model} comprise three terms: Firstly, a single-species Hamiltonian, secondly an interaction between fields $\phivec_1,\phivec_2$ and thirdly a conserved noise with strength $\Gamma_i$. By inspection of \Erefs{Model} and \eref{noises}, both order parameter fields are globally conserved. The Hamiltonians $\Hamiltonian_1,\Hamiltonian_2$ contain  single-species terms and are defined as
\begin{equation} \elabel{def_Hamiltonian}
    \Hamiltonian[i] = \int \ddint{x} \left( \frac{r_i}{2} \phivec_i^2 + \frac{D_i}{2} (\nabla \phivec_i)^2 + \frac{u_i}{4!} (\phivec_i\cdot \phivec_i)^2\right)
\end{equation}
with 
\begin{equation}
(\nabla \phivec_i)^2
=\sum_{a=1}^n\sum_{\mu=1}^d 
  \left(\frac{\partial}{\partial x_\mu}
  \phi_i^a
  \right)^2
\end{equation}
where we use the Greek index $\mu$ for the $d$ components of $\nabla$. The dot-product $(\phivec_i\cdot \phivec_i)$ multiplies two vectors in the ``spin-space'' $\Rset^n$ and is henceforth reserved for projections in the spin-space. Computing the functional derivatives acting on the fields in \Erefs{Model}, we arrive at the following Langevin equations:
\begin{subequations} \elabel{Model2}
    \begin{gather}
        \dot{\phivec_1} = \nablasquared\left[\rone - \Done \nablasquared + \frac{\uone}{3!} \phivec_1 \cdot \phivec_1 + \frac{\gonetwo}{2} \phivec_2 \cdot \phivec_2 \right]\phivec_1 + \sqrt{2\Gamma_1} \xivec_1 \\ 
        \dot{\phivec_2} = \nablasquared\left[\rtwo - \Dtwo \nablasquared + \frac{\utwo}{3!} \phivec_2 \cdot \phivec_2 + \frac{\gtwoone}{2} \phivec_1 \cdot \phivec_1 \right]\phivec_2 + \sqrt{2\Gamma_2} \xivec_2
    \end{gather}
\end{subequations}

In the Hamiltonians \Eref{def_Hamiltonian} the mass $r_i$ and the non-linearity $u_i$ quantify the harmonic and the quartic part of the local potential of the $\phivec_i$ field. After Fourier-transforming, $u_i$ will mediate the single species interaction between the modes $\kvec$ of $\phivec_i$. The $D_i$ characterise the diffusivity of the fields. Individually, these systems are stable only for $u_i>0$. Both Hamiltonians are isotropic as $\Hamiltonian[i]$ do not change under global rotations $O(n)$ of $\phivec_i$. In fact, $\dot{\phivec}_1$
in \Eref{eom_one} is invariant under rotations of $\phivec_2$, because $\phivec_2$ enters into $\dot{\phivec}_1$ only by magnitude. Similarly, $\dot{\phivec}_2$ of \Eref{eom_two} is invariant under rotations of $\phivec_1$. Furthermore, the statistics of the pair of local order parameter fields $\phivec_1,\phivec_2$ is unchanged under global rotations $O(n)\otimes O(n)$.

For particular choices of the parameters, \Erefs{Model} describe an equilibrium system \cite{multicritical_ising, ft_of_bicritical_moser_1, ft_of_bicritical_moser_2, PhysRevE.88.042141}. The condition for equilibrium is most easily identified by considering the Boltzmann distribution $\exp{-\SuperHamiltonian}$ such that 
\begin{subequations}\elabel{Langevin_from_SuperHamiltonian}
    \begin{align}
    \dot{\phivec}_1(\xvec,t) &= \Gamma_1\nabla^2 \frac{\delta \SuperHamiltonian}{\delta \phivec_1(\xvec,t)}+ \sqrt{2 \Gamma_1} \xivec_1(\xvec,t)\\
    \dot{\phivec}_2(\xvec,t) &= \Gamma_2\nabla^2 \frac{\delta \SuperHamiltonian}{\delta \phivec_2(\xvec,t)} + \sqrt{2 \Gamma_2} \xivec_2(\xvec,t)
    \end{align}
\end{subequations}
which implies
\begin{equation}\elabel{def_SuperHamiltonian}
    \SuperHamiltonian = \frac{\Hamiltonian[1]}{\Gamma_1}
    +
    \frac{\Hamiltonian[2]}{\Gamma_2}
    + 
    \int\ddint{x} \frac{\gonetwo}{4\Gamma_1}\phivec_1^2\phivec_2^2
\end{equation}
using \Eref{def_Hamiltonian}, \Erefs{Langevin_from_SuperHamiltonian} and \eref{def_SuperHamiltonian} immediately reproduce \Eref{eom_one}, but \Eref{eom_two} if and only if $\gonetwo\Gamma_2=\gtwoone \Gamma_1$. This condition is trivially fulfilled if $\gonetwo$ and $\gtwoone$ both vanish, such that the two fields $\phi_1$ and $\phi_2$ do not interact. In that case, their dynamics corresponds to that of two \emph{isolated} systems, possibly at different temperatures $\Gamma_{1,2}$. When $\gonetwo$ does not vanish, equilibrium is obtained only when the two systems have effectively, namely concerning their coupling, the same temperature $\Gamma_2=\Gamma_1\gtwoone/\gonetwo$. To capture the equilibrium condition more succinctly, we introduce the ratio $\sigma$, such that 
\begin{equation}\elabel{def_sigma}
    \frac{\gtwoone}{\Gamma_2} = \sigma\frac{\gonetwo}{\Gamma_1} 
\end{equation}
\citeauthor{young_nonequilibrium_2024} introduced a re-parameterisation that renders $\gtwoone/\gonetwo\in\{-1,0,1\}$, making $\sigma$ a ratio of $\Gamma_1$ and $\Gamma_2$ or simply $0$, but the terms parameterised by $\gonetwo$ and $\gtwoone$ in \Eref{Model} generally acquire different corrections under renormalisation, breaking effectively $\gtwoone/\gonetwo\in\{-1,0,1\}$ beyond the tree level.

We expect qualitatively different behaviour depending on the signs of $\gtwoone$ and $\gonetwo$. If they have the same sign, then the phenomenology might not be too far from equilibrium. If one coupling vanishes, the corresponding field displays standard Model~B behaviour, while the other is still subject to its fluctuations. However, if they have opposite signs, then one field reduces and one field increases the mass of the other field. Such a ``truly non-reciprocal setup'' is not commonly studied in the literature \cite{saha_scalar_2020, NRCahnHilliarddefect, johnsrud_phase_2025, doi:10.1073/pnas.2407705121,doi:10.1073/pnas.2010318117,marchetti_nr_pattern_formation,fruchart_non-reciprocal_2021}, where the non-reciprocity enters linearly and thereby induces run-and-chase dynamics. The critical dynamics of such models has been analysed in \cite{tauber_perturbative_2014}, where the effective dynamics are governed by a complex Landau-Ginzburg equation \cite{garcia-morales_complex_2012}. It was shown that the critical dynamics of that model is effectively thermal, a situation also seen in \cite{migliorini_critical_2008, risler_universal_2004, risler_universal_2005}.

\subsection{Critical Points of the Model} \label{sec:meanfieldcritical}
In this section, we perform a mean-field analysis to unveil the phase diagram of the model. Typically, a mean-field analysis consists of finding stable homogenous configurations $\phivec_i(\xvec,t) =  \varphivec_i$ that solve \Erefs{Model2} in the limit of vanishing noise. However, due to the $\nablasquared$ prefactor, \emph{any} homogeneous configuration will trivially solve \Erefs{Model2} in the zero-noise limit. 

Furthermore, due to the conserved nature of the dynamics, the total field
\begin{equation}
    \Phivec_i(t) = \int \ddint{x} \phivec_i(\xvec,t),
\end{equation}
is set by the initialisation and does not change in time, 
which readily follows from integrating \Erefs{Model2} over all space. In the following mean-field analysis we choose $\Phivec_i =  \zerovec$, physically corresponding to an initial configuration with zero overall ``magnetisation'' at all times. Any spontaneous symmetry breaking will need to happen locally: the field will ``phase separate'', whereby different regions of space will have different magnetisation direction joined together by possibly broad ``interfaces'' while the total global magnetisation stays at $\Phivec_i =  \zerovec$.

We will thus consider in the following a groundstate of the form
\begin{equation} \label{eq:mean_field_ansatz}
    \phivec_i(\xvec,t) = \varphi_i \dvec_i(\xvec),
\end{equation}
where $\varphi_i \in \Rset^{0,+}$ is the global magnitude of the magnetisation and $\dvec_i(\xvec)$ is the director, a unit vector characterising the spatial modulation of the magnetisation.
This director $\dvec_i(\xvec)$ must satisfy
\begin{equation}\elabel{integral_dvec}
    \int \ddint{x} \dvec_i(\xvec) = \zerovec, 
\end{equation}
to ensure that the total magnetisation $\Phivec_i$ vanishes. 

The nature of the configuration, essentially defined by the director, is dependent on the number of field components $n$. For $n>1$, the system can sustain spin waves, which have minimal energetic cost. However, for $n=1$, such a smooth ``interface'' is no longer possible and local phases are separated by a sharp boundaries, also known as a domain walla. In this case, $d_i(\xvec)$ may be the sign of one component of $\xvec$, say $d_i(\xvec) = \text{sgn}(x^{1})$ which satisfies \Eref{integral_dvec}. Though the nature of the stable configurations for $n=1$ seems different to that of $n>1$, which is what we will focus on in the following, the subsequent mean-field analysis performed will hold true for all $n\ge1$. 

The phases of the model, and consequently the critical behaviour upon crossing different phases, is controlled by the values of $r_1$ and $r_2$. 
To characterise the stable ground states of the model \Erefs{Model}, we consider solutions of the form \Eref{mean_field_ansatz}. 
Specifically, for a finite system of length $L$ and volume $V = L^d$, they are given by spin waves with the smallest wave-vector $|\qvec| = 2\pi/L$, satisfying $\nabla^2 \dvec_i(\xvec) = -(2\pi/L)^2 \dvec_i(\xvec)$. Inserting this into \Erefs{Model2} and using the unit vector property, we find that the $\varphi_i$ satisfy:
\begin{subequations} \label{eq:Mean_field_equations}
    \begin{align}
        \left(\frac{2\pi}{L} \right)^2 \left[ r_1  + D_1\left(\frac{2\pi}{L} \right)^2  + \frac{u_1}{6} \varphi_1^2 + \frac{g_{12}}{2}\varphi_2^2\right]\varphi_1 \dvec_1(\xvec) = 0 \\
        \left(\frac{2\pi}{L} \right)^2 \left[ r_2  + D_2\left(\frac{2\pi}{L} \right)^2  + \frac{u_2}{6} \varphi_2^2 + \frac{g_{21}}{2}\varphi_1^2\right]\varphi_2 \dvec_2(\xvec) = 0
    \end{align}
\end{subequations}
We see that the existence of the spatial variation of the director adds an energetic cost that scales like $D_i/L^2$, but vanishes in the thermodynamics limit, $L \to \infty$. We are interested in the values of $\varphi_i$ in the thermodynamic limit. The analysis simplifies if we assume $u_i>0$, which is justified as otherwise the local potential allows for configurations where at least one of the $\varphi_i$ diverges. There is no such constraint on $g_{12}$ and $g_{21}$. Defining $\varphitilde_i = \sqrt{u_i/6} \varphi_i$, as well as $g_1 = 3g_{12}/u_2$ and $g_2 = 3g_{21}/u_1$, \Eref{Mean_field_equations} gives in the thermodynamic limit
\begin{subequations} \label{eq:Mean_field_equations_simplified}
    \begin{align}
       \left( r_1  + \varphitilde_1^2 + g_1\varphi_2^2\right)\varphitilde_1  &= 0 \\
       \left( r_2  + \varphitilde_2^2 + g_2\varphi_1^2\right)\varphitilde_2  &= 0
\ .
    \end{align}
\end{subequations} 
Solving \Erefs{Mean_field_equations_simplified} yields several possible configurations for the fields. However, not all of them are viable as they are unstable against small perturbations. To find the stable solutions, we perform a linear stability analysis whereby we perturb  the solutions by a small amount, $\varphitilde_i \to \varphitilde_i + \delta \varphitilde_i$, and analyse the resulting linear force. Plugging this perturbation in \Erefs{Mean_field_equations_simplified} and retaining terms to linear order in $\delta \varphitilde_i$, we find that the stability of the solutions are governed by the matrix
\begin{equation} \label{eq:stability_mat}
    \mathcal{M} = \begin{pmatrix}
        r_1 + 3\varphitilde_1^2 + g_1 \varphitilde^2_2 & 2g_1 \varphitilde_1 \varphitilde_2 \\
        2g_2 \varphitilde_1 \varphitilde_2 & r_2 + 3\varphitilde_2^2 + g_2 \varphitilde^2_1
    \end{pmatrix}\ .
\end{equation}
For the solutions to be stable, we require the matrix $\mathcal{M}$ to have only positive eigenvalues. 

The $\varphitilde_i$ solving \Erefs{Mean_field_equations_simplified} can be found easily. There are four cases in total, depending on how many and which of the $\varphitilde_i$ vanish,
\begin{subequations} \label{eq:solutions_mean_field}
    \begin{align}
        \text{(I):}&~~~ \varphitilde_1 = 0, ~~~  \varphitilde_2 = 0 \\
        \text{(II):}&~~~ \varphitilde_1 = \pm \sqrt{-r_1}, ~~~  \varphitilde_2 = 0 \\
        \text{(III):}&~~~ \varphitilde_1 = 0, ~~~  \varphitilde_2 = \pm \sqrt{-r_2} \\
        \text{(IV):}&~~~ \varphitilde_1 = \pm\sqrt{\frac{g_1 r_2 - r_1}{1-g_1g_2}}, ~~~  \varphitilde_2 = \pm\sqrt{\frac{g_2 r_1 - r_2}{1-g_1g_2}}
    \end{align}
\end{subequations}
Without recourse to the stability, we can immediately deduce that cases~(II) and (III) require negative masses $r_1$ and $r_2$, respectively. Similarly, the final case (IV) requires $(g_1 r_2 - r_1)/(1-g_1 g_2)$ and $(g_2 r_1 - r_2)/(1-g_1 g_2)$ to be simultaneously positive. To determine the viable solutions for a given $r_1$ and $r_2$, we compute the eigenvalues of the matrix $\mathcal{M}$ associated with each cases and determine the regions, parametrised by $r_1$ and $r_2$, where each case yields positive eigenvalues for the matrix $\mathcal{M}$. After some straightforward algebra, we find: 
\begin{equation} \label{eq:phase_diagram}
    (\varphitilde_1, \varphitilde_2) = \begin{cases}
        \text{(I)}, ~\text{if}~~r_1 > 0~~ \text{and} ~~r_2  >0 \\
        \text{(II)}, ~\text{if}~~r_1 <0~~ \text{and} ~~r_2  > g_2 r_1\\
        \text{(III)}, ~\text{if}~~r_1 > g_1 r_2~~ \text{and} ~~r_2  <0\\
        \text{(IV)}, ~\text{if}~~r_1 < g_1 r_2~~ \text{and} ~~r_2  < g_2 r_1 ~~ \text{and} ~~1-g_1 g_2 > 0 \ .
    \end{cases}
\end{equation}
If both masses are positive, $r_1, r_2 >0$, the trivial solution~(I) is stable, whereas cases~(II) and (III) require at least one negative $r_i$. Though not immediately obvious, the stability condition of case~(IV) cannot be met when $r_1,r_2 > 0$. Therefore, non-trivial solutions start appearing when at least one of the masses $r_i$ is negative. \Eref{phase_diagram} also suggests that the topology of the phase diagram depend on the values of $g_1$ and $g_2$. Indeed, there are $6$ cases to consider, depending on the sign of each $g_1$, $g_2$ and $1-g_1 g_2$, as illustrated in \Fref{phase_diagram}. We find that if the fields are coupled non-reciprocally, \ie $g_1g_2<0$, the phase defined by case~(IV) always exists, as opposed to equilibrium, where its existence is confined to particular values of $g_1g_2$.
\begin{figure}
    \centering
    \includegraphics[width=0.9\linewidth]{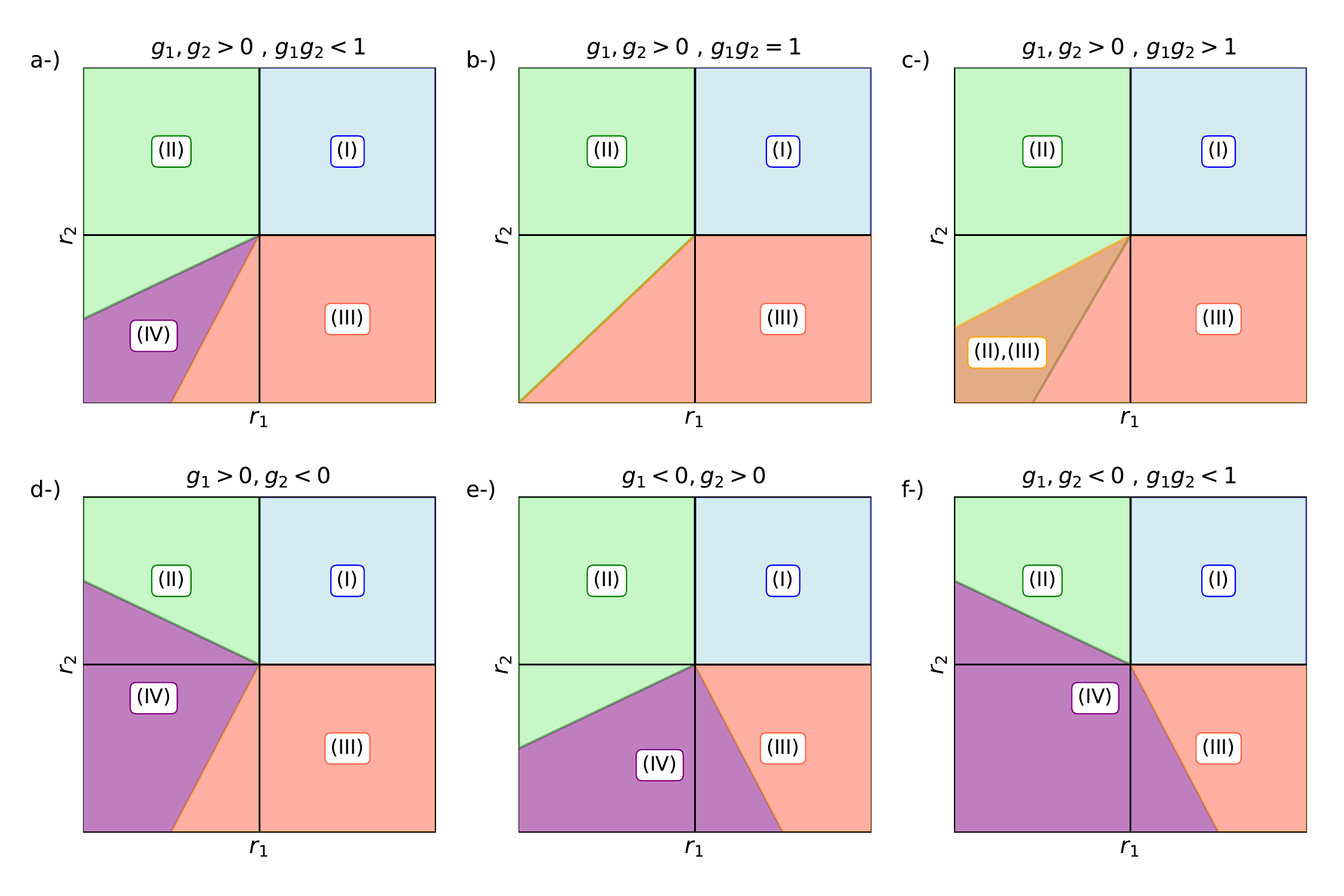}
    \caption{The mean-field phase diagram of the model for different values of $g_1$ and $g_2$ where the Roman numerals correspond to the phases defined by \Erefs{solutions_mean_field}. The qualitative features of the phase diagram depend on the conditions satisfied by $g_1$ and $g_2$, \Erefs{phase_diagram}. Phase (IV) of both fields non-vanishing, exists ony when for $g_1 g_2 < 1$. }
    \label{fig:phase_diagram}
\end{figure}

The phase diagrams in \Fref{phase_diagram} each display four phase boundaries (two of which coalesce to one when $g_1 g_2 = 1$, \Fref{phase_diagram}b), with locations that depend on the precise values of $g_1$ and $g_2$, all meeting at the tricritical point $r_1 = r_2 = 0$. Therefore, we expect there to be $5$ overall continuous phase transitions, namely the traversal of each phase boundary and the tricritical point. Except for the tricritical point, at each of the phase boundaries, exactly one field displays ordering dynamics in the presence of another field (ordered or not), exerting additional, finite-ranged noise. For all these transitions, the ordering dynamics is covered by the equilibrium Model~B behaviour of Halperin and Hohenberg's classification \cite{hohenberg_theory_1977}. This is easily seen for $r_1 \to 0$ and $r_2 > 0$ or vice versa. 
The dynamics near this critical point is described by
\begin{subequations} 
    \begin{gather}
        \dot{\phivec_1}(\xvec,t) = \nablasquared\left[\rone - \Done \nablasquared + \frac{\uone}{3!} \phivec_1(\xvec,t) \cdot \phivec_1(\xvec,t) \right]\phivec_1(\xvec,t) + \sqrt{2\Gamma_1} \xivec_1(\xvec,t) \\ 
        \phivec_2(\xvec,t) = \zerovec \ .
    \end{gather}
\end{subequations}
At this phase boundary, the short and long-ranged field decouple and the latter undergoes equilibrium Model~B dynamics. This scenario also holds at the phase boundaries where one of the fields, say $\phivec_2$, has already ordered, $r_2 < 0$, as in case~(III), \Eref{phase_diagram}. In such a regime, the $\mathcal{O}(n)$ symmetry associated with the ordered field is spontaneously broken and the effective theory is given for the Goldstone modes whose direction lie on the the $n$-dimensional sphere. On sufficiently large scales, 
we may therefore assume constant $\phivec_2(\xvec,t)\cdot \phivec_2(\xvec,t)$. 
We can then replace any instance of $|\phivec_2|^2$ in the equation of motion of $\phivec_1$ by the mean-field value $|\phivec_2|^2=-3g_{12}r_2/u_2 = -g_1r_2$, yielding
\begin{equation}
    \dot{\phivec_1}(\xvec,t) = \nablasquared\left[\left(\rone - r_2g_1 \right) - \Done \nablasquared + \frac{\uone}{3!} \phivec_1(\xvec,t) \cdot \phivec_1(\xvec,t) \right]\phivec_1(\xvec,t) + \sqrt{2\Gamma_1} \xivec_1(\xvec,t) 
\end{equation}
which displays standard Model~B behaviour when $r_1$ is tuned to its critical value, $r_{1,c} \equiv r_2g_1$. 

The only phase transition that is not covered by Model~B is therefore the tricritical point where \emph{both} fields undergo ordering dynamics and display non-Gaussian fluctuations. 
In what follows, we use renormalised field theory with minimal subtraction and dimensional regularisation with $d = \duc-\epsilon$ \cite{tauber_critical_2014, zinn-justin_quantum_2002, bellac_quantum_1992, amit_field_2005} to characterise it. To identify the upper critical dimension, $\duc=4$, of this model, we perform the usual power counting \cite{tauber_critical_2014}. 
Choosing $\dimx = L$ to be the unit of length, we demand fluctuations and noise remain invariant under spatial rescaling, \ie $[\Gamma_i] = B_i$ and $[D_i] = A_i$, \Erefs{Model} and \eref{noises}, so that the unit of time becomes $[t]= L^4/A_i$, which means that $A_1=A_2=A$. The fields then have dimensions $[\phivec_i(\xvec,t)] = L^{\frac{2-d}{2}} B_i^{1/2} A^{-1/2}$ and thus $[u_1]  = [g_{21}] = L^{d-4} A^2 B_1^{-1}$ and $[u_2]  = [g_{12}] = L^{d-4} A^2 B_2^{-1}$. For $d > \duc = 4$, these couplings become irrelevant and the critical theory is described by the mean field exponents. Otherwise, they become relevant.

\subsection{Field Theory} \label{sec:FieldTheory}
In this sub-section, we derive the field theory and elucidate its structure, following the structure long established in the literature \cite{janssen_lagrangean_1976, bausch_renormalized_1976, janssen_renormalized_1977, janssen_renormalized_1992,tauber_critical_2014, MSR, Dominicis1976TECHNIQUESDR}.
The action $\action = \action_{0} + \action_{I}$ can be separated into the harmonic part
\begin{equation} \label{eq:actionharmonic}
        \action_{0} = \sum_{i = 1}^2 \xint \tint \phitildeveci \cdot \left[ \left( \partial_t - \nablasquared(r_i - D_i \nablasquared) \right) \phiveci + \Gamma_i \nablasquared \phitildeveci\right] 
\end{equation}
and the perturbative part
\begin{equation} \label{eq:actionperturbative}
    \action_{I} = -\sum_{i = 1}^{2} \frac{u_i}{3!} \xint \tint  (\phiveci \cdot \phiveci)( \phiveci \cdot \nablasquared \phitildeveci) - \xint \tint \left[ \frac{\gonetwo}{2}  (\phivec_2 \cdot \phivec_2)( \phivec_1 \cdot \nablasquared \phitildevecone) + \frac{\gtwoone}{2}  (\phivec_1 \cdot \phivec_1) (\phivec_2 \cdot \nablasquared \phitildevectwo) \right]
\end{equation}
where the $\nablasquared$ of the Langevin \Eref{Model} has been moved to act (only) on the auxiliary fields $\phitildevec_i$ by the use of Gauss' theorem and the action enters into the path integral in the form $\exp{-\action}$. To diagonalise the harmonic part of the action, we Fourier transform the fields 
\begin{subequations} \label{eq:fourierconvention}
\begin{gather}
    \phitildeveci(\xvec,t) = \kint \omegaint e^{\imag(\kvec \cdot \xvec - \omega t)} \phitildeveci(\kvec, \omega) \\
    \phiveci(\xvec,t) = \kint \omegaint e^{\imag(\kvec \cdot \xvec - \omega t)} \phiveci(\kvec, \omega) 
\end{gather}
\end{subequations}
where $\dbar^d k = d^d k/(2\pi)^d$ and $\dbar \omega = d \omega/ (2\pi)$. Correspondingly we use $\deltabar(\kvec) \defequal (2\pi)^{d}\delta(\kvec)$ and $\deltabar(\omega) \defequal 2\pi \delta(\omega)$.
The harmonic part of the action thus becomes
\begin{subequations}
\elabel{bare_propagators}
    \begin{align}
        \action_{0} &= \sum_{i = 1}^2 \kint \omegaint \phitildeveci(-\kvec,-\omega) \cdot \left[ \left( -\imag\omega + \kvecsquared(r_i + D_i \kvecsquared) \right) \phiveci(\kvec,\omega) - \Gamma_i \kvecsquared \phitildeveci(\kvec,\omega)\right] \\ 
        &= \frac{1}{2}\sum_{i = 1}^2 \kint \omegaint \begin{pmatrix}
    \phitildeveci(-\kvec,-\omega) & \phiveci(-\kvec, -\omega) \end{pmatrix} \begin{pmatrix}
-2\Gamma_i \kvecsquared & -\imag\omega + \deltaik \\
\elabel{harmonic_action_with_matrix}
\imag\omega + \deltaik & 0 
\end{pmatrix}
\begin{pmatrix}
    \phitildeveci(\kvec,\omega) \\
    \phiveci(\kvec,\omega)
\end{pmatrix}
    \end{align}
\end{subequations}
where we have introduced $\deltaik \defequal \kvecsquared (r_i + D_i \kvecsquared)$. Inverting the matrix in \Eref{harmonic_action_with_matrix}, we can find the bare propagators and the noise vertices
\begin{subequations} \label{eq:diagramsharmonic}
\begin{gather}
  \begin{tikzpicture}[baseline = (o.base)]
  \begin{feynman}
    \vertex (o) at (0,0) {\(a\)};
    \vertex (i) at (1.8,0) {\(b\)};
    \diagram* {
      (o) -- [anti fermion, edge label={\(\kvec, \omega\)}] (i)
    };
  \end{feynman}
\end{tikzpicture}
\corresponds G_{1,0}(\kvec, \omega)\kroneckerdelta{} \defequal \frac{\kroneckerdelta{}}{-\imag \omega + \Delta_{1}(\kvec)}, \qquad
\begin{tikzpicture}[baseline = (o.base)]
  \begin{feynman}
    \vertex (o) at (0,0) {\(a\)};
    \vertex (i) at (1.8,0) {\(b\)};
    \diagram* {
      (o) -- [anti charged scalar, edge label={\(\kvec, \omega\)}] (i)
    };
  \end{feynman}
\end{tikzpicture}
\corresponds G_{2,0}(\kvec, \omega)\kroneckerdelta{}  \defequal \frac{\kroneckerdelta{}}{-\imag \omega + \Delta_{2}(\kvec)}, \qquad \label{eq:barepropdiagram}\\
\begin{tikzpicture}[baseline = (o.base)]
  \begin{feynman}
    \vertex (o) at (0,0) ;
    \vertex [above left=0.6 of o] (i1) {\(a\)};
    \vertex [below left=0.6 of o] (i2) {\(b\)};
    \diagram* {
      (o) -- [plain, decoration={markings, mark=at position 0.5 with {\draw[-] (0,-3pt) -- (0,3pt);}}, postaction={decorate},edge label'={\(\kvec, \omega\)}] (i1),
      (o) -- [plain, decoration={markings, mark=at position 0.5 with {\draw[-] (0,-3pt) -- (0,3pt);}}, postaction={decorate}, edge label={\(-\kvec, -\omega\)}] (i2)
    };
  \end{feynman}
\end{tikzpicture}
\defequal 2\Gamma_1 \kvecsquared \kroneckerdelta{}, ~~
\begin{tikzpicture}[baseline = (o.base)]
  \begin{feynman}
    \vertex (o) at (0,0) ;
    \vertex [above left=0.6 of o] (i1) {\(a\)};
    \vertex [below left=0.6 of o] (i2) {\(b\)};
    \diagram* {
      (o) -- [scalar,decoration={markings, mark=at position 0.5 with {\draw[-] (0,-3pt) -- (0,3pt);}}, postaction={decorate}, edge label'={\(\kvec, \omega\)}] (i1),
      (o) -- [scalar, decoration={markings, mark=at position 0.5 with {\draw[-] (0,-3pt) -- (0,3pt);}}, postaction={decorate}, edge label={\(-\kvec, -\omega\)}] (i2)
    };
  \end{feynman}
\end{tikzpicture}
\defequal 2\Gamma_2 \kvecsquared \kroneckerdelta{}, \label{eq:noisevertexdiagram}
\end{gather}
\end{subequations}
where the Kronecker $\kroneckerdelta{}$ is a consequence of rotational invariance. Although \Erefs{bare_propagators} are the bare propagators, proportionality to $\kroneckerdelta{}$ of the propagators holds to all orders, i.e. different components never ``mix''. This is in fact an exact result. Here and in the following shortened legs indicate amputated diagrams, such as the two noise vertices in \Erefs{noisevertexdiagram}. The dash on the emerging lines of these diagrams denote the multiplication by the $\kvec$ carried by them, corresponding to a gradient in real space. 

The correlation function, $C_i(\kvec,\omega) \deltabar(\kvec + \qvec) \deltabar(\omega + \nu) \kroneckerdelta{} \equiv \langle\phi_i^{a}(\kvec,\omega) \phi_i^{b}(\qvec,\nu)\rangle$ can be built from the noise vertices by attaching two propagators. At bare level, they are given by
\begin{subequations} \label{eq:barecorrfuncs}
\begin{gather}
    \begin{tikzpicture}[baseline = (o.base)]
  \begin{feynman}
    \vertex (o) at (0,0) ;
    \vertex [above left=of o] (i1) {\(a\)};
    \vertex [below left=of o] (i2) {\(b\)};
    \diagram* {
      (o) -- [fermion, decoration={markings, mark=at position 0.25 with {\draw[-] (0,-3pt) -- (0,3pt);}}, postaction={decorate},edge label'={\(\kvec, \omega\)}] (i1),
      (o) -- [fermion, decoration={markings, mark=at position 0.25 with {\draw[-] (0,-3pt) -- (0,3pt);}}, postaction={decorate}, edge label={\(-\kvec, -\omega\)}] (i2)
    };
  \end{feynman}
\end{tikzpicture}
\corresponds C_{1,0}(\kvec,\omega) \kroneckerdelta{} \defequal \frac{2\Gamma_1 \kvecsquared \kroneckerdelta{}}{\left( -\imag \omega + \deltaonek\right)\left( \imag \omega + \Delta_1(-\kvec)\right)}, ~~ \\
\begin{tikzpicture}[baseline = (o.base)]
  \begin{feynman}
    \vertex (o) at (0,0) ;
    \vertex [above left=of o] (i1) {\(a\)};
    \vertex [below left=of o] (i2) {\(b\)};
    \diagram* {
      (o) -- [charged scalar,decoration={markings, mark=at position 0.25 with {\draw[-] (0,-3pt) -- (0,3pt);}}, postaction={decorate}, edge label'={\(\kvec, \omega\)}] (i1),
      (o) -- [charged scalar, decoration={markings, mark=at position 0.25 with {\draw[-] (0,-3pt) -- (0,3pt);}}, postaction={decorate}, edge label={\(-\kvec, -\omega\)}] (i2)
    };
  \end{feynman}
\end{tikzpicture}
\corresponds C_{2,0}(\kvec,\omega) \kroneckerdelta{} \defequal \frac{2\Gamma_2 \kvecsquared \kroneckerdelta{}}{\left( -\imag \omega + \deltatwok\right)\left( \imag \omega + \Delta_2(-\kvec)\right)}
\end{gather}
\end{subequations}
To find the interaction vertices, we rewrite the perturbative part of the action \Eref{actionperturbative} in terms of the Fourier transformed fields and we take the opportunity to render the summation over the components of the fields symmetric, say
\begin{multline}
\int \ddintbar{k}_{1,2,3,4}\,\dintbar{\omega}_{1,2,3,4}
    \ \delta^{ab}\delta^{cd} 
    \phitildeany[a](\kvec_1,\omega_1) 
    \phiany[b](\kvec_2,\omega_2) 
    \phiany[c](\kvec_3,\omega_3) 
    \phiany[d](\kvec_4,\omega_4) 
\\
=
\int \ddintbar{k}_{1,2,3,4}\,\dintbar{\omega}_{1,2,3,4}
    \ \frac{1}{3}\big(
    \delta^{ab}\delta^{cd} 
    +\delta^{ac}\delta^{bd} 
    +\delta^{ad}\delta^{bc} 
    \big)
    \phitildeany[a](\kvec_1,\omega_1) 
    \phiany[b](\kvec_2,\omega_2) 
    \phiany[c](\kvec_3,\omega_3) 
    \phiany[d](\kvec_4,\omega_4) 
\end{multline}
using the Einstein convention. The perturbative part of the action then becomes
    \begin{align} \label{eq:pertactioninfourierspace}
        \action_{I} & = \int \ddintbar{k} \ddintbar{q} \ddintbar{q'} \ddintbar{q''} \int \dintbar{\omega} \dintbar{\nu} \dintbar{\nu'} \dintbar{\nu''} \deltabar\left( \kvec + \qvec + \qvec'+\qvec''\right) \deltabar\left( \omega +\nu +\nu' +\nu''\right) \\
        &\ \times \kvecsquared \Bigg[ 
        \frac{\uone}{3 \times 3!} \Qtensor{} \phitildeone[a](\kvec,\omega) \phione[b](\qvec,\nu) \phione[c](\qvec',\nu') \phione[d](\qvec'', \nu'') +  \frac{\utwo}{3 \times 3!} \Qtensor{} \phitildetwo[a](\kvec,\omega) \phitwo[b](\qvec,\nu) \phitwo[c](\qvec',\nu') \phitwo[d](\qvec'',\nu'') \nonumber  \\
        \nonumber
        &\qquad\ + \frac{\gonetwo}{2} \kroneckerdelta{} \kroneckerdelta{cd} \phitildeone[a](\kvec,\omega) \phione[b](\qvec,\nu) \phitwo[c](\qvec',\nu') \phitwo[d](\qvec'',\nu'') + \frac{\gtwoone}{2} \kroneckerdelta{} \kroneckerdelta{cd} \phitildetwo[a](\kvec,\omega) \phitwo[b](\qvec,\nu) \phione[c](\qvec',\nu') \phione[d](\qvec'',\nu'') \Bigg]
    \end{align}
where $\Qtensor{} \defequal \kroneckerdelta{ab} \kroneckerdelta{cd} + \kroneckerdelta{ac}\kroneckerdelta{bd} + \kroneckerdelta{ad}\kroneckerdelta{bc}$.

We read off the interaction vertices from \Eref{pertactioninfourierspace}, 
\begin{subequations} \label{eq:diagramsperturbative}
\begin{gather}
  \begin{tikzpicture}[baseline = (o.base), scale = 1.2]
  \begin{feynman}
    \vertex (origin) at (0,0) ;
    \vertex (o) at (-0.6,0) {\(a\)};
    \vertex (i1) at (0.6,0) {\(c\)};
    \vertex (i2) at (0.6,0.5) {\(b\)};
    \vertex (i3) at (0.6,-0.5) {\(d\)};
    \diagram* {
      (origin) -- [plain, decoration={markings, 
  mark=at position 0.20 with {\draw[-] (0,-3pt) -- (0,3pt);},
  mark=at position 0.25 with {\draw[-] (0,-3pt) -- (0,3pt);}
}, postaction={decorate}, edge label'={\(\kvec\)}] (o),
      (origin) -- [plain] (i1),
      (origin) -- [plain] (i2),
      (origin) -- [plain] (i3),
    };
  \end{feynman}
\end{tikzpicture}
\corresponds-\kvecsquared \frac{\uone}{3}\Qtensor{} , ~~
\begin{tikzpicture}[baseline = (o.base), scale = 1.2]
  \begin{feynman}
    \vertex (origin) at (0,0) ;
    \vertex (o) at (-0.6,0) {\(a\)};
    \vertex (i1) at (0.6,0) {\(c\)};
    \vertex (i2) at (0.6,0.5) {\(b\)};
    \vertex (i3) at (0.6,-0.5) {\(d\)};
    \diagram* {
      (origin) -- [scalar, decoration={markings, 
  mark=at position 0.20 with {\draw[-] (0,-3pt) -- (0,3pt);},
  mark=at position 0.25 with {\draw[-] (0,-3pt) -- (0,3pt);}
}, postaction={decorate}, edge label'={\(\kvec\)}] (o),
      (origin) -- [scalar] (i1),
      (origin) -- [scalar] (i2),
      (origin) -- [scalar] (i3),
    };
  \end{feynman}
\end{tikzpicture}
\corresponds-\kvecsquared \frac{\utwo}{3}\Qtensor{} , \label{eq:singlespeciesquarticvertices}\\
\begin{tikzpicture}[baseline = (o.base), scale = 1.2]
  \begin{feynman}
    \vertex (origin) at (0,0) ;
    \vertex (o) at (-0.6,0) {\(a\)};
    \vertex (i1) at (0.6,0) {\(c\)};
    \vertex (i2) at (0.6,0.5) {\(b\)};
    \vertex (i3) at (0.6,-0.5) {\(d\)};
    \diagram* {
      (origin) -- [plain, decoration={markings, 
  mark=at position 0.20 with {\draw[-] (0,-3pt) -- (0,3pt);},
  mark=at position 0.25 with {\draw[-] (0,-3pt) -- (0,3pt);}
}, postaction={decorate}, edge label'={\(\kvec\)}] (o),
      (origin) -- [scalar] (i1),
      (origin) -- [plain] (i2),
      (origin) -- [scalar] (i3),
    };
  \end{feynman}
\end{tikzpicture}
\corresponds-\kvecsquared \gonetwo \kroneckerdelta{} \kroneckerdelta{cd},  ~~
\begin{tikzpicture}[baseline = (o.base), scale = 1.2]
  \begin{feynman}
    \vertex (origin) at (0,0) ;
    \vertex (o) at (-0.6,0) {\(a\)};
    \vertex (i1) at (0.6,0) {\(c\)};
    \vertex (i2) at (0.6,0.5) {\(b\)};
    \vertex (i3) at (0.6,-0.5) {\(d\)};
    \diagram* {
      (origin) -- [scalar, decoration={markings, 
  mark=at position 0.20 with {\draw[-] (0,-3pt) -- (0,3pt);},
  mark=at position 0.25 with {\draw[-] (0,-3pt) -- (0,3pt);}
}, postaction={decorate}, edge label'={\(\kvec\)}] (o),
      (origin) -- [plain] (i1),
      (origin) -- [scalar] (i2),
      (origin) -- [plain] (i3),
    };
  \end{feynman}
\end{tikzpicture}
\corresponds-\kvecsquared \gtwoone \kroneckerdelta{} \kroneckerdelta{cd},\label{eq:speciescouplingvertices}
\end{gather}
\end{subequations}

\section{Renormalisation Group Procedure} \label{sec:RG}
In this section, we carry out the Renormalisation Group (RG) programme for the field theory introduced in the previous section. We focus on the tricritical point, $r_1 \to 0$ and  $r_2 \to 0$ simultaneously, where both fields display non-Gaussian fluctuations.

\subsection{Renormalisation Scheme} \label{sec:renormoverview}

In field theoretic RG, the aim is to regularise the logarithmic UV-divergences that appear in the perturbative expansion, most easily implemented via dimensional regularisation. Upon regularising the perturbative expansion and through the use of the Callan-Symanzik equation (see \Sref{scalingexp}) these divergences determine the critical behaviour, in particular the critical exponents of the theory.

This regularisation is done most efficiently by considering \emph{vertex functions} associated with various correlation functions \cite{tauber_critical_2014, amit_field_2005, bellac_quantum_1992}. These vertex functions are, firstly the two-point vertices
\begin{subequations} \label{eq:vertexfuncdefn}
\begin{align}
   \invpropi{ab}{\kvec, \omega} &= \frac{\kroneckerdelta{}}{G_i(\kvec,\omega)},\\
   ~\vertexnoisei{ab}{\kvec, \omega} &=\kroneckerdelta{} \frac{C_i(\kvec,\omega)}{G_{i}(\kvec,\omega)G_{i}(-\kvec,-\omega)}
\end{align}
namely the inverse propagator for $i\in\{1,2\}$ and the noise vertex, again for $i\in\{1,2\}$. Secondly, the 
four-point vertices
\begin{align}
        \vertexquartic{abcd}{\kvec,\omega; \qvec, \nu, \qvec', \nu', \qvec'', \nu''} &\deltabar\left( \kvec + \qvec + \qvec'+ \qvec'' \right) \deltabar\left( \omega + \nu + \nu'+ \nu'' \right) = \nonumber \\
        &\frac{\langle \phi_i^{a}(\kvec,\omega) \phitilde_i^{b}(\qvec,\nu) \phitilde_j^{c}(\qvec',\nu') \phitilde_j^{d}(\qvec'',\nu'')\rangle_c}{G_i(\kvec,\omega) G_i(-\qvec,-\nu) G_j(-\qvec',-\nu') G_j(-\qvec'',-\nu'')},
\end{align}
\end{subequations}
which the effective non-linearities,  $i,j\in\{1,2\}$, corresponding to $u_i$ for $i=j$ and $g_{ij}$ for $i\ne j$.
In the definitions \Erefs{vertexfuncdefn} $G_i(\kvec,\omega)$ and $C_i(\kvec,\omega)$ are the full propagators and correlation functions, respectively. The subscript ``c" denotes the connected diagrams. At tree-level, the vertex functions, \Erefs{vertexfuncdefn}, can be read off from the action,  \Erefs{actionharmonic} and \eref{actionperturbative}:
\begin{subequations} \label{eq:treelevelvertexfuncs}
    \begin{align}
        \invpropi{ab}{\kvec, \omega} &= \kroneckerdelta{}\left( -\imag \omega + \kvecsquared(r_i + D_i \kvecsquared)  \right) + \mathcal{O}(\lambda),\\
        \vertexnoisei{ab}{\kvec, \omega}& = 2\Gamma_i \kvecsquared\kroneckerdelta{} + \mathcal{O}(\lambda), \\
        \vertexui{abcd}{\kvec,\omega; \qvec, \nu, \qvec', \nu', \qvec'', \nu''} &= - \kvecsquared \frac{u_i}{3} Q^{abcd} + \mathcal{O}(\lambda^2), \\
        \vertexquartic{abcd}{\kvec,\omega; \qvec, \nu, \qvec', \nu', \qvec'', \nu''} &= -\kvec^2 g_{ij} \kroneckerdelta{} \kroneckerdelta{cd} + \mathcal{O}(\lambda^2),
    \end{align}
\end{subequations}
where $i \neq j$ in the final line, $\lambda \in \{u_1,u_2, \gonetwo, \gtwoone\}$ denotes the non-linear coupling constants of the theory, so that $\mathcal{O}(\lambda)$ and $\mathcal{O}(\lambda^2)$ in \Erefs{treelevelvertexfuncs} contains only loops, which will be accounted as explained in the following.

During the RG procedure, higher order vertices or higher order frequency and wave-vector dependence in the vertices in Eq. \eqref{eq:treelevelvertexfuncs} will be generated. However, these have no bearing on the critical behaviour of the theory as they are RG \emph{irrelevant}, according to their engineering dimensions, \Sref{meanfieldcritical}. Thus, in what follows, we consider the corrections by loop diagrams to the bare parameters in \Erefs{treelevelvertexfuncs}. Generally, the RG relevant or marginal parts of the vertex functions can be written as
\begin{subequations} \elabel{fullvertexfuncs}
    \begin{align}
        \invpropi{}{\kvec, \omega} &= \kroneckerdelta{}\left( -\imag \omega \diagramcontribution{\omega_i} + \kvecsquared(\diagramcontribution{\tau_i} \tau_i + \diagramcontribution{D_i} D_i \kvecsquared )\right)
        \elabel{inverse_propagator}
        , \\ \vertexnoisei{}{\kvec} &= 2\Gamma_i\diagramcontribution{\Gamma_i} \kvecsquared\kroneckerdelta{}, \\
        \vertexui{}{\kvec} &= - \kvecsquared \frac{u_i}{3} \diagramcontribution{u_i}Q^{abcd}, \\
        \vertexquartic{}{\kvec} &= -\kvec^2 g_{ij} \diagramcontribution{g_{ij}}\kroneckerdelta{} \kroneckerdelta{cd},
    \end{align}
\end{subequations}
where once again $i \neq j$. In \Erefs{fullvertexfuncs} we have introduced the loop corrections $\diagramcontribution{p}$ with $p \in \{\omega_i, D_i, r_i, \Gamma_i, u_i, \gonetwo, \gtwoone\}$, which will be evaluated at $\kvec=\zerovec$ and $\omega=0$. The mass of the inverse propagator, \Eref{inverse_propagator}, is now rewritten as $\tau_i=r_i-r_{i,c}$.

We have further defined
\begin{subequations}
\elabel{defs_vertex_shorthands}
\begin{align}
\vertexnoisei{}{\kvec} & = \vertexnoisei{}{\kvec, 0}, \\
\vertexui{}{\kvec} & = \vertexui{}{\kvec, 0; -\kvec/3, 0, -\kvec/3, 0, -\kvec/3, 0}, \\
\vertexquartic{}{\kvec} &=\vertexquartic{}{\kvec, 0; -\kvec/3, 0, -\kvec/3, 0, -\kvec/3, 0}, 
\end{align}
\end{subequations}
for notational ease. The choice of $\kvec/3$ on the right-hand side of \Erefs{defs_vertex_shorthands} is not the subtraction point \cite{bellac_quantum_1992}, as we will take $\kvec=\zerovec$ below, but rather a convenient choice to define the effective renormalised couplings below.

To perform the RG procedure, we first identify the tricritical point of the theory where the correlation lengths of both fields diverge. At mean-field level, this tricritical point is 
$r_{1,c}=r_{2,c}=0$.
However, as is well understood \cite{tauber_critical_2014}, this value will get shifted due to the fluctuations of the fields. The critical point $(r_{1,c}, r_{2,c})$ is determined by the simultaneous roots of:
\begin{equation} \label{eq:criticalpointdefn}
    \left.\frac{ \partial \invpropone{}{\kvec , \omega }}{\partial \kvecsquared}\right|_{\kvec = \zerovec, \omega = 0} = \left.\frac{ \partial \invproptwo{}{\kvec , \omega }}{\partial \kvecsquared}\right|_{\kvec = \zerovec, \omega = 0} = 0
\end{equation}
This ensures that the fluctuations of the fields are massless and hence have infinite correlation length. Upon identifying the critical point of the model, we can parameterise the system in terms of $\tauone = \rone-r_{1,c}$ and $\tautwo = r - r_{2,c}$ which measure the distance from the tricritical point. We can now proceed with eliminating the divergences of our theory. This elimination is done at an arbitrary inverse length scale $\mu$. To absorb the UV-divergences that arise in the perturbative expansion, we introduce the following $Z$-factors and renormalised quantities, indicated by the subscript $R$: Firstly the field renormalisation,
\begin{equation} \elabel{def_field_renormalisation}
        \phivec_{i,R} = Z_{\phivec_i} \phivec_i, \hspace{0.2cm} \phitildevec_{i,R} = Z_{\phitildevec_i} \phitildevec_i, \end{equation}
and secondly the renormalisation of the couplings,
\begin{subequations} \elabel{Zfactorsdefn}
    \begin{gather}
        \tau_{i,R} = Z_{\tau_i} \tau_i \mu^{-2}, D_{i,R} = Z_{D_i} D_i, \elabel{Zfactorsdefn_propagators}\\
          \Gamma_{i,R} = Z_{\Gamma_i} \Gamma_i, \\ u_{i,R} = Z_{u_i} u_i \mu^{\epsilon}, \hspace{0.2cm} \\
        g_{ij,R} = Z_{g_{ij}} g_{ij} \mu^{\epsilon} \ .
    \end{gather}
\end{subequations}
Any renormalised coupling has been made spatially dimensionless using $\epsilon = 4- d$ for the non-linearities. With this adjustment, renormalised couplings remain finite and any physical observable will be given in terms of the renormalised quantities which are the well-defined objects in the field theory. 

We choose the normalisation point (NP) of the renormalisation scheme at a finite $\tau_{1,R}$ and $\tau_{2,R}$, approaching the tricritical point as $\mu\to0$. Under renormalisation, the ratio $\tauRat=\tau_1/\tau_2$ might flow away from its bare value, which we will need to characterise in principle, however, it turns out that $\tauRat$ enters solely into itself. 

Furthermore, we impose, adhering to the minimal subtraction scheme \cite{bellac_quantum_1992}, the following normalisation conditions on the renormalised vertices to leading order in the divergence in $1/\epsilon$:
\begin{subequations} \elabel{renormconds}
    \begin{align}
        \left. \invpropiren{}{\kvec, \omega} \right.|_{\rm NP} &= \kroneckerdelta{}\left( -\imag \omega + \kvecsquared(\mu^2\tau_{i,R} + D_{i,R} \kvecsquared )\right), \label{eq:renormconds_invprop}\\
        \left. \vertexnoiseiren{}{\kvec} \right.|_{\rm NP}&= 2\Gamma_{i,R} \kvecsquared \kroneckerdelta{}, \elabel{renormconds_noise}\\
        \left. \vertexuiren{}{\kvec} \right.|_{\rm NP} &= - \kvecsquared \frac{u_{i,R}\muepspos}{3} Q^{abcd} \elabel{renormconds_ui} \\
       \left.  \vertexquarticren{}{\kvec} \right.|_{\rm NP} &= -\kvec^2 g_{ij,R}\muepspos \kroneckerdelta{} \kroneckerdelta{cd}, \elabel{renormconds_gij}
    \end{align}
\end{subequations}
where $i \neq j$ in the last line. The renormalised vertices may be written in terms of the renormalised fields as
\begin{subequations} \label{eq:Renvertexfuncdefn}
\begin{align}
   \invpropiren{ab}{\kvec, \omega} &= \frac{\kroneckerdelta{}}{G_{i,R}(\kvec,\omega)}, \\
   \vertexnoiseiren{ab}{\kvec, \omega} &= \kroneckerdelta{}\frac{C_{i,R}(\kvec,\omega)}{|G_{i,R}(\kvec,\omega)|^2}, \\
        \vertexquarticren{abcd}{\kvec,\omega; \qvec, \nu, \qvec', \nu', \qvec'', \nu''} &\deltabar\left( \kvec + \qvec + \qvec'+ \qvec'' \right) \deltabar\left( \omega + \nu + \nu'+ \nu'' \right) = \nonumber \\
        &\frac{\langle \phi_{i,R}^{a}(\kvec,\omega) \phitilde_{i,R}^{b}(\qvec,\nu) \phitilde_{j,R}^{c}(\qvec',\nu') \phitilde_{j,R}^{d}(\qvec'',\nu'')\rangle_c}{G_{i,R}(\kvec,\omega) G_{i,R}(-\qvec,-\nu) G_{j,R}(-\qvec',-\nu') G_{j,R}(-\qvec'',-\nu'')}.
\end{align}
\end{subequations}
where $\langle \phiiR[a](\kvec, \omega) \phitildeiR[b](\qvec, \nu)\rangle = \kroneckerdelta{} \deltabar(\kvec+\qvec) \deltabar(\omega +\nu) G_{i,R}(\kvec, \omega)$ and $\langle \phiiR[a](\kvec, \omega) \phiiR[b](\qvec, \nu)\rangle = \kroneckerdelta{} \deltabar(\kvec+\qvec) \deltabar(\omega +\nu) C_{i,R}(\kvec, \omega)$. 
Using the definition of the renormalised fields, \Eref{def_field_renormalisation} in \Erefs{renormconds} and comparing to \Eref{vertexfuncdefn}, it follows that 
\begin{subequations} \elabel{renvertextonormalvertex}
\begin{align}
    \invpropiren{ab}{\kvec, \omega} &= \left(Z_{\phivec_i} Z_{\phitildevec_i} \right)^{-1}\invpropi{}{} \elabel{renvertextonormalvertex_invprop},\\
    \vertexnoiseiren{ab}{\kvec, \omega} &= Z_{\phitildevec_i}^{-2}\vertexnoisei{}{}, \elabel{renvertextonormalvertex_noise} \\
        \vertexuiren{abcd}{\kvec}  &= Z_{\phitildevec_i}^{-1} Z_{\phivec_i}^{-3}  \vertexui{abcd}{\kvec}, \\
        \vertexquarticren{}{\kvec} &= Z_{\phitildevec_i}^{-1} Z_{\phivec_i}^{-1} Z_{\phivec_j}^{-2} \vertexquartic{}{\kvec}
\end{align}
\end{subequations}
where $i \neq j$ in the last line. Using \Erefs{renvertextonormalvertex} in the normalisation condition \Eref{renormconds} and comparing to the parameterisation \Erefs{fullvertexfuncs} allows us to express 
the loop-corrections $\diagramcontribution{p}$ in terms of
the $Z$-factors,
\begin{subequations}
\elabel{I_inTermsOf_Z}
\begin{align}
    \diagramcontribution{\omega_i} &= Z_{\phitildevec_i} Z_{\phivec_i} \elabel{I_inTermsOf_Z_omega} \\
    \diagramcontribution{\tau_i} &= Z_{\phitildevec_i} Z_{\phivec_i}  Z_{\tau_i} \elabel{I_inTermsOf_Z_tau} \\
    \diagramcontribution{D_i} &= Z_{\phivec_i} Z_{\phitildevec_i} Z_{D_i} \elabel{I_inTermsOf_Z_D}\\
    \diagramcontribution{\Gamma_i} &=  Z^2_{\phitildevec_i} Z_{\Gamma_i} \elabel{I_inTermsOf_Z_gamma}\\
    \diagramcontribution{u_i} &= Z_{\phitildevec_i}Z^3_{\phivec_i}  Z_{u_i}  \elabel{I_inTermsOf_Z_ui}\\
    \diagramcontribution{g_{ij}} &=  Z_{\phitildevec_i} Z_{\phivec_i} Z^2_{\phivec_j} Z_{g_{ij}} \elabel{I_inTermsOf_Z_gij}
\end{align}
\end{subequations}
and thus conversely the $Z$-factors, which are of the form $Z=1+\mathcal{O}(\lambda)$, in terms of the loop corrections $\diagramcontribution{p}$ of the same form. We determine them in \Sref{1loop}, after deriving some exact results, in particular  $\Zphitildeone \Zphione= \Zphitildetwo \Zphitwo=1$, \Eref{Zfactorrelation1}, which significantly simplifies \Erefs{I_inTermsOf_Z}.

\subsection{Ward Identities}
In this section, we derive Ward Identities, which amount to exact results about some $Z-$factors. The most obvious symmetry is the conserved nature of the dynamics, as every deterministic term in \Erefs{Model} is preceded by a $\nabla^2$ and the noise is conserved. Perturbatively, this results in corrections to the propagator, or any other correlation function for that matter, to vanish at $\kvec=\zerovec$. Non-perturbatively, the integral over all space of the propagator is unity,
\begin{equation} \label{eq:response_integral}
    \int\ddint{x} 
    \ave{\phi^a_i(\xvec,t)\phitilde^b_j(\zerovec,t')} = \theta(t-t') \delta_{ij}\delta^{ab} \ .
\end{equation}
At the level of the field theory, this imposes Ward identities which, in turn, yield certain relations among the $Z$-factors, most easily identified by writing \Eref{response_integral} as
\begin{equation} \elabel{WardIdentity1}
    \invpropi{}{\kvec = \zerovec, \omega} = -\imag \omega \kroneckerdelta{} \ ,
\end{equation}
which correspondingly has a unit amplitude \cite{tauber_critical_2014}. 
Evaluating \Eref{WardIdentity1} with \Erefs{renormconds_invprop} and \eref{renvertextonormalvertex_invprop} produces
\begin{equation} \label{eq:Zfactorrelation1}
    \Zphitildeone \Zphione = \Zphitildetwo \Zphitwo = 1,
\end{equation}
so that $ \diagramcontribution{\omega_i}=1$, \Eref{I_inTermsOf_Z_omega}.
Corrections to the conserved noise vertex acquire at least a factor $\kvec^4$, so that
\begin{equation} \label{eq:WardIdentity2}
    \left. \frac{\partial \vertexnoisei{}{}}{\partial \kvecsquared}\right|_{\kvec = \zerovec, \omega = 0} = -2\Gamma_i \kroneckerdelta{} 
\end{equation}
to all orders and therefore with \Erefs{renormconds_noise} and \eref{renvertextonormalvertex_noise}
\begin{equation} \label{eq:Zfactorrelation2}
    Z_{\Gamma_i} Z_{\phitildevec_i}^2 = 1 \ ,
\end{equation}
so that $\diagramcontribution{\Gamma_i}=1$, \Eref{I_inTermsOf_Z_gamma}.
Together with \Eref{Zfactorrelation1} it follows that
\begin{equation} \elabel{FDRZfactors}
    Z_{\Gamma_i} = Z_{\phivec_i}/Z_{\phitildevec_i}
\end{equation}
which in equilibrium is obtained from the Ward identity following from a fluctuation-dissipation relation \cite{tauber_critical_2014}. In the present case, however, we have a distinctly non-equilibrium system with no apparent fluctuation-dissipation relation. Yet, the conserved dynamics gives sufficient constraints to produce the same effect \Eref{FDRZfactors} of a standard fluctuation-dissipation relation. Furthermore, we show in \Sref{scalingexp} that \Erefs{Zfactorrelation1} and \eref{Zfactorrelation2} are sufficient to ensure that the anomalous exponents of the correlation and response functions are the same. Of course, this does not imply that the present system sustains a ``hidden'' fluctuation-dissipation relation.

\subsection{1-Loop Renormalisation} \seclabel{1loop}
In this section, we calculate the $Z$-factors \Erefs{Zfactorsdefn} imposing  \Erefs{renormconds}, or, equivalently, by calculating the loop corrections and reading off from \Erefs{fullvertexfuncs} the $\diagramcontribution{p}$, which generate the $Z$-factors by \Erefs{I_inTermsOf_Z}. The actual evaluation of the loop diagrams is relegated to \APref{DetailsLoop}. Here, we merely quote the results. 

Starting with the renormalisation of the inverse propagators, $\invpropi{}{}$, to one loop \begin{subequations} \label{eq:invprop_oneloop}
\begin{gather}
    \invpropone{}{} = \left(-\imag \omega + \kvecsquared\left( D_1 \kvecsquared +r_1\right) \right) \kroneckerdelta{} - 
    \begin{tikzpicture}[baseline = (o.base), scale = 1.2]
   \begin{feynman}
     \vertex (o) at (0,0) {\(a\)};
    \vertex (a) at (0.8,0);
    \vertex (b) at (1.7,1);
    \vertex (i) at (1.4,0) {\(b\)};
    \diagram* {
      (a) -- [anti fermion, decoration={markings, mark=at position 0.65 with {\draw[-] (0,-3pt) -- (0,3pt);}}, postaction={decorate}, bend left=40] (b),
      (o) -- [plain , edge label = {\(\kvec,\omega\)}](a),
      (a) -- [anti fermion, decoration={markings, mark=at position 0.65 with {\draw[-] (0,-3pt) -- (0,3pt);}}, postaction={decorate}, bend right=40] (b),
      (o) -- [plain, decoration={markings, 
  mark=at position 0.20 with {\draw[-] (0,-3pt) -- (0,3pt);},
  mark=at position 0.25 with {\draw[-] (0,-3pt) -- (0,3pt);}
}, postaction={decorate}, ] (i)
    };
  \end{feynman}
\end{tikzpicture}
- \begin{tikzpicture}[baseline = (o.base), scale = 1.2]
   \begin{feynman}
    \vertex (o) at (0,0) {\(a\)};
    \vertex (a) at (0.8,0);
    \vertex (b) at (1.7,1);
    \vertex (i) at (1.4,0) {\(b\)};
    \diagram* {
      (a) -- [anti charged scalar, decoration={markings, mark=at position 0.65 with {\draw[-] (0,-3pt) -- (0,3pt);}}, postaction={decorate}, bend left=40] (b),
      (o) -- [plain , edge label = {\(\kvec,\omega\)}](a),
      (a) -- [anti charged scalar, decoration={markings, mark=at position 0.65 with {\draw[-] (0,-3pt) -- (0,3pt);}}, postaction={decorate}, bend right=40] (b),
      (o) -- [plain, decoration={markings, 
  mark=at position 0.20 with {\draw[-] (0,-3pt) -- (0,3pt);},
  mark=at position 0.25 with {\draw[-] (0,-3pt) -- (0,3pt);}
}, postaction={decorate}] (i)
    };
  \end{feynman}
\end{tikzpicture}
+ \OC(\lambda^2)
    \\
    \invproptwo{}{} = \left(-\imag \omega + \kvecsquared\left( D_2 \kvecsquared +r_2\right) \right) \kroneckerdelta{} -
    \begin{tikzpicture}[baseline = (o.base), scale = 1.2]
   \begin{feynman}
   \vertex (o) at (0,0) {\(a\)};
    \vertex (a) at (0.8,0);
    \vertex (b) at (1.7,1);
    \vertex (i) at (1.4,0) {\(b\)};
    \diagram* {
      (a) -- [anti charged scalar, decoration={markings, mark=at position 0.65 with {\draw[-] (0,-3pt) -- (0,3pt);}}, postaction={decorate}, bend left=40] (b),
      (o) -- [scalar, edge label = {\(\kvec,\omega\)}] (a),
      (a) -- [anti charged scalar, decoration={markings, mark=at position 0.65 with {\draw[-] (0,-3pt) -- (0,3pt);}}, postaction={decorate}, bend right=40] (b),
      (o) -- [scalar, decoration={markings, 
  mark=at position 0.20 with {\draw[-] (0,-3pt) -- (0,3pt);},
  mark=at position 0.25 with {\draw[-] (0,-3pt) -- (0,3pt);}
}, postaction={decorate}, ] (i)
    };
  \end{feynman}
\end{tikzpicture}
 - \begin{tikzpicture}[baseline = (o.base), scale = 1.2]
   \begin{feynman}
    \vertex (o) at (0,0) {\(a\)};
    \vertex (a) at (0.8,0);
    \vertex (b) at (1.7,1);
    \vertex (i) at (1.4,0) {\(b\)};
    \diagram* {
      (a) -- [anti fermion, decoration={markings, mark=at position 0.65 with {\draw[-] (0,-3pt) -- (0,3pt);}}, postaction={decorate}, bend left=40] (b),
      (o) -- [scalar, edge label={\(\kvec,\omega\)}] (a),
      (a) -- [anti fermion, decoration={markings, mark=at position 0.65 with {\draw[-] (0,-3pt) -- (0,3pt);}}, postaction={decorate}, bend right=40] (b),
      (o) -- [scalar, decoration={markings, 
  mark=at position 0.20 with {\draw[-] (0,-3pt) -- (0,3pt);},
  mark=at position 0.25 with {\draw[-] (0,-3pt) -- (0,3pt);}
}, postaction={decorate}, ] (i)
    };
  \end{feynman}
\end{tikzpicture}
+ \OC(\lambda^2)
\end{gather}
\end{subequations}
where the first set of terms is the bare theory and the minus sign is due to Dyson-summing one-loop diagrams. Rather than performing an additive renormalisation \cite{tauber_critical_2014}, we replace $r_i$ by $\tau_i= \mu^2\tau_{i,R}+\OC(\lambda)$, \Eref{Zfactorsdefn_propagators}, regularising the infrared (IR). Where we encounter IR divergences of the form $\tau_i^{-\epsilon/2}$, we may replace them by $\mu^{-\epsilon/2}$ for both $i=1$ and $i=2$ to leading order in small $\epsilon$ as $\tau_{i,R}/D_{i,R}$ are finite. A ratio of the form $\tau_2/\tau_1$, however, is $\tau_{1,R}/\tau_{2,R}+\OC(\lambda)$. As for the ultraviolet (UV), we use the analytic continuation of the UV-divergence captured by $\Gamma(1-d/2)=-\Gamma(\epsilon/2)+\mathcal{O}(\epsilon^0)$ to write \Eref{invprop_oneloop} as

\begin{subequations} \elabel{invprop_oneloop_algebra}
    \begin{align}
         \left.\invpropone{}{\kvec,\omega}\right|_{NP}  &= \left( -\imag \omega + \Done\kvec^4 +     \frac{\mu^{2} \tau_{1,R}}{\Zrone}  \kvecsquared  \left[1 -  \frac{\numcomponent+2}{6} \frac{\uone \Gamma_1 }{\Done^2} \loopfactor \muepsneg  - \frac{\numcomponent}{2}\frac{\gonetwo \Gamma_2 }{\Dtwo^2} \frac{\tau_2}{\tau_1}\loopfactor \muepsneg \right] \right) \kroneckerdelta{}\\
         \left.\invproptwo{}{\kvec,\omega}\right|_{NP}  &= \left( -\imag \omega + \Dtwo \kvec^4 +  \frac{\mu^{2} \tau_{2,R}}{\Zrtwo}\kvecsquared   \left[1 -  \frac{\numcomponent+2}{6} \frac{\utwo \Gamma_2 }{\Dtwo^2} \loopfactor \muepsneg  - \frac{\numcomponent}{2}\frac{\gtwoone \Gamma_1 }{  \Done^2} \frac{\tau_1}{\tau_2}\loopfactor \muepsneg \right] \right) \kroneckerdelta{}
    \end{align}
\end{subequations}
where 
$\epsilon=4-d$.
As $Z_{\phivec_i} Z_{\phitildevec_i}=1$, \Eref{Zfactorrelation1}, the loop integrals $\diagramcontribution{\tau_i}$ equal the $Z$-factors $Z_{\tau_i}$, \Erefs{I_inTermsOf_Z_tau} and \eref{I_inTermsOf_Z_D}, and are readily identified as the square-brackets in \Erefs{invprop_oneloop_algebra}, so that
\begin{subequations} \label{eq:Zfactorpropagator}
    \begin{align}
        \diagramcontribution{\tau_1}&=\Zrone = 1 - \left( \frac{\numcomponent+2}{6} \frac{ \uone \Gamma_1}{\Done^2}   +  \frac{\numcomponent}{2}\frac{\gonetwo \Gamma_2 }{ \Dtwo^2} \frac{\tau_2}{\tau_1} \right) \loopfactor \muepsneg  \label{eq:Ztauone_bare}\\
        \diagramcontribution{\tau_2} &= \Zrtwo = 1 -  \left( \frac{\numcomponent+2}{6} \frac{ \utwo \Gamma_2 }{\Dtwo^2} + \frac{\numcomponent}{2}\frac{ \gtwoone \Gamma_1 }{ \Done^2} \frac{\tau_1}{\tau_2} \right) \loopfactor \muepsneg \label{eq:Ztautwo_bare}\\
        \diagramcontribution{D_1}&=\ZDone = 1 \label{eq:ZDone_bare}\\
        \diagramcontribution{D_2}&=\ZDtwo = 1 \label{eq:ZDtwo_bare}\ .
    \end{align} 
\end{subequations}

We now consider the renormalisation of the single species quartic vertices $\vertexui{}{\kvec}$. To one loop, the (amputated) diagrams are given by 
\begin{subequations}
    \begin{gather}
        \vertexuone{}{\kvec} = 
        \begin{tikzpicture}[baseline = (o.base), scale = 1.2]
  \begin{feynman}
    \vertex (origin) at (0,0) ;
    \vertex (o) at (-0.6,0) {\(a\)};
    \vertex (i1) at (0.6,0) {\(c\)};
    \vertex (i2) at (0.6,0.6) {\(b\)};
    \vertex (i3) at (0.6,-0.6) {\(d\)};
    \diagram* {
      (origin) -- [plain, decoration={markings, 
  mark=at position 0.20 with {\draw[-] (0,-3pt) -- (0,3pt);},
  mark=at position 0.25 with {\draw[-] (0,-3pt) -- (0,3pt);}
}, postaction={decorate}, edge label'={\(\kvec\)}] (o),
      (origin) -- [plain] (i1),
      (origin) -- [plain] (i2),
      (origin) -- [plain] (i3),
    };
  \end{feynman}
\end{tikzpicture}
+
\begin{tikzpicture}[baseline = (o.base),scale = 1.2]
\begin{feynman}
\vertex (o) at (0,0) {\(a\)};
\vertex (a) at (0.8,0);
\vertex (i3) at ($(a) + (0.75,0)$) {\(d\)};
\vertex (b) at ($(a) + (0.33,1)$);     
\vertex (c) at ($(a) + (1,0.4)$);    
\vertex (i1) at ($(b) + (0.5,0.6)$) {\(b\)};  
\vertex (i2) at ($(b) + (0.5,0.3)$) {\(c\)};  
\diagram* {
(o) -- [plain, edge label = \(\kvec\), decoration={markings, 
  mark=at position 0.75 with {\draw[-] (0,-3pt) -- (0,3pt);},
  mark=at position 0.80 with {\draw[-] (0,-3pt) -- (0,3pt);}
}, postaction={decorate},] (a),
(a) -- [plain] (i3),
(a) -- [anti fermion, decoration={markings, 
  mark=at position 0.80 with {\draw[-] (0,-3pt) -- (0,3pt);},
  mark=at position 0.75 with {\draw[-] (0,-3pt) -- (0,3pt);}
} ] (b),
(b) -- [anti fermion, decoration={markings, mark=at position 0.65 with {\draw[-] (0,-3pt) -- (0,3pt);}},] (c),
(a) -- [anti fermion, decoration={markings, mark=at position 0.65 with {\draw[-] (0,-3pt) -- (0,3pt);}},] (c),
(b) -- [plain] (i1),
(b) -- [plain] (i2),
};
\end{feynman}
\end{tikzpicture}
~~~ + \sym 
~+~
\begin{tikzpicture}[baseline = (o.base), scale = 1.2]
\begin{feynman}
\vertex (o) at (0,0) {\(a\)};
\vertex (a) at (0.8,0);
\vertex (i3) at ($(a) + (0.75,0)$) {\(d\)};
\vertex (b) at ($(a) + (0.33,1)$);     
\vertex (c) at ($(a) + (1,0.4)$);    
\vertex (i1) at ($(b) + (0.5,0.6)$) {\(b\)};  
\vertex (i2) at ($(b) + (0.5,0.3)$) {\(c\)};  
\diagram* {
(o) -- [plain, edge label = \(\kvec\), decoration={markings, 
  mark=at position 0.75 with {\draw[-] (0,-3pt) -- (0,3pt);},
  mark=at position 0.8 with {\draw[-] (0,-3pt) -- (0,3pt);}
}, postaction={decorate},] (a),
(a) -- [plain] (i3),
(a) -- [anti charged scalar, decoration={markings, 
  mark=at position 0.80 with {\draw[-] (0,-3pt) -- (0,3pt);},
  mark=at position 0.75 with {\draw[-] (0,-3pt) -- (0,3pt);}
} ] (b),
(b) -- [anti charged scalar, decoration={markings, mark=at position 0.65 with {\draw[-] (0,-3pt) -- (0,3pt);}}, postaction={decorate}] (c),
(a) -- [anti charged scalar, decoration={markings, mark=at position 0.65 with {\draw[-] (0,-3pt) -- (0,3pt);}}, postaction={decorate}] (c),
(b) -- [plain] (i1),
(b) -- [plain] (i2),
};
\end{feynman}
\end{tikzpicture}
~~~ + \sym 
\\
        \vertexutwo{}{\kvec} = 
        \begin{tikzpicture}[baseline = (o.base), scale = 1.2]
  \begin{feynman}
    \vertex (origin) at (0,0) ;
    \vertex (o) at (-0.6,0) {\(a\)};
    \vertex (i1) at (0.6,0) {\(c\)};
    \vertex (i2) at (0.6,0.6) {\(b\)};
    \vertex (i3) at (0.6,-0.6) {\(d\)};
    \diagram* {
      (origin) -- [scalar, decoration={markings, 
  mark=at position 0.25 with {\draw[-] (0,-3pt) -- (0,3pt);},
  mark=at position 0.30 with {\draw[-] (0,-3pt) -- (0,3pt);}
}, postaction={decorate}, edge label'={\(\kvec\)}] (o),
      (origin) -- [scalar] (i1),
      (origin) -- [scalar] (i2),
      (origin) -- [scalar] (i3),
    };
  \end{feynman}
\end{tikzpicture}
+ 
\begin{tikzpicture}[baseline = (o.base), scale = 1.2]
\begin{feynman}
\vertex (o) at (0,0) {\(a\)};
\vertex (a) at (0.8,0);
\vertex (i3) at ($(a) + (0.75,0)$) {\(d\)};
\vertex (b) at ($(a) + (0.33,1)$);     
\vertex (c) at ($(a) + (1,0.4)$);    
\vertex (i1) at ($(b) + (0.5,0.6)$) {\(b\)};  
\vertex (i2) at ($(b) + (0.5,0.3)$) {\(c\)};  
\diagram* {
(o) -- [scalar, edge label = \(\kvec\), decoration={markings, 
  mark=at position 0.80 with {\draw[-] (0,-3pt) -- (0,3pt);},
  mark=at position 0.75 with {\draw[-] (0,-3pt) -- (0,3pt);}
}, postaction={decorate},] (a),
(a) -- [scalar] (i3),
(a) -- [anti charged scalar, decoration={markings, 
  mark=at position 0.80 with {\draw[-] (0,-3pt) -- (0,3pt);},
  mark=at position 0.75 with {\draw[-] (0,-3pt) -- (0,3pt);}
} ] (b),
(b) -- [anti charged scalar, decoration={markings, mark=at position 0.65 with {\draw[-] (0,-3pt) -- (0,3pt);}},] (c),
(a) -- [anti charged scalar, decoration={markings, mark=at position 0.65 with {\draw[-] (0,-3pt) -- (0,3pt);}},] (c),
(b) -- [scalar] (i1),
(b) -- [scalar] (i2),
};
\end{feynman}
\end{tikzpicture}
~~~ + \sym 
~+~
\begin{tikzpicture}[baseline = (o.base), scale = 1.2]
\begin{feynman}
\vertex (o) at (0,0) {\(a\)};
\vertex (a) at (0.8,0);
\vertex (i3) at ($(a) + (0.75,0)$) {\(d\)};
\vertex (b) at ($(a) + (0.33,1)$);     
\vertex (c) at ($(a) + (1,0.4)$);    
\vertex (i1) at ($(b) + (0.5,0.6)$) {\(b\)};  
\vertex (i2) at ($(b) + (0.5,0.3)$) {\(c\)};  
\diagram* {
(o) -- [scalar, edge label = \(\kvec\), decoration={markings, 
  mark=at position 0.80 with {\draw[-] (0,-3pt) -- (0,3pt);},
  mark=at position 0.75 with {\draw[-] (0,-3pt) -- (0,3pt);}
}, postaction={decorate},] (a),
(a) -- [scalar] (i3),
(a) -- [anti fermion, decoration={markings, 
  mark=at position 0.80 with {\draw[-] (0,-3pt) -- (0,3pt);},
  mark=at position 0.75 with {\draw[-] (0,-3pt) -- (0,3pt);}
} ] (b),
(b) -- [anti fermion, decoration={markings, mark=at position 0.65 with {\draw[-] (0,-3pt) -- (0,3pt);}}, postaction={decorate}] (c),
(a) -- [anti fermion, decoration={markings, mark=at position 0.65 with {\draw[-] (0,-3pt) -- (0,3pt);}}, postaction={decorate}] (c),
(b) -- [scalar] (i1),
(b) -- [scalar] (i2),
};
\end{feynman}
\end{tikzpicture}
~~~ + \sym 
\end{gather}
\end{subequations}
where $\sym$ refers to the symmetrisation of the diagrams over the incoming legs, such as 
\[
\begin{tikzpicture}[baseline = (o.base), scale = 1.2]
\begin{feynman}
\vertex (o) at (0,0) {\(a\)};
\vertex (a) at (0.8,0);
\vertex (i3) at ($(a) + (0.75,0)$) {\(c\)};
\vertex (b) at ($(a) + (0.33,1)$);     
\vertex (c) at ($(a) + (1,0.4)$);    
\vertex (i1) at ($(b) + (0.5,0.6)$) {\(d\)};  
\vertex (i2) at ($(b) + (0.5,0.3)$) {\(b\)};  
\diagram* {
(o) -- [plain, edge label = \(\kvec\), decoration={markings, 
  mark=at position 0.75 with {\draw[-] (0,-3pt) -- (0,3pt);},
  mark=at position 0.80 with {\draw[-] (0,-3pt) -- (0,3pt);}
}, postaction={decorate},] (a),
(a) -- [plain] (i3),
(a) -- [anti fermion, decoration={markings, 
  mark=at position 0.80 with {\draw[-] (0,-3pt) -- (0,3pt);},
  mark=at position 0.75 with {\draw[-] (0,-3pt) -- (0,3pt);}
} ] (b),
(b) -- [anti fermion, decoration={markings, mark=at position 0.65 with {\draw[-] (0,-3pt) -- (0,3pt);}},] (c),
(a) -- [anti fermion, decoration={markings, mark=at position 0.65 with {\draw[-] (0,-3pt) -- (0,3pt);}},] (c),
(b) -- [plain] (i1),
(b) -- [plain] (i2),
};
\end{feynman}
\end{tikzpicture}
~+~
\begin{tikzpicture}[baseline = (o.base), scale = 1.2]
\begin{feynman}
\vertex (o) at (0,0) {\(a\)};
\vertex (a) at (0.8,0);
\vertex (i3) at ($(a) + (0.75,0)$) {\(b\)};
\vertex (b) at ($(a) + (0.33,1)$);     
\vertex (c) at ($(a) + (1,0.4)$);    
\vertex (i1) at ($(b) + (0.5,0.6)$) {\(c\)};  
\vertex (i2) at ($(b) + (0.5,0.3)$) {\(d\)};  
\diagram* {
(o) -- [plain, edge label = \(\kvec\), decoration={markings, 
  mark=at position 0.75 with {\draw[-] (0,-3pt) -- (0,3pt);},
  mark=at position 0.80 with {\draw[-] (0,-3pt) -- (0,3pt);}
}, postaction={decorate},] (a),
(a) -- [plain] (i3),
(a) -- [anti fermion, decoration={markings, 
  mark=at position 0.80 with {\draw[-] (0,-3pt) -- (0,3pt);},
  mark=at position 0.75 with {\draw[-] (0,-3pt) -- (0,3pt);}
} ] (b),
(b) -- [anti fermion, decoration={markings, mark=at position 0.65 with {\draw[-] (0,-3pt) -- (0,3pt);}},] (c),
(a) -- [anti fermion, decoration={markings, mark=at position 0.65 with {\draw[-] (0,-3pt) -- (0,3pt);}},] (c),
(b) -- [plain] (i1),
(b) -- [plain] (i2),
};
\end{feynman}
\ .
\end{tikzpicture}
\]
Using \Erefs{vertex1300}-\eref{vertex0013}, the vertex at the normalisation point to one loop order takes the form
\begin{subequations}
    \begin{align}
        \left. \vertexuone{}{\kvec} \right|_{NP} &= -\kvecsquared \frac{\Qtensor{}}{3}  \uone \left[  1 - \uone \frac{\numcomponent + 8}{6} \frac{\Gamma_1}{\Done^2} \loopfactor \muepsneg - \frac{\gonetwo \gtwoone}{\uone} \frac{3\numcomponent}{2} \frac{\Gamma_2}{\Dtwo^2} \loopfactor \muepsneg \right],\\
        \left.  \vertexutwo{}{\kvec} \right|_{NP} &= -\kvecsquared \frac{\Qtensor{}}{3}  \utwo \left[ 1 - \utwo \frac{\numcomponent + 8}{6} \frac{\Gamma_2}{\Dtwo^2} \loopfactor \muepsneg - \frac{\gonetwo \gtwoone}{\utwo} \frac{3\numcomponent}{2} \frac{\Gamma_1}{\Done^2} \loopfactor \muepsneg \right] \ ,
    \end{align}
\end{subequations}
which allows us to read off $\diagramcontribution{u_i}$ as of \Erefs{renormconds_ui} and \eref{I_inTermsOf_Z_ui}
\begin{subequations} \label{eq:Zfactoru}
    \begin{align}
        \diagramcontribution{u_1}&=Z_{\Gamma_1} \Zuone = 1 - \left(\uone \frac{\numcomponent + 8}{6} \frac{\Gamma_1}{\Done^2} + \frac{\gonetwo \gtwoone}{\uone} \frac{3\numcomponent}{2} \frac{\Gamma_2}{\Dtwo^2}  \right) \loopfactor \muepsneg \label{eq:Zuone_bare}\\
        \diagramcontribution{u_2}&=Z_{\Gamma_2} \Zutwo = 1 - \left(\utwo \frac{\numcomponent + 8}{6} \frac{\Gamma_2}{\Dtwo^2}  + \frac{\gonetwo \gtwoone}{\utwo} \frac{3\numcomponent}{2} \frac{\Gamma_1}{\Done^2}\right) \loopfactor \muepsneg \label{eq:Zutwo_bare}
    \end{align}
\end{subequations}
using that $Z_{\phitildevec_i}Z^3_{\phivec_i}= Z^{-2}_{\phitildevec_i}=Z_{\Gamma_i}$ as $ Z_{\phitildevec_i}Z_{\phivec_i}= 1$, \Eref{Zfactorrelation1}, and $Z_{\phitildevec_i}^2=1/Z_{\Gamma_i}$, \Eref{Zfactorrelation2}.

We finally consider the renormalisation of the cross-species quartic vertices $\vertexui{}{\kvec}$. To one loop, the (amputated) diagrams are given by 
\begin{subequations} \label{eq:gvertexdiagrams}
\allowdisplaybreaks[4]
    \begin{align}
        \vertexgone{}{\kvec} &= 
        \begin{tikzpicture}[baseline = (o.base), scale = 1.2]
  \begin{feynman}
    \vertex (origin) at (0,0) ;
    \vertex (o) at (-0.6,0) {\(a\)};
    \vertex (i1) at (0.6,0) {\(c\)};
    \vertex (i2) at (0.6,0.6) {\(b\)};
    \vertex (i3) at (0.6,-0.6) {\(d\)};
    \diagram* {
      (origin) -- [plain, decoration={markings, 
  mark=at position 0.20 with {\draw[-] (0,-3pt) -- (0,3pt);},
  mark=at position 0.25 with {\draw[-] (0,-3pt) -- (0,3pt);}
}, postaction={decorate}, edge label'={\(\kvec\)}] (o),
      (origin) -- [scalar] (i1),
      (origin) -- [plain] (i2),
      (origin) -- [scalar] (i3),
    };
  \end{feynman}
\end{tikzpicture}
~+~
\begin{tikzpicture}[baseline = (o.base), scale = 1.2]
\begin{feynman}
\vertex (o) at (0,0) {\(a\)};
\vertex (a) at (0.8,0);
\vertex (i3) at ($(a) + (0.75,0)$) {\(b\)};
\vertex (b) at ($(a) + (0.33,1)$);     
\vertex (c) at ($(a) + (1,0.4)$);    
\vertex (i1) at ($(b) + (0.5,0.6)$) {\(c\)};  
\vertex (i2) at ($(b) + (0.5,0.3)$) {\(d\)};  
\diagram* {
(o) -- [plain, edge label = \(\kvec\), decoration={markings, 
  mark=at position 0.75 with {\draw[-] (0,-3pt) -- (0,3pt);},
  mark=at position 0.8 with {\draw[-] (0,-3pt) -- (0,3pt);}
}, postaction={decorate},] (a),
(a) -- [plain] (i3),
(a) -- [anti fermion, decoration={markings, 
  mark=at position 0.80 with {\draw[-] (0,-3pt) -- (0,3pt);},
  mark=at position 0.75 with {\draw[-] (0,-3pt) -- (0,3pt);}
} ] (b),
(b) -- [anti fermion, decoration={markings, mark=at position 0.65 with {\draw[-] (0,-3pt) -- (0,3pt);}}, postaction={decorate}] (c),
(a) -- [anti fermion, decoration={markings, mark=at position 0.65 with {\draw[-] (0,-3pt) -- (0,3pt);}}, postaction={decorate}] (c),
(b) -- [scalar] (i1),
(b) -- [scalar] (i2),
};
\end{feynman}
\end{tikzpicture}
+ \sym 
+ 
\begin{tikzpicture}[baseline = (o.base),scale = 1.2]
\begin{feynman}
\vertex (o) at (0,0) {\(a\)};
\vertex (a) at (0.8,0);
\vertex (i3) at ($(a) + (0.75,0)$) {\(b\)};
\vertex (b) at ($(a) + (0.33,1)$);     
\vertex (c) at ($(a) + (1,0.4)$);    
\vertex (i1) at ($(b) + (0.5,0.6)$) {\(c\)};  
\vertex (i2) at ($(b) + (0.5,0.3)$) {\(d\)};  
\diagram* {
(o) -- [plain, edge label = \(\kvec\), decoration={markings, 
  mark=at position 0.75 with {\draw[-] (0,-3pt) -- (0,3pt);},
  mark=at position 0.80 with {\draw[-] (0,-3pt) -- (0,3pt);}
}, postaction={decorate},] (a),
(a) -- [plain] (i3),
(a) -- [anti charged scalar, decoration={markings, 
  mark=at position 0.80 with {\draw[-] (0,-3pt) -- (0,3pt);},
  mark=at position 0.75 with {\draw[-] (0,-3pt) -- (0,3pt);}
} ] (b),
(b) -- [anti charged scalar, decoration={markings, mark=at position 0.65 with {\draw[-] (0,-3pt) -- (0,3pt);}},] (c),
(a) -- [anti charged scalar, decoration={markings, mark=at position 0.65 with {\draw[-] (0,-3pt) -- (0,3pt);}},] (c),
(b) -- [scalar] (i1),
(b) -- [scalar] (i2),
};
\end{feynman}
\end{tikzpicture}
+ \sym 
 + 
\begin{tikzpicture}[baseline = (o.base),scale = 1.2]
\begin{feynman}
\vertex (o) at (0,0) {\(a\)};
\vertex (a) at (0.8,0);
\vertex (i3) at ($(a) + (0.75,0)$) {\(c\)};
\vertex (b) at ($(a) + (0.33,1)$);     
\vertex (c) at ($(a) + (1,0.4)$);    
\vertex (i1) at ($(b) + (0.5,0.6)$) {\(b\)};  
\vertex (i2) at ($(b) + (0.5,0.3)$) {\(d\)};  
\diagram* {
(o) -- [plain, edge label = \(\kvec\), decoration={markings, 
  mark=at position 0.75 with {\draw[-] (0,-3pt) -- (0,3pt);},
  mark=at position 0.80 with {\draw[-] (0,-3pt) -- (0,3pt);}
}, postaction={decorate},] (a),
(a) -- [scalar] (i3),
(a) -- [anti fermion, decoration={markings, 
  mark=at position 0.80 with {\draw[-] (0,-3pt) -- (0,3pt);},
  mark=at position 0.75 with {\draw[-] (0,-3pt) -- (0,3pt);}
} ] (b),
(b) -- [anti charged scalar, decoration={markings, mark=at position 0.65 with {\draw[-] (0,-3pt) -- (0,3pt);}},] (c),
(a) -- [anti charged scalar, decoration={markings, mark=at position 0.65 with {\draw[-] (0,-3pt) -- (0,3pt);}},] (c),
(b) -- [plain] (i1),
(b) -- [scalar] (i2),
};
\end{feynman}
\end{tikzpicture} 
+ \sym \nonumber  \\
& + 
\begin{tikzpicture}[baseline = (o.base),scale = 1.2]
\begin{feynman}
\vertex (o) at (0,0) {\(a\)};
\vertex (a) at (0.8,0);
\vertex (i3) at ($(a) + (0.75,0)$) {\(c\)};
\vertex (b) at ($(a) + (0.33,1)$);     
\vertex (c) at ($(a) + (1,0.4)$);    
\vertex (i1) at ($(b) + (0.5,0.6)$) {\(b\)};  
\vertex (i2) at ($(b) + (0.5,0.3)$) {\(d\)};  
\diagram* {
(o) -- [plain, edge label = \(\kvec\), decoration={markings, 
  mark=at position 0.75 with {\draw[-] (0,-3pt) -- (0,3pt);},
  mark=at position 0.80 with {\draw[-] (0,-3pt) -- (0,3pt);}
}, postaction={decorate},] (a),
(a) -- [scalar] (i3),
(a) -- [anti charged scalar, decoration={markings, 
  mark=at position 0.80 with {\draw[-] (0,-3pt) -- (0,3pt);},
  mark=at position 0.75 with {\draw[-] (0,-3pt) -- (0,3pt);}
} ] (b),
(b) -- [anti fermion, decoration={markings, mark=at position 0.65 with {\draw[-] (0,-3pt) -- (0,3pt);}},] (c),
(a) -- [anti fermion, decoration={markings, mark=at position 0.65 with {\draw[-] (0,-3pt) -- (0,3pt);}},] (c),
(b) -- [plain] (i1),
(b) -- [scalar] (i2),
};
\end{feynman}
\end{tikzpicture} 
+ \sym,  \label{eq:diagramsvertexgonetwo}
\\
    \vertexgtwo{}{\kvec} &= 
    \begin{tikzpicture}[baseline = (o.base), scale = 1.2]
  \begin{feynman}
    \vertex (origin) at (0,0) ;
    \vertex (o) at (-0.6,0) {\(a\)};
    \vertex (i1) at (0.6,0) {\(c\)};
    \vertex (i2) at (0.6,0.6) {\(b\)};
    \vertex (i3) at (0.6,-0.6) {\(d\)};
    \diagram* {
      (origin) -- [scalar, decoration={markings, 
  mark=at position 0.25 with {\draw[-] (0,-3pt) -- (0,3pt);},
  mark=at position 0.30 with {\draw[-] (0,-3pt) -- (0,3pt);}
}, postaction={decorate}, edge label'={\(\kvec\)}] (o),
      (origin) -- [plain] (i1),
      (origin) -- [scalar] (i2),
      (origin) -- [plain] (i3),
    };
  \end{feynman}
\end{tikzpicture}
+ 
\begin{tikzpicture}[baseline = (o.base), scale = 1.2]
\begin{feynman}
\vertex (o) at (0,0) {\(a\)};
\vertex (a) at (0.8,0);
\vertex (i3) at ($(a) + (0.75,0)$) {\(b\)};
\vertex (b) at ($(a) + (0.33,1)$);     
\vertex (c) at ($(a) + (1,0.4)$);    
\vertex (i1) at ($(b) + (0.5,0.6)$) {\(c\)};  
\vertex (i2) at ($(b) + (0.5,0.3)$) {\(d\)};  
\diagram* {
(o) -- [scalar, edge label = \(\kvec\), decoration={markings, 
  mark=at position 0.80 with {\draw[-] (0,-3pt) -- (0,3pt);},
  mark=at position 0.75 with {\draw[-] (0,-3pt) -- (0,3pt);}
}, postaction={decorate},] (a),
(a) -- [scalar] (i3),
(a) -- [anti fermion, decoration={markings, 
  mark=at position 0.80 with {\draw[-] (0,-3pt) -- (0,3pt);},
  mark=at position 0.75 with {\draw[-] (0,-3pt) -- (0,3pt);}
} ] (b),
(b) -- [anti fermion, decoration={markings, mark=at position 0.65 with {\draw[-] (0,-3pt) -- (0,3pt);}},] (c),
(a) -- [anti fermion, decoration={markings, mark=at position 0.65 with {\draw[-] (0,-3pt) -- (0,3pt);}},] (c),
(b) -- [plain] (i1),
(b) -- [plain] (i2),
};
\end{feynman}
\end{tikzpicture}
 + \sym 
+
\begin{tikzpicture}[baseline = (o.base), scale = 1.2]
\begin{feynman}
\vertex (o) at (0,0) {\(a\)};
\vertex (a) at (0.8,0);
\vertex (i3) at ($(a) + (0.75,0)$) {\(b\)};
\vertex (b) at ($(a) + (0.33,1)$);     
\vertex (c) at ($(a) + (1,0.4)$);    
\vertex (i1) at ($(b) + (0.5,0.6)$) {\(c\)};  
\vertex (i2) at ($(b) + (0.5,0.3)$) {\(d\)};  
\diagram* {
(o) -- [scalar, edge label = \(\kvec\), decoration={markings, 
  mark=at position 0.80 with {\draw[-] (0,-3pt) -- (0,3pt);},
  mark=at position 0.75 with {\draw[-] (0,-3pt) -- (0,3pt);}
}, postaction={decorate},] (a),
(a) -- [scalar] (i3),
(a) -- [anti charged scalar, decoration={markings, 
  mark=at position 0.80 with {\draw[-] (0,-3pt) -- (0,3pt);},
  mark=at position 0.75 with {\draw[-] (0,-3pt) -- (0,3pt);}
} ] (b),
(b) -- [anti charged scalar, decoration={markings, mark=at position 0.65 with {\draw[-] (0,-3pt) -- (0,3pt);}}, postaction={decorate}] (c),
(a) -- [anti charged scalar, decoration={markings, mark=at position 0.65 with {\draw[-] (0,-3pt) -- (0,3pt);}}, postaction={decorate}] (c),
(b) -- [plain] (i1),
(b) -- [plain] (i2),
};
\end{feynman}
\end{tikzpicture}
 + \sym +
\begin{tikzpicture}[baseline = (o.base),scale = 1.2]
\begin{feynman}
\vertex (o) at (0,0) {\(a\)};
\vertex (a) at (0.8,0);
\vertex (i3) at ($(a) + (0.75,0)$) {\(d\)};
\vertex (b) at ($(a) + (0.33,1)$);     
\vertex (c) at ($(a) + (1,0.4)$);    
\vertex (i1) at ($(b) + (0.5,0.6)$) {\(c\)};  
\vertex (i2) at ($(b) + (0.5,0.3)$) {\(b\)};  
\diagram* {
(o) -- [scalar, edge label = \(\kvec\), decoration={markings, 
  mark=at position 0.75 with {\draw[-] (0,-3pt) -- (0,3pt);},
  mark=at position 0.80 with {\draw[-] (0,-3pt) -- (0,3pt);}
}, postaction={decorate},] (a),
(a) -- [plain] (i3),
(a) -- [anti fermion, decoration={markings, 
  mark=at position 0.80 with {\draw[-] (0,-3pt) -- (0,3pt);},
  mark=at position 0.75 with {\draw[-] (0,-3pt) -- (0,3pt);}
} ] (b),
(b) -- [anti charged scalar, decoration={markings, mark=at position 0.65 with {\draw[-] (0,-3pt) -- (0,3pt);}},] (c),
(a) -- [anti charged scalar, decoration={markings, mark=at position 0.65 with {\draw[-] (0,-3pt) -- (0,3pt);}},] (c),
(b) -- [plain] (i1),
(b) -- [scalar] (i2),
};
\end{feynman}
\end{tikzpicture} 
+ \sym \nonumber  \\
& + 
\begin{tikzpicture}[baseline = (o.base),scale = 1.2]
\begin{feynman}
\vertex (o) at (0,0) {\(a\)};
\vertex (a) at (0.8,0);
\vertex (i3) at ($(a) + (0.75,0)$) {\(d\)};
\vertex (b) at ($(a) + (0.33,1)$);     
\vertex (c) at ($(a) + (1,0.4)$);    
\vertex (i1) at ($(b) + (0.5,0.6)$) {\(c\)};  
\vertex (i2) at ($(b) + (0.5,0.3)$) {\(b\)};  
\diagram* {
(o) -- [scalar, edge label = \(\kvec\), decoration={markings, 
  mark=at position 0.75 with {\draw[-] (0,-3pt) -- (0,3pt);},
  mark=at position 0.80 with {\draw[-] (0,-3pt) -- (0,3pt);}
}, postaction={decorate},] (a),
(a) -- [plain] (i3),
(a) -- [anti charged scalar, decoration={markings, 
  mark=at position 0.80 with {\draw[-] (0,-3pt) -- (0,3pt);},
  mark=at position 0.75 with {\draw[-] (0,-3pt) -- (0,3pt);}
} ] (b),
(b) -- [anti fermion, decoration={markings, mark=at position 0.65 with {\draw[-] (0,-3pt) -- (0,3pt);}},] (c),
(a) -- [anti fermion, decoration={markings, mark=at position 0.65 with {\draw[-] (0,-3pt) -- (0,3pt);}},] (c),
(b) -- [plain] (i1),
(b) -- [scalar] (i2),
};
\end{feynman}
\end{tikzpicture} 
+ \sym \ . \label{eq:diagramsvertexgtwoone}
\end{align}
\end{subequations}
We observe that the loop-diagrams are identical in \Erefs{diagramsvertexgonetwo} and \eref{diagramsvertexgtwoone}, further qualified below. Using \Erefs{vertex0211}-\Eref{vertex1102}, the cross-species quartic vertices to one loop are
\begin{subequations}
\elabel{cross-vertex_algebra}
    \begin{gather}
        \left. \vertexgone{}{\kvec} \right|_{NP} = -\kvec^2 \kroneckerdelta{} \kroneckerdelta{cd} \gonetwo \left[ 1 - \left( u_1\frac{n+2}{6}\frac{\Gamma_1}{D_1^2} + \utwo \frac{\numcomponent + 2}{6}\frac{\Gamma_2}{\Dtwo^2}  + \frac{2 \gonetwo \Gamma_2}{\Dtwo(\Done + \Dtwo)}  + \frac{2 \gtwoone \Gamma_1}{\Done(\Done + \Dtwo)}  \right)\loopfactor \muepsneg \right]
       \\
        \left. \vertexgtwo{}{\kvec} \right|_{NP} = -\kvec^2 \kroneckerdelta{} \kroneckerdelta{cd} \gtwoone \left[ 1 - \left( u_1\frac{n+2}{6}\frac{\Gamma_1}{D_1^2} + \utwo \frac{\numcomponent + 2}{6}\frac{\Gamma_2}{\Dtwo^2}  + \frac{2 \gonetwo \Gamma_2}{\Dtwo(\Done + \Dtwo)}  + \frac{2 \gtwoone \Gamma_1}{\Done(\Done + \Dtwo)}  \right)\loopfactor \muepsneg \right]
    \end{gather}
\end{subequations}
with identical expressions in the square brackets, so that both vertices are being renormalised identically. \Erefs{cross-vertex_algebra}
allow us to read off $\diagramcontribution{g_{ij}}$ as of \Erefs{renormconds_gij} and \eref{I_inTermsOf_Z_gij}
\begin{equation} \label{eq:Zfactorcrosscoupling}
        \diagramcontribution{g_{12}} = \diagramcontribution{g_{21}} = Z_{\Gamma_2} Z_{\gonetwo} = Z_{\Gamma_1} Z_{\gtwoone} =  1 - \left( u_1\frac{n+2}{6}\frac{\Gamma_1}{D_1^2} + \utwo \frac{\numcomponent + 2}{6}\frac{\Gamma_2}{\Dtwo^2}  + \frac{2 \gonetwo \Gamma_2}{\Dtwo(\Done + \Dtwo)}  + \frac{2 \gtwoone \Gamma_1}{\Done(\Done + \Dtwo)} \right)\loopfactor \muepsneg
\end{equation}
using that $Z_{\phitildevec_i}Z_{\phivec_i} Z^2_{\phivec_j}= Z^{-2}_{\phitildevec_j}=Z_{\Gamma_j}$ as $ Z_{\phitildevec_i}Z_{\phivec_i}= 1$, \Eref{Zfactorrelation1}, and $Z_{\phitildevec_i}^2=1/Z_{\Gamma_i}$, \Eref{Zfactorrelation2}.

All $Z$-factors, \Erefs{Zfactorpropagator}, \eref{Zfactoru} and \eref{Zfactorcrosscoupling} are expressed in terms of ten bare parameters of the theory, namely $\tau_i$, $u_i$, $D_i$, $\Gamma_i$, $\gonetwo$ and $\gtwoone$. However, we can define new dimensionless expansion parameters to reduce their number to seven,
\begin{equation} \label{eq:expansionparameterdefn}
        \utildei = \frac{u_i \Gamma_i}{(4 \pi)^{d/2}D_i^2}, ~ w = \frac{D_1}{D_2}, ~\gtildeonetwo = \frac{\gonetwo \Gamma_2}{(4 \pi)^{d/2}\Done \Dtwo}, ~ \gtildetwoone = \frac{\gtwoone \Gamma_1}{(4 \pi)^{d/2}\Done \Dtwo}.
\end{equation}
We further define their renormalised counterparts as
\begin{equation} \label{eq:expansionparametersrenorm}
        \utildeiR = Z_{\utildei} \utildei \muepsneg, ~ w_R = Z_{w} w ,~
        \gtildeonetwoR = Z_{\gtildeonetwo} \gtildeonetwo \muepsneg, ~ \gtildetwooneR = Z_{\gtildetwoone} \gtildetwoone \muepsneg
\end{equation}
With these definitions in place, we derive their $Z$ factors in terms of $\{\utildeiR, \gtildeonetwoR, \gtildetwooneR, w_R\}$ to leading order in $1/\epsilon$ and $\lambda_R$. Using \Erefs{expansionparameterdefn} and \eref{expansionparametersrenorm} as well as $\Gamma(\epsilon/2) \sim 2/\epsilon$ for small $\epsilon$, we have together with \Eref{Ztauone_bare} and \eref{Ztautwo_bare}
\begin{subequations} \elabel{Zfactors}
\allowdisplaybreaks[4]
    \begin{align}
         \Zrone &= 1 -  \frac{\numcomponent+2}{3} \frac{ \utildeoneR}{\epsilon}  -  \numcomponent \frac{\gtildeonetwoR w_R }{ \epsilon} \frac{\tau_{2,R}}{\tau_{1,R}} +\mathcal{O}(\lambda_R^2) \label{eq:Zfactors_tauone}\\
        \Zrtwo &= 1 -  \frac{\numcomponent+2}{3} \frac{ \utildetwoR }{\epsilon}  -  \numcomponent \frac{\gtildetwooneR }{ \epsilon w_R} \frac{\tau_{1,R}}{\tau_{2,R}} +\mathcal{O}(\lambda_R^2) \label{eq:Zfactors_tautwo}
\ ,
\end{align}
together with \Erefs{ZDone_bare} and \eref{ZDtwo_bare}
    \begin{align}
        Z_w &= 1 +\mathcal{O}(\lambda_R^2)
\ ,
\end{align}
together with \Erefs{Zuone_bare}, \eref{Zutwo_bare}, \eref{ZDone_bare} and \eref{ZDtwo_bare}
    \begin{align}
        Z_{\utildeone} &= 1 -  \frac{\numcomponent + 8}{3} \frac{\utildeoneR }{\epsilon}   - 3 \numcomponent \frac{\gtildeonetwoR \gtildetwooneR }{\utildeoneR \epsilon} + \mathcal{O}(\lambda_R^2)   \\
         Z_{\utildetwo} &= 1 -  \frac{\numcomponent + 8}{3} \frac{\utildetwoR }{\epsilon}   - 3 \numcomponent \frac{\gtildeonetwoR \gtildetwooneR }{\utildetwoR \epsilon}  +\mathcal{O}(\lambda_R^2)
\ ,
\end{align}
and finally with \Erefs{Zfactorcrosscoupling}, \eref{ZDone_bare} and \eref{ZDtwo_bare}
    \begin{align}
         Z_{\gtildeonetwo} &= Z_{\gtildetwoone} = 1 -   \frac{\numcomponent + 2}{3}\frac{\utildeoneR }{\epsilon} -  \frac{\numcomponent + 2}{3}\frac{\utildetwoR }{\epsilon} - 4\frac{w_R}{1 + w_R}\frac{\gtildeonetwoR }{\epsilon} - \frac{4}{1+w_R} \frac{\gtildetwooneR }{\epsilon} +\mathcal{O}(\lambda_R^2) \ .
    \end{align}
\end{subequations}
where we used $\ZDone = \ZDtwo=1 + \mathcal{O}(\lambda^2)$, \Erefs{ZDone_bare} and \eref{ZDtwo_bare}, to obtain the $Z-$factors for $\utildei$, $\Zgonetwo$ and $\Zgtwoone$. To leading order in $\lambda_R$ we simply replaced the bare couplings by their renormalised counterparts using \Eref{expansionparametersrenorm}. In the next section, we use these $Z$-factors \Erefs{Zfactors} to characterise the fixed points of the theory. 

\subsection{Fixed Points of the RG flow} \seclabel{criticalpoints}
\Erefs{Zfactors} show all $Z$-factors in terms of the new effective expansion parameters 
$\utildeiR$, $\gtildeonetwoR$, $\gtildetwooneR$ and $w_R$. 

To identify the fixed points of our theory, we first calculate the $\beta$-functions for these couplings using their definitions \Erefs{expansionparametersrenorm}:
\begin{subequations} \elabel{betafuncdefn}
    \begin{align}
        \betautildei &= \mu \frac{\plaind \tilde{u}_{i,R}}{\plaind \mu} = \utildeiR \left(-\epsilon + \mu \frac{\plaind \logzutildei}{\plaind \mu} \right) \ , 
        \qquad
        &\betaw &= \mu \frac{\plaind w_R}{\plaind \mu} = w_R \left( \mu\frac{\plaind \log Z_{w}}{\plaind \mu} \right) \\
        \betagtildeonetwo &= \mu \frac{\partial \gtildeonetwoR}{\partial \mu} = \gtildeonetwoR \left(-\epsilon + \mu \frac{\plaind \logzgtildeonetwo}{\plaind \mu} \right) \ , 
        \qquad 
        &\betagtildetwoone &= \mu \frac{\plaind \gtildetwooneR}{\plaind \mu} = \gtildetwooneR \left(-\epsilon + \mu \frac{\plaind \logzgtildetwoone}{\plaind \mu} \right)
\ .
    \end{align}
\end{subequations}
To determine them explicitly, we use \Erefs{Zfactors}, replacing $\mu \plaind \lambda_R/\plaind \mu$ by their leading orders $-\epsilon\lambda_R$ according to \Erefs{betafuncdefn}, 
\begin{subequations} \elabel{Flow}
\allowdisplaybreaks[4]
\begin{align}
    \betautildeone &= -\epsilon \utildeoneR + \frac{\numcomponent + 8}{3} \utildeoneR^2 + 3 \numcomponent \gtildeonetwoR \gtildetwooneR \elabel{Flow_uone}\\
    \betautildetwo &= -\epsilon \utildetwoR + \frac{\numcomponent + 8}{3} \utildetwoR^2 + 3 \numcomponent \gtildeonetwoR \gtildetwooneR \elabel{Flow_utwo}\\
    \betagtildeonetwo &= \gtildeonetwoR \left( -\epsilon + 4\frac{w \gtildeonetwoR + \gtildetwooneR}{1 + w} + \frac{\numcomponent + 2}{3} \utildeoneR + \frac{\numcomponent + 2}{3} \utildetwoR \right) \elabel{Flow_gonetwo}\\
    \betagtildetwoone &= \gtildetwooneR \left( -\epsilon + 4\frac{w \gtildeonetwoR + \gtildetwooneR}{1 + w} + \frac{\numcomponent + 2}{3} \utildeoneR + \frac{\numcomponent + 2}{3} \utildetwoR \right) \elabel{Flow_gtwoone}\\
    \betaw &= 0 \elabel{Flow_w}
\end{align} 
\end{subequations}
where \Erefs{Flow_gonetwo} and \eref{Flow_gtwoone} have been written to indicate that $\gtildeonetwoR=0$ and $\gtildetwooneR=0$ respectively are roots of these $\beta$-functions. Indeed, $\gtildeonetwoR=\gtildetwooneR=0$ correspond to the equilibrium Model~B and neither of the two couplings can be generated under the RG from other non-linearities. As the brackets of \Erefs{Flow_gonetwo} and \eref{Flow_gtwoone} are identical to one loop order, it follows immediately that $\betagtildeonetwo/\gtildeonetwoR = \betagtildetwoone/\gtildetwooneR$. At the level of the vertex functions, this follows from $\vertexgone{}{}/\gonetwo = \vertexgtwo{}{}/\gtwoone$. As a consequence, the ratio $\sigma_R=\gtildeonetwoR/\gtildetwooneR$, \Erefs{def_sigma} and \eref{expansionparameterdefn}, does not flow to one-loop order and will thus be set by the bare values. Similarly, $w_R$, \Erefs{expansionparameterdefn} and \eref{Flow_w}, does not flow to one loop. The fixed points of the present model  and potentially the scaling exponents, will therefore depend on the bare values of $w$ and $\sigma$ to one loop. 
This ``non-universality'' resolves at two loops, however, calculating all vertices to two loops is beyond the scope of the present work. For a model similar to ours, Young \etal \cite{young_nonequilibrium_2020,young_nonequilibrium_2024} decided to consider the flow only of $w_R$ as well as $v = \Gamma_2/\Gamma_1$, which does not flow in the present model to any order, but leave the variable corresponding to $\gtwoone/\gonetwo=\sigma v$ at the bare value.

In the following, we present the analysis of the one-loop flow functions, in particular the dependence of the fixed points and their stability on the ratio between the inter-species coupling $\sigma = \gtildetwoone/\gtildeonetwo$ and the ratio of diffusivity $w=D_1/D_2$. In equilibrium, where $\sigma=1$, \Erefs{Langevin_from_SuperHamiltonian}, \Eref{def_SuperHamiltonian} and \Eref{def_sigma}, corresponding models with bi-quadratic couplings display a rich phase diagram where the stability of the fixed points depend strongly on the number of the field components $n$ \cite{ft_of_bicritical_moser_1,ft_of_bicritical_moser_2,multicritical_ising}. However, it was shown Young \etal \cite{young_nonequilibrium_2020, young_nonequilibrium_2024} that for $\sigma \in \{-1,0\}$, the fixed point structure is far simpler. Below, we show how the number of stable fixed points depends on the sign of $\sigma$.

To ease the search for the fixed points, we reproduce in \Erefs{Flowred} three of the five flow functions in \Erefs{Flow}, dropping $\beta_w\equiv0$ as well as $\betagtildetwoone$, which is essentially identical to $\betagtildeonetwo$. To one loop, we will need to content ourselves with the presence of the non-universal couplings $\sigma$ and $w$.

Restating \Erefs{Flow}, we will need to find the roots of
\begin{subequations} \label{eq:Flowred}
    \begin{align}
            \betautildeone &= -\epsilon \utildeoneR + \frac{n + 8}{3} \utildeoneR^2 + 3 \sigma n \gtildeonetwoR^2 \label{eq:betautildeone}\\
    \betautildetwo &= -\epsilon \utildetwoR + \frac{n + 8}{3} \utildetwoR^2 + 3 \sigma n \gtildeonetwoR^2 \label{eq:betautildetwo}\\
    \betagtildeonetwo &= \gtildeonetwoR \left( -\epsilon + 4\frac{w  + \sigma}{w + 1} \gtildeonetwoR + \frac{n + 2}{3} \utildeoneR + \frac{n + 2}{3} \utildetwoR \right) \label{eq:betagtildeonetwo}
\ ,
    \end{align}
\end{subequations}
and determine the stability of these fixed points, verifying that at a given fixed point the matrix 
\begin{equation} \label{eq:stabilitymat}
    \Lambda_{ab} = \left. \frac{\partial \beta_{\tilde{\lambda}_a}}{\partial \tilde{\lambda}_b} \right|_{\tilde{\lambda}_{b} = \tilde{\lambda}^{\star}_b} \quad \text{so that} \quad
    \bm{\Lambda} = \begin{pmatrix}
        -\epsilon + \frac{2(n+8)}{3} \utildeonestar & 0 & 6 n \sigma \gtildeonetwostar \\
         0 & -\epsilon + \frac{2(n+8)}{3} \utildetwostar & 6 n \sigma \gtildeonetwostar \\
        \frac{1}{3} (n+2) \gtildeonetwostar & \frac{1}{3} (n+2) \gtildeonetwostar & -\epsilon + 8\frac{w+\sigma}{w+1}\gtildeonetwostar + \frac{(n+2)}{3}(\utildeonestar + \utildetwostar)
    \end{pmatrix}    
\end{equation}
for $\tilde{\lambda}_a \in \{\utildeoneR, \utildetwoR, \gtildeonetwoR\}$ is positive definite to render it IR stable.

A detailed discussion of the different fixed points $\{\utildeonestar, \utildetwostar, \gtildeonetwostar\}$ and their stability is relegated to \APref{AnalysisFlow}. We state the relevant results here. Firstly, 
\begin{equation} \elabel{first_fixed_points}
    (\utildeonestar \scalebox{1.5}{,} \utildetwostar \scalebox{1.5}{,} \gtildeonetwostar) = \begin{cases}
        \left(\frac{4-n}{4(n+2)} \frac{h(\sigma,w,n)}{h(\sigma,w,n) + \frac{6}{n+2}}\epsilon + \frac{\epsilon}{4}\scalebox{1.5}{,} \frac{4-n}{4(n+2)} \frac{h(\sigma,w,n)}{h(\sigma,w,n) + \frac{6}{n+2}}\epsilon + \frac{\epsilon}{4}\scalebox{1.5}{,} \frac{4-n}{4(n+2)} \frac{w+1}{w+\sigma} \frac{\epsilon}{\frac{6}{n+2} + h(\sigma,w,n)}\right), ~~ n<4 \\
        \left(\frac{3 \epsilon}{n+8}\scalebox{1.5}{,} \frac{3 \epsilon}{n+8}\scalebox{1.5}{,}0 \right), ~ n> 4
    \end{cases}
\end{equation}
with 
\begin{equation} \label{eq:hfactor}
    h(\sigma,w,n) \defequal \sqrt{1 - \sigma \frac{n(4-n)(w+1)^2}{4(w+\sigma)^2}}. 
\end{equation}
is a stable solution for all $\sigma$ and $w$, provided only the solution \Eref{first_fixed_points} is real. As in equilibrium \cite{ft_of_bicritical_moser_1,ft_of_bicritical_moser_2,multicritical_ising}, the stability of the fixed points depends on the number of components $n$. For $n>4$ and any $\sigma$, at the stable fixed point \Eref{first_fixed_points} the two order parameter fields $\phivec_1$ and $\phivec_2$ decouple, as $\gtildeonetwostar=0$, so that the present model reduces into two independent Model~B. For $n <4$, the fixed points depend non-trivially on the bare couplings $\sigma$ and $w$. 

Secondly, for $\sigma \in  (-\sigma_{+}, -\sigma_{-})$, where
\begin{equation}
    \sigma_{\pm} \defequal \left( w + \frac{n(n+2)^2}{8(n+8)} (w+1)^2\right)\left( 1 \pm \sqrt{1 - \left(\frac{w}{ w + \frac{n(n+2)^2}{8(n+8)} (w+1)^2} \right)^2}\right)\,
\end{equation}
another stable fixed point exists for all $n \in \Nset^{+}$
\begin{equation} \label{eq:second_fixed_point}
    (\utildeonestar, \utildetwostar, \gtildeonetwostar) =
        \left(\frac{4-n}{4(n+2)} \frac{h(\sigma,w,n)}{h(\sigma,w,n) - \frac{6}{n+2}}\epsilon + \frac{\epsilon}{4}\scalebox{1.5}{,} \frac{4-n}{4(n+2)} \frac{h(\sigma,w,n)}{h(\sigma,w,n) - \frac{6}{n+2}}\epsilon + \frac{\epsilon}{4}\scalebox{1.5}{,} \frac{4-n}{4(n+2)} \frac{w+1}{w+\sigma} \frac{\epsilon}{\frac{6}{n+2} - h(\sigma,w,n)}\right).
\end{equation}
For all physical parameters values, $w \in \Rset^{+}$ and $n \in \Nset^+$, $\sigma_{\pm}$ is greater than unity so that the region $(-\sigma_{+}, -\sigma_{-})$ always comes to lie on the negative half line. The existence of the additional fixed point, \Eref{second_fixed_point} suggests that non-reciprocity alters the topology of the one-loop fixed points, inducing other stable points of the RG flow.

\subsection{Scaling Exponents} \label{sec:scalingexp}
The physics at the fixed points is scale invariant and consequently physical observables typically have a power-law dependence. The scaling exponents of these observables define the universality class of the model and can be used to characterise the system. As usual, we are interested in the scaling behaviour of the response and correlation functions. We expect them to take the following form at the critical point \cite{tauber_critical_2014}:
\begin{subequations} \label{eq:scalingfunctions}
\begin{align}
    \chi_i(\kvec, \omega) &= |\kvec|^{-2 + \frac{\eta_i +\etatilde_i}{2}} \hat{\chi}_i\left( \frac{\omega}{|\kvec|^{z_i}}, \frac{r_1}{D_1 |\kvec|^{1/\nu_{1}}}, \frac{r_2}{D_2|\kvec|^{1/\nu_{2}}}, \ldots \right), \label{eq:scalingformpropagator}\\
    C_i(\kvec, \omega) &= |\kvec|^{-2 - z_i + \eta_i}\hat{C}_i\left( \frac{\omega}{|\kvec|^{z_i}},  \frac{r_1}{D_1|\kvec|^{1/\nu_{1}}}, \frac{r_2}{D_2|\kvec|^{1/\nu_{2}}}, \ldots \right), \label{eq:scalingformcorrelator}
\end{align}
\end{subequations}
where $\chi_i(\kvec, \omega)$ and $C_i(\kvec, \omega)$ are the response and correlation functions for the species $i \in \{1,2\}$, respectively. The universal exponents $\eta_i$, $z_i$, $\etatilde_i$ and $\nu_i$ have the usual meaning: $z_i$ is the dynamical exponent for the species $i$, controlling the relative scaling of space and time, $\nu_i$ controls the divergence of the correlation length as a function of $r_i$, and $\eta_i$, $\etatilde_i$ are anomalous exponents. In the absence of a fluctuation-dissipation theorem, there is no reason why the latter two should equal each other.

In the following, we derive the exponents defined in \Eref{scalingfunctions} from the Wilson $\gamma$-functions at the fixed points found \Sref{1loop}. To this end, we write the correlation functions \Eref{scalingfunctions} in terms of vertex functions \cite{tauber_critical_2014}, 
\begin{subequations} \elabel{vertextofuncs}
    \begin{align}
        \chi_{i}(\kvec, \omega) & = \kvecsquared/\invpropifunc{} \\
        C_i(\kvec, \omega) & = \vertexnoiseifunc{}/|\invpropifunc{}|^2 \label{eq:Corr_to_vertex}
    \end{align}
\end{subequations}
where we have removed the dependence of the vertices and the correlation functions on the component indices $a$ and $b$, \eg \Eref{invprop_oneloop_algebra}, because vertices and correlation functions are diagonal and the correlation functions draw on the matrix-inverse of the vertex functions, \Eref{vertexfuncdefn}.

As a first step, we extract the scaling of the renormalised vertex functions \Erefs{Renvertexfuncdefn} and \eref{renvertextonormalvertex} by solving the Callan-Symanzik equation. To this end, we need to make explicit the parameters that the renormalised and the bare vertex functions depend on, \Eref{renvertextonormalvertex}. It turns out that writing the vertex functions with an explicit dependence on $\tau_{1}$ and $\tau_2$, as well as the corresponding renormalised $\tau_{1,R}$ and $\tau_{2,R}$, 
\Eref{Zfactorsdefn_propagators},
is not convenient, as their flow is coupled,
\begin{equation} \label{eq:flowoftau}
    \mu \frac{\plaind}{\plaind\mu} 
    \begin{pmatrix}
        \tau_{1,R}(\mu)\\
        \tau_{2,R}(\mu)
    \end{pmatrix}
     = 
     \begin{pmatrix}
         -2 + \frac{n+2}{3} \utildeonestar & n w \gtildeonetwostar \\
         n \frac{\sigma}{w} \gtildeonetwostar&  -2 + \frac{n+2}{3} \utildetwostar
     \end{pmatrix}
     \begin{pmatrix}
        \tau_{1,R}(\mu)\\
        \tau_{2,R}(\mu)
    \end{pmatrix}
\end{equation}
from 
\Erefs{Zfactorsdefn_propagators}, \eref{Zfactors_tauone} and \eref{Zfactors_tautwo}, where the couplings are evaluated at the fixed points of the beta functions \Erefs{Flowred}. To ease notation, we use that $\utildeonestar=\utildetwostar$ henceforth.

Denoting the left eigenvectors of the matrix in \Eref{flowoftau} by $\bra{t_1}$ and $\bra{t_2}$ with eigenvalues 
\begin{equation}\elabel{gamma_tautilde_actually}
\gamma_{\tautilde_1}= -2 + \frac{n+2}{3} \utildeonestar + n  \sqrt{\sigma} \gtildeonetwostar~
\text{ and } 
~
\gamma_{\tautilde_2}= -2 + \frac{n+2}{3} \utildeonestar - n \sqrt{\sigma} \gtildeonetwostar \ ,
\end{equation}
and correspondingly the right eigenvectors by $\ket{t_1}$ and $\ket{t_2}$, we introduce
\begin{subequations}
\begin{align}
    \tautilde_{1,R}(\mu)&=\bra{t_1}
    \begin{pmatrix}
        \tau_{1,R}(\mu)\\
        \tau_{2,R}(\mu)
    \end{pmatrix} \\
    \tautilde_{2,R}(\mu)&=\bra{t_2}
    \begin{pmatrix}
        \tau_{1,R}(\mu)\\
        \tau_{2,R}(\mu)
    \end{pmatrix}
\end{align}
\end{subequations}
and arrive at the scaling form
\begin{subequations}
\begin{align}
     \tautilde_{1,R}(\mu) & = \tautilde_{1,R}(\mu_0) \ell^{\gamma_{\tautilde_1}}\\
     \tautilde_{2,R}(\mu) & = \tautilde_{2,R}(\mu_0) \ell^{\gamma_{\tautilde_2}}
\end{align}
\end{subequations}
with $\ell=\mu/\mu_0$ henceforth.

Introducing the vertex functions with the full set of parameters along the lines of \Eref{defs_vertex_shorthands} and \eref{renvertextonormalvertex}, say
\begin{multline}\elabel{GammaR_from_Gamma}
            Z_{\phitildevec_i}(\mu) Z_{\phivec_i}(\mu) 
\invpropiren{ab}{
            \tautilde_{1,R}(\mu),
            \tautilde_{2,R}(\mu),
            D_{1,R}(\mu),
            D_{2,R}(\mu),
            \Gamma_{1,R}(\mu),
            \Gamma_{2,R}(\mu),
            \utilde_{1,R}(\mu),
            \utilde_{2,R}(\mu),
            \gtildeonetwoR(\mu),
            \sigma_R(\mu);
            \{\kvec\},\{\omega\}; 
            \mu} \\         
= \invpropi{ab}{
             \tau_1,
             \tau_2,
             D_1,
             D_2,
             \Gamma_1,
             \Gamma_2,
             u_1,
             u_2,
             \gonetwo,
             \gtwoone;
            \{\kvec\},\{\omega\}},
\end{multline}
with some of the parameters on the left defined via \Eref{expansionparameterdefn}. 
\Eref{GammaR_from_Gamma} leads us to the Callan-Symanzik equation \cite{bellac_quantum_1992} by differentiation with respect to $\mu$ on both sides.  It is solved by the method of characteristics and leads to 
\begin{multline}\elabel{CSeqn_soln}
                \ell^{\gammaphitildeistar + \gammaphiistar}
\Gamma^{ab}_{R, \phitilde_i \phi_i}\big(
            \ell^{\gammatautildeonestar} \tautilde_{1,R}(\mu_0),
            \ell^{\gammatautildetwostar} \tautilde_{2,R}(\mu_0),
            \ell^{\gamma^{*}_{D_1}} D_{1,R}(\mu_0),
            \ell^{\gamma^{*}_{D_2}} D_{2,R}(\mu_0), \\
            \ell^{\gammaGammaonestar} \Gamma_{1,R}(\mu_0),
            \ell^{\gammaGammatwostar} \Gamma_{2,R}(\mu_0),
            \utildestar_{1,R},
            \utildestar_{2,R},
            \gtildeonetwostar,
            \sigmastar;
            \{\kvec\},\{\omega\}; 
            \ell \mu_0 \big)\\         
=
\invpropiren{ab}{
            \tautilde_{1,R}(\mu_0),
            \tautilde_{2,R}(\mu_0),
            D_{1,R}(\mu_0),
            D_{2,R}(\mu_0),
            \Gamma_{1,R}(\mu_0),
            \Gamma_{2,R}(\mu_0),
            \utilde_{1,R}(\mu_0),
            \utilde_{2,R}(\mu_0),
            \gtildeonetwoR(\mu_0),
            \sigma_R(\mu_0);
            \{\kvec\},\{\omega\}; 
            \mu_0}         
\end{multline}
close to the critical point, where
\begin{subequations} \label{eq:defnwilson}
    \begin{align}
        \gamma_{\phi_i} &= \mu \frac{\partial \log Z_{\phi_i} }{\partial \mu}  \qquad \gamma_{\phitilde_i} = \mu \frac{\partial \log Z_{\phitilde_i} }{\partial \mu}, \label{eq:wilsonfield} \\ 
        %
        \gamma_{D_i} &= \mu \frac{\partial \log D_{i,R}}{\partial \mu} = \mu \frac{\partial \log Z_{D_i}}{\partial \mu} \label{eq:wilsonD}\\
        \gamma_{\Gamma_i} &= \mu \frac{\partial \log \Gamma_{i,R}}{\partial \mu} = \mu \frac{\partial \log Z_{\Gamma_i}}{\partial \mu} \label{eq:wilsonGamma}
        %
    \end{align}
\end{subequations}
where $\gamma_{\tautilde_i}$ are stated in \Eref{gamma_tautilde_actually} and
$\gamma_{\phi_i}$, $\gamma_{\phitilde_i}$ and $\gamma_{\Gamma_i}$ are determined only in the form 
\begin{subequations} \label{eq:WardtoWilson}
    \begin{align}
        \gamma_{\phi_i} + \gamma_{\phitilde_i} &= 0 \label{eq:WardtoWilsonone}\\
        \gamma_{\Gamma_i} + 2\gamma_{\phitilde_i} &= 0 \label{eq:WardtoWilsontwo}
    \end{align}
\end{subequations}
due to the over-parameterisation that allows us to introduce two independent dimensions $B_1$ and $B_2$ above.

\Eref{CSeqn_soln} relates the renormalised vertex function at different scales, $\mu=\ell\mu_0$ on the left and $\mu_0$ on the right. We may think of $\mu_0$ at being a scale at which we can evaluate our field theory safely by a perturbation theory. Close to the critical point the three couplings $\utildestar_{1,R}$, $\utildestar_{2,R}$ and $\gtildeonetwostar$ attain their fixed point values, \Sref{criticalpoints}, and similarly $\wstar$ and $\sigmastar$, which to one loop do not flow and therefore keep their bare values.

Based on the dimensional analysis in \Sref{meanfieldcritical} we further have that
\begin{multline}
    \invpropiren{ab}{
            \tautilde_{1,R},
            \tautilde_{2,R},
            D_{1,R},
            D_{2,R},
            \Gamma_{1,R},
            \Gamma_{2,R},
            \utildestar_{1,R},
            \utildestar_{2,R},
            \gtildeonetwostar,
            \sigmastar;
            \{\kvec\},\{\omega\}; 
            \mu}
\\ =
A L^{-4}
    \invpropiren{ab}{
            \frac{\tautilde_{1,R}}{A},
            \frac{\tautilde_{2,R}}{A},
            \frac{D_{1,R}}{A},
            \frac{D_{2,R}}{A},
            \frac{\Gamma_{1,R}}{B_1},
            \frac{\Gamma_{2,R}}{B_2},
            \utildestar_{1,R},
            \utildestar_{2,R},
            \gtildeonetwostar,
            \sigmastar;
            \{\kvec L\},
            \left\{
            \frac{\omega L^4}{A}
            \right\}; 
            \mu L}
\end{multline}
holds for any $A,B_1,B_2,L\in\Rset^+$. Using this on the left-hand side of \Eref{CSeqn_soln} with 
diffusive scale $A=D_{i,R}(\mu_0) \ell^{\gamma^{*}_{D_i}}$, 
noise scale $B_i=\Gamma_{i,R}(\mu_0)\ell^{\gamma^{*}_{\Gamma_i}}$ and length scale $L=1/\ell$ then gives the desired scaling form
\begin{multline}\elabel{scaling_form_of_inverse_prop}
\invpropiren{ab}{
            \tautilde_{1,R}(\mu_0),
            \tautilde_{2,R}(\mu_0),
            D_{1,R}(\mu_0), 
            D_{2,R}(\mu_0),
            \Gamma_{1,R}(\mu_0),
            \Gamma_{2,R}(\mu_0),
            \utilde_{1,R}(\mu_0),
            \utilde_{2,R}(\mu_0),
            \gtildeonetwoR(\mu_0),
            \sigma_R(\mu_0);
            \{\kvec\},\{\omega\}; 
            \mu_0}\\
            = D_{i,R}(\mu_0) 
                    \ell^{4 + \gammaphitildeistar + \gammaphiistar +  \gamma^{*}_{D_i}}
 \Gamma^{ab}_{R, \phitilde_i \phi_i}\big(
            \ell^{\gammatautildeonestar-\gamma^{*}_{D_i}} 
            \frac{\tautilde_{1,R}(\mu_0)}{D_{i,R}(\mu_0)},
            \ell^{\gammatautildetwostar-\gamma^{*}_{D_i}} 
            \frac{\tautilde_{2,R}(\mu_0)}{D_{i,R}(\mu_0)},
            \ell^{\gammaDonestar-\gamma^{*}_{D_i}}\frac{D_{1,R}(\mu)}{D_{i,R}(\mu)},
            \ell^{\gammaDtwostar-\gamma^{*}_{D_i}}\frac{D_{2,R}(\mu)}{D_{i,R}(\mu)}, \\
            1,
            1,
            \utildestar_{1,R},
            \utildestar_{2,R},
            \gtildeonetwostar,
            \sigmastar;
            \{\ell^{-1} \kvec \},\left\{ \frac{\omega}{D_{i,R}(\mu_0) \ell^{4+\gamma^{*}_{D_i}}}\right\}; 
            \mu_0 \big) \ .      
\end{multline}
The choice $A= D_{i,R}(\mu_0)\ell^{\gammaDonestar}$ makes obvious that the two fields are characterised by two different dynamical exponents, \Eref{scalingformpropagator}
\begin{equation} \label{eq:dynamicalscalingexponent}
z_i=4+\gamma^{*}_{D_{i}}=4 + \mathcal{O}(\epsilon^2)
\end{equation} 
with \Eref{wilsonD}, as can be gleaned from the scaling of $\omega$ with $\ell$. Further, from the prefactor $\ell^{4+\gammaphitildeistar + \gammaphiistar +  \gamma^{*}_{D_i}}=\ell^{4 - \eta_i/2-\etatilde_i/2}$ and \Erefs{wilsonD} and \eref{WardtoWilsonone} we have that $\eta_i+\etatilde_i= -2\gamma^{*}_{D_i} = 0 + \mathcal{O}(\epsilon^2)$. Even though $\eta_i+\etatilde_i$ is zero at one-loop level, we find that it is directly dependent on the dynamical scaling exponent
\begin{equation} \label{eq:z_to_anomalous}
    z_i = 4-\frac{\eta_i + \etatilde_i}{2},
\end{equation}
a property shared by models with conserved dynamics. As for the divergence of the correlation lengths, in principle there are four exponents, but once a particular scaling form is assumed, we may reduce them to two. For the present scaling forms, \Eref{scalingformpropagator}, we have
$1/\nu_{1}=-\gamma^{*}_{\tautilde_1} +\gamma^{*}_{D_1}$ and $1/\nu_{2}=-\gamma^{*}_{\tautilde_2} +\gamma^{*}_{D_2}$. To one loop, however, $\gamma^{*}_{D_1}=\gamma^{*}_{D_2}=0$, so that
\begin{subequations} \label{eq:divergence_corrlength}
\begin{align}
    \nu_{1}&=-\frac{1}{\gamma^{*}_{\tautilde_1}} = \frac{1}{2} + \frac{n+2}{12} \utildeonestar + \frac{n \sqrt{\sigma}}{4} \gtildeonetwostar\\
    \nu_{2}&=-\frac{1}{\gamma^{*}_{\tautilde_2}} = \frac{1}{2} + \frac{n+2}{12} \utildeonestar - \frac{n \sqrt{\sigma}}{4} \gtildeonetwostar
\end{align}
\end{subequations}
from \Eref{gamma_tautilde_actually} to be evaluated with $\utildeonestar$, $\utildetwostar$, $\gtildeonetwostar$ from \Erefs{first_fixed_points} or \eref{second_fixed_point} and with $\sigma$ to one loop determined at bare level.

A similar analysis can be performed for the correlation function, $\langle \phi^{a}_i(\kvec,\omega)  \phi^{b}_i(\kvec',\omega') \rangle = \kroneckerdelta{} \deltabar(\kvec + \kvec') \deltabar(\omega + \omega') C_{i}(\kvec,\omega)$, to determine $\eta_i$ and $\etatilde_i$ independently. Similar to \Eref{GammaR_from_Gamma}, we introduce the renormalised correlation function \Eref{Corr_to_vertex}
with the full set of renormalised parameters
\begin{multline}\elabel{CorrR_from_Corr}
            Z^{-2}_{\phivec_i}(\mu)
C_{i,R}(
            \tautilde_{1,R}(\mu),
            \tautilde_{2,R}(\mu),
            D_{1,R}(\mu),
            D_{2,R}(\mu),
            \Gamma_{1,R}(\mu),
            \Gamma_{2,R}(\mu),
            \utilde_{1,R}(\mu),
            \utilde_{2,R}(\mu),
            \gtildeonetwoR(\mu),
            \sigma_R(\mu);
            \{\kvec\},\{\omega\}; 
            \mu) \\         
= C_{i}(
             \tau_1,
             \tau_2,
             D_1,
             D_2,
             \Gamma_1,
             \Gamma_2,
             u_1,
             u_2,
             \gonetwo,
             \gtwoone;
            \{\kvec\},\{\omega\}),
\end{multline}
with some of the arguments of the renormalised correlation function defined in \Eref{expansionparameterdefn}. Differentiating the LHS of \Eref{CorrR_from_Corr} with respect to $\mu$ leads us to another Callan-Symanzik equation, now for the correlation function, which can also be solved by method of characteristics as before, leading to
\begin{multline}\elabel{CSeqn_soln_corr}
                \ell^{-2\gammaphiistar}
C_{i,R}\big(
            \ell^{\gammatautildeonestar} \tautilde_{1,R}(\mu_0),
            \ell^{\gammatautildetwostar} \tautilde_{2,R}(\mu_0),
            \ell^{\gamma^{*}_{D_1}} D_{1,R}(\mu_0),
            \ell^{\gamma^{*}_{D_2}} D_{2,R}(\mu_0), \\
            \ell^{\gammaGammaonestar} \Gamma_{1,R}(\mu_0),
            \ell^{\gammaGammatwostar} \Gamma_{2,R}(\mu_0),
            \utildestar_{1,R},
            \utildestar_{2,R},
            \gtildeonetwostar,
            \sigmastar;
            \{\kvec\},\{\omega\}; 
            \ell \mu_0 \big)\\         
=
C_{i,R}\big(
            \tautilde_{1,R}(\mu_0),
            \tautilde_{2,R}(\mu_0),
            D_{1,R}(\mu_0),
            D_{2,R}(\mu_0),
            \Gamma_{1,R}(\mu_0),
            \Gamma_{2,R}(\mu_0),
            \utilde_{1,R}(\mu_0),
            \utilde_{2,R}(\mu_0),
            \gtildeonetwoR(\mu_0),
            \sigma_R(\mu_0);
            \{\kvec\},\{\omega\}; 
            \mu_0)        
\end{multline}
close to the critical point. There, the couplings $\utildeonestar$, $\utildetwostar$ and $\gtildeonetwostar$ reach their fixed point values, while $\wstar$ and $\sigmastar$ retain their bare values as they do not flow to one loop, so that the Wilson gamma functions, \Eref{defnwilson}, can be treated as constants.

Based on dimensional analysis in \Sref{meanfieldcritical} we further have that
\begin{multline}
    C_{i,R}(
            \tautilde_{1,R},
            \tautilde_{2,R},
            D_{1,R},
            D_{2,R},
            \Gamma_{1,R},
            \Gamma_{2,R},
            \utildestar_{1,R},
            \utildestar_{2,R},
            \gtildeonetwostar,
            \sigmastar;
            \{\kvec\},\{\omega\}; 
            \mu)
\\ =
 B_i A^{-2} L^{6} 
    C_{i,R}\bigg(
            \frac{\tautilde_{1,R}}{A},
            \frac{\tautilde_{2,R}}{A},
            \frac{D_{1,R}}{A},
            \frac{D_{2,R}}{A},
            \frac{\Gamma_{1,R}}{B_1},
            \frac{\Gamma_{2,R}}{B_2},
            \utildestar_{1,R},
            \utildestar_{2,R},
            \gtildeonetwostar,
            \sigmastar;
            \{\kvec L\},
            \left\{
            \frac{\omega L^4}{A}
            \right\}; 
            \mu L\bigg)
\end{multline}
which holds for any $A, B_1, B_2,L \in \Rset^{+}$. Once again, using this on the LHS of \Eref{CSeqn_soln_corr} with the diffusive scale $A = D_{i,R}(\mu_0) \ell^{\gamma^{*}_{D_i}}$, noise scale $B_i = \Gamma_{i,R}(\mu_0) \ell^{\gamma^{*}_{\Gamma_i}}$ and length scale $L = 1/\ell$ gives the desired scaling form
\begin{multline}
C_{i,R}(
            \tautilde_{1,R}(\mu_0),
            \tautilde_{2,R}(\mu_0),
            D_{1,R}(\mu_0), 
            D_{2,R}(\mu_0),
            \Gamma_{1,R}(\mu_0),
            \Gamma_{2,R}(\mu_0),
            \utilde_{1,R}(\mu_0),
            \utilde_{2,R}(\mu_0),
            \gtildeonetwoR(\mu_0),
            \sigma_R(\mu_0);
            \{\kvec\},\{\omega\}; 
            \mu_0)\\
            = \frac{\Gamma_{i,R}(\mu_0) }{D^2_{i,R}(\mu_0)}
                    \ell^{-6 + \gamma^{*}_{\Gamma_i} - 2 \gamma^{*}_{\phivec_i} -2 \gamma^{*}_{D_i}}
 C_{i,R}\bigg(
            \ell^{\gammatautildeonestar-\gamma^{*}_{D_i}} 
            \frac{\tautilde_{1,R}(\mu_0)}{D_{i,R}(\mu_0)},
            \ell^{\gammatautildetwostar-\gamma^{*}_{D_i}} 
            \frac{\tautilde_{2,R}(\mu_0)}{D_{i,R}(\mu_0)},
            \ell^{\gammaDonestar-\gamma^{*}_{D_i}}\frac{D_{1,R}(\mu)}{D_{i,R}(\mu)},
            \ell^{\gammaDtwostar-\gamma^{*}_{D_i}}\frac{D_{2,R}(\mu)}{D_{i,R}(\mu)}, \\
            1,
            1,
            \utildestar_{1,R},
            \utildestar_{2,R},
            \gtildeonetwostar,
            \sigmastar;
            \{\ell^{-1} \kvec \},\left\{ \frac{\omega}{D_{i,R}(\mu_0) \ell^{4+\gamma^{*}_{D_i}}}\right\}; 
            \mu_0 \bigg) \ .      
\end{multline}
The prefactor $\ell^{-6 + \gamma^{*}_{\Gamma_i} - 2\gamma^{*}_{\phivec_i} -2 \gamma^{*}_{D_i}}$ can be simplified by employing \Erefs{WardtoWilson}, whereby $\gamma^{*}_{\Gamma_i} - 2\gamma^{*}_{\phivec_i} = 0$ and we can individually identify $\eta_i$ and $\etatilde_i$ by comparing the prefactor, $\ell^{-6 + \gamma^{*}_{\Gamma_i} - 2\gamma^{*}_{\phivec_i} -2 \gamma^{*}_{D_i}}$, with \Eref{scalingformcorrelator} and using the value of the dynamical scaling exponent \Eref{dynamicalscalingexponent}
\begin{equation}  \label{eq:anomalousexponents}
    \eta_{i} = \etatilde_{i} = -\gamma^{*}_{D_i} = 0+\mathcal{O}(\epsilon^2).
\end{equation}
Even though the present non-reciprocal Model~B breaks detailed balance, we therefore find that $\etatilde_i = \eta_i$ to all orders in the perturbative expansion, as $-2-z_i+\eta_i=-6-2\gamma^*_{D_i}$, $\eta_i+\etatilde_i=-2\gamma^*_{D_i}$ and $z_i=4-(\eta_i+\etatilde_i)/2$, \Eref{z_to_anomalous}. The equality $\eta_i=\etatilde_i$ thus goes back to \Eref{WardtoWilson} and thus to \Eref{FDRZfactors}, \ie the conserved nature of the dynamics, while in equilibrium \Eref{FDRZfactors} is a result of a fluctuation-dissipation relation.
The equality of $\etatilde$ and $\eta$ may thus be considered a generic feature of models with conserved dynamics, regardless of whether the underlying dynamics obeys detailed balance. 

\section{Conclusion} \label{sec:conclusion}
In this paper, we have considered the critical dynamics of two conserved order parameter fields with non-reciprocal interactions. Unlike typical implementations of non-reciprocity \cite{fruchart_non-reciprocal_2021,alston_irreversibility,cocconi_bound_states,PhysRevX.12.010501,marchetti_nr_pattern_formation,doi:10.1073/pnas.2010318117,saha_scalar_2020,doi:10.1073/pnas.2407705121,risler_universal_2005,risler_universal_2004,tauber_perturbative_2014} which induce dynamical steady states and spatio-temporal patterns, we have chosen an implementation where the non-reciprocity enters only through the non-linear coupling between the two order parameter fields. After casting the model into an MSR field theory, we have constructed the perturbative expansion and used it to carry out the field-theoretic RG procedure to one loop. The results of the RG calculation reveal the exponents to first order in $\epsilon=4-d$, namely $z_i$ \Erefs{dynamicalscalingexponent}, $\nu_i$ \eref{divergence_corrlength} and $\eta_i,\etatilde_i$ \eref{anomalousexponents}. A two-loop analysis is needed to determine the non-trivial behaviour of $z_i$ and $\eta_i,\etatilde_i$ for $n<4$,
as several couplings, $w$ and $\sigma$, do not receive any correction and retain their bare value under the RG. For $n>4$, however, at one stable fixed point \Eref{first_fixed_points} the fields decouple and they individually undergo Model~B equilibrium dynamics of Hohenberg and Halperin's classification \cite{halperin_calculation_1972}.
Finally, by using the conservation of the order parameter fields, we have shown that even if the detailed balance is broken through non-reciprocal couplings, the scaling of correlation and response functions of each field will be characterised by a single anomalous exponent $\eta_i=\etatilde_i$ \Eref{anomalousexponents}, that is also linearly dependent on the dynamical scaling exponent, $z_i$, \Eref{z_to_anomalous}. 

The present work opens a number of interesting new avenues: Firstly, it is necessary to perform a \emph{full} two-loop analysis and verify the existence of the proposed non-equilibrium fixed points of \cite{young_nonequilibrium_2020, young_nonequilibrium_2024}. Secondly, one may generalise the present setup to the fields having different number of components, $\phivec_1 \in \mathbb{R}^{n_1}$ and $\phivec_2 \in \mathbb{R}^{n_2}$. \citeauthor{young_nonequilibrium_2024} observed this to break the strong dynamical scaling $z_1=z_2$ and cause non-trivial scaling exponents. Finally, for $n_1 = n_2 = n$, it would be intriguing to introduce non-reciprocity between the two fields at the \emph{linear} level, providing a way to promote the underlying symmetry from $O(n) \otimes O(n)$ to $U(n)$, and explore the effect on the Wilson-Fisher fixed point. \citeauthor{tauber_perturbative_2014} showed that for $n=1$ such systems can elegantly be written in terms of a single \emph{scalar} complex order parameter field. Applying their approach to the present setup, one would repeat our RG analysis with \emph{complex} order parameter vector fields where the non-reciprocity is encoded in the coupling constants acquiring a imaginary component. 

\section*{Acknowledgments}
E.S. would like to thank Giulia Pisegna and Martin Johnsrud for interesting conversations. E.S. is supported by Roth PhD scholarships funded by the Department of Mathematics at Imperial College London.

\bibliography{NRcoupledModelB}

\appendix

\section{Details of the Loop Calculation} \label{app:DetailsLoop}
In this section, we list and evaluate the amputated loop diagrams that renormalise the relevant vertices. The field theory defined by Eqs. \eqref{eq:actionharmonic} and \eqref{eq:actionperturbative} is closed under RG flow, meaning no terms that are not already present in the action will be generated by the RG procedure. In what follows, we focus on the one-loop diagrams. 
Overall, there are $16$ such diagrams. However, we need to calculate only eight of them since the other half can be calculated easily by swapping the indices of the species $1$ and $2$. We proceed vertex-by-vertex. 

We start with $\invpropone{}{}$. To one loop, we have the following diagrams:
\begin{equation} \label{eq:invpropone_app}
    \invpropone{}{} = \left(-\imag \omega + \kvecsquared\left( D_1 \kvecsquared +r_1\right) \right) \kroneckerdelta{} - 
    \begin{tikzpicture}[baseline = (o.base), scale = 1.2]
   \begin{feynman}
     \vertex (o) at (0,0) {\(a\)};
    \vertex (a) at (0.8,0);
    \vertex (b) at (1.7,1);
    \vertex (i) at (1.4,0) {\(b\)};
    \diagram* {
      (a) -- [anti fermion, decoration={markings, mark=at position 0.65 with {\draw[-] (0,-3pt) -- (0,3pt);}}, postaction={decorate}, bend left=40] (b),
      (o) -- [plain , edge label = {\(\kvec,\omega\)}](a),
      (a) -- [anti fermion, decoration={markings, mark=at position 0.65 with {\draw[-] (0,-3pt) -- (0,3pt);}}, postaction={decorate}, bend right=40] (b),
      (o) -- [plain, decoration={markings, 
  mark=at position 0.20 with {\draw[-] (0,-3pt) -- (0,3pt);},
  mark=at position 0.25 with {\draw[-] (0,-3pt) -- (0,3pt);}
}, postaction={decorate}, ] (i)
    };
  \end{feynman}
\end{tikzpicture}
- \begin{tikzpicture}[baseline = (o.base), scale = 1.2]
   \begin{feynman}
    \vertex (o) at (0,0) {\(a\)};
    \vertex (a) at (0.8,0);
    \vertex (b) at (1.7,1);
    \vertex (i) at (1.4,0) {\(b\)};
    \diagram* {
      (a) -- [anti charged scalar, decoration={markings, mark=at position 0.65 with {\draw[-] (0,-3pt) -- (0,3pt);}}, postaction={decorate}, bend left=40] (b),
      (o) -- [plain , edge label = {\(\kvec,\omega\)}](a),
      (a) -- [anti charged scalar, decoration={markings, mark=at position 0.65 with {\draw[-] (0,-3pt) -- (0,3pt);}}, postaction={decorate}, bend right=40] (b),
      (o) -- [plain, decoration={markings, 
  mark=at position 0.20 with {\draw[-] (0,-3pt) -- (0,3pt);},
  mark=at position 0.25 with {\draw[-] (0,-3pt) -- (0,3pt);}
}, postaction={decorate}] (i)
    };
  \end{feynman}
\end{tikzpicture}
\end{equation}
where the minus sign in front of the loop corrections comes from the Dyson-summation performed. We label the diagrams $\diagram_{\massrenormone[0.2]}(\kvec,\omega)$ and $\diagram_{\massrenormtwo[0.2]}(\kvec, \omega)$, respectively, whereby \Eref{invpropone_app} becomes
\begin{equation} \label{eq:invpropone_oneloop}
    \invpropone{}{} = \left\{-\imag \omega + \kvecsquared\left( D_1 \kvecsquared +r_1\right)  \kroneckerdelta{} +\kvecsquared u_1 \frac{n+2}{6} \diagram_{\massrenormone[0.4]}(\kvec,\omega) + \kvecsquared g_{12} \frac{n}{2} \diagram_{\massrenormtwo[0.4]} (\kvec,\omega) \right\} \kroneckerdelta{}
\end{equation}
where the diagrams are defined as
\begin{subequations}\elabel{mass_loops}
\begin{align}
    \diagram_{\massrenormone[0.4]}(\kvec,\omega) & \defequal \int \ddintbar{k_1} \int \dintbar{\omega_1} \frac{2 \Gamma_1 \kvecsquared_1}{(-\imag\omega_1 + \Delta_1(\kvec_1))(\imag\omega_1 + \Delta_1(\kvec_1))}, \nonumber \\
    &= \frac{\Gamma_1}{\Done} \int \ddintbar{k_1} \frac{1}{\kvecsquared_1 + r_1/\Done} ,\\
    \diagram_{\massrenormtwo[0.4]}(\kvec,\omega) & \defequal \int \ddintbar{k}_1 \int \dintbar{\omega_1} \frac{2 \Gamma_2 \kvecsquared_1}{(-\imag\omega_1 +\Delta_2(\kvec_1))(\imag\omega_1 + \Delta_2(\kvec_1))} , \nonumber\\
    & = \frac{\Gamma_2}{\Dtwo} \int \ddintbar{k_1} \frac{1}{\kvecsquared_1 + r_2/\Dtwo} ,
\end{align}
\end{subequations}
using the propagators in \Erefs{diagramsharmonic} with $\deltaik=\kvecsquared (r_i + D_i \kvecsquared)$. To one loop, as generically found in field theories with only quartic non-linearities, only the mass and not the diffusion constant get corrected. The quadratic divergences in \Erefs{mass_loops} require an additive renormalisation, so that the critical point $r_{1,c}$ moves away from the mean-field value. Demanding $\left. \frac{\partial \invpropone{}{\kvec, \omega}}{\partial \kvecsquared} \right|_{\kvec = 0, \omega = 0} = 0$ at $r_1=r_{1,c}$ in \Eref{invpropone_oneloop}, self-consistency requires
\begin{align} \label{eq:critical_point_defn}
    r_{1,c} &= - \uone\frac{\numcomponent + 2}{6} \frac{\Gamma_1}{\Done} \kint_1 \frac{1}{\kvecsquared_1 + r_{1,c}/\Done} - \gonetwo \frac{\numcomponent}{2}\frac{\Gamma_2}{\Dtwo} \kint_1 \frac{1}{\kvecsquared_1 + r_{2,c}/\Dtwo} \nonumber \\ 
    &= -\uone\frac{\numcomponent + 2}{6} \frac{\Gamma_1}{\Done} \kint_1 \frac{1}{\kvecsquared_1 } - \gonetwo \frac{\numcomponent}{2}\frac{\Gamma_2}{\Dtwo} \kint_1 \frac{1}{\kvecsquared_1} + \mathcal{O}(\lambda^2) \ .
\end{align}
Parameterising the field theory in terms of $\tau_i = r_i - r_{i,c}$, $\invpropone{}{}$ to one loop can be written as
\label{eq:vertex1100}
    \begin{align}
        & \invpropone{}{} \nonumber \\
        &= \left[ -\imag \omega + \Done \kvec^4 + \kvec^2 \left(\tau_1 + r_{1,c} + \uone\frac{\numcomponent + 2}{6} \frac{\Gamma_1}{\Done} \kint_1 \frac{1}{\kvecsquared_1 + (\tau_1 + r_{1,c})/\Done} + \gonetwo \frac{\numcomponent}{2}\frac{\Gamma_2}{\Dtwo} \kint_1 \frac{1}{\kvecsquared_1 + (\tau_2 + r_{2,c})/\Dtwo}\right)  \right] \kroneckerdelta{} \nonumber \\
        & = \left[ -\imag \omega + \Done \kvec^4 + \tau_1\kvec^2 \left(1 - \uone\frac{\numcomponent + 2}{6} \frac{\Gamma_1}{\Done^2} \kint_1 \frac{1}{\kvecsquared_1(\kvecsquared_1 + \tau_1/\Done)} - \gonetwo \frac{\numcomponent}{2}\frac{\Gamma_2}{\Dtwo^2} \frac{\tau_2}{\tau_1} \kint_1 \frac{1}{\kvecsquared_1(\kvecsquared_1 + \tau_2/\Dtwo)} \right)  \right] \kroneckerdelta{} \nonumber \\
        & = \left[ -\imag \omega + \Done \kvec^4 + \tau_1 \kvec^2 \left(1 - \uone\frac{\numcomponent + 2}{6} \frac{\Gamma_1}{\Done^2} \loopfactor \left(1-\frac{\epsilon}{2}\right)^{-1}
        \left( \frac{\tau_1}{\Done}\right)^{-\epsilon/2} 
        - 
        \gonetwo \frac{\numcomponent}{2}\frac{\Gamma_2}{\Dtwo^2} \frac{\tau_2}{\tau_1} \loopfactor \left(1-\frac{\epsilon}{2}\right)^{-1}
        \left( \frac{\tau_2}{\Dtwo }\right)^{-\epsilon/2} 
        \right)  \right] \kroneckerdelta{}
    \end{align}
where we have used the definition of $r_{1,c}$ \Eref{critical_point_defn} from the second to the third line to leading order in the non-linearities, thereby omitting $r_{1,c}$ in the integrands.

Applying the same procedure to $\invproptwo{}{}$, 
\begin{equation}
        \invproptwo{}{} = \left(-\imag \omega + \kvecsquared\left( D_2 \kvecsquared +r_2\right) \right) \kroneckerdelta{} -
    \begin{tikzpicture}[baseline = (o.base), scale = 1.2]
   \begin{feynman}
   \vertex (o) at (0,0) {\(a\)};
    \vertex (a) at (0.8,0);
    \vertex (b) at (1.7,1);
    \vertex (i) at (1.4,0) {\(b\)};
    \diagram* {
      (a) -- [anti charged scalar, decoration={markings, mark=at position 0.65 with {\draw[-] (0,-3pt) -- (0,3pt);}}, postaction={decorate}, bend left=40] (b),
      (o) -- [scalar, edge label = {\(\kvec,\omega\)}] (a),
      (a) -- [anti charged scalar, decoration={markings, mark=at position 0.65 with {\draw[-] (0,-3pt) -- (0,3pt);}}, postaction={decorate}, bend right=40] (b),
      (o) -- [scalar, decoration={markings, 
  mark=at position 0.20 with {\draw[-] (0,-3pt) -- (0,3pt);},
  mark=at position 0.25 with {\draw[-] (0,-3pt) -- (0,3pt);}
}, postaction={decorate}, ] (i)
    };
  \end{feynman}
\end{tikzpicture}
 - \begin{tikzpicture}[baseline = (o.base), scale = 1.2]
   \begin{feynman}
    \vertex (o) at (0,0) {\(a\)};
    \vertex (a) at (0.8,0);
    \vertex (b) at (1.7,1);
    \vertex (i) at (1.4,0) {\(b\)};
    \diagram* {
      (a) -- [anti fermion, decoration={markings, mark=at position 0.65 with {\draw[-] (0,-3pt) -- (0,3pt);}}, postaction={decorate}, bend left=40] (b),
      (o) -- [scalar, edge label={\(\kvec,\omega\)}] (a),
      (a) -- [anti fermion, decoration={markings, mark=at position 0.65 with {\draw[-] (0,-3pt) -- (0,3pt);}}, postaction={decorate}, bend right=40] (b),
      (o) -- [scalar, decoration={markings, 
  mark=at position 0.20 with {\draw[-] (0,-3pt) -- (0,3pt);},
  mark=at position 0.25 with {\draw[-] (0,-3pt) -- (0,3pt);}
}, postaction={decorate}, ] (i)
    };
  \end{feynman}
\end{tikzpicture}
\end{equation}
we arrive at
    \begin{align} \label{eq:vertex0011}
        &\invproptwo{}{} = \nonumber \\
        &= \left[ -\imag \omega + \Dtwo \kvec^4 + \tau_2 \kvec^2 \left(1 - \utwo\frac{\numcomponent + 2}{6} \frac{\Gamma_2}{\Dtwo^2} \loopfactor \left(1-\frac{\epsilon}{2}\right)^{-1}
        \left( \frac{\tau_2}{\Dtwo}\right)^{-\epsilon/2} 
        - 
        \gtwoone \frac{\numcomponent}{2}\frac{\Gamma_1}{\Done^2} \frac{\tau_1}{\tau_2} \loopfactor \left(1-\frac{\epsilon}{2}\right)^{-1}
        \left( \frac{\tau_1}{\Done }\right)^{-\epsilon/2} 
        \right)  \right] \kroneckerdelta{}
    \end{align}
As for the renormalisation of the "usual" quartic vertex $\vertexuone{}{}$, the diagrams contributing are 
\begin{equation}
    \vertexuone{}{\kvec} = 
        \begin{tikzpicture}[baseline = (o.base), scale = 1.2]
  \begin{feynman}
    \vertex (origin) at (0,0) ;
    \vertex (o) at (-0.6,0) {\(a\)};
    \vertex (i1) at (0.6,0) {\(c\)};
    \vertex (i2) at (0.6,0.6) {\(b\)};
    \vertex (i3) at (0.6,-0.6) {\(d\)};
    \diagram* {
      (origin) -- [plain, decoration={markings, 
  mark=at position 0.20 with {\draw[-] (0,-3pt) -- (0,3pt);},
  mark=at position 0.25 with {\draw[-] (0,-3pt) -- (0,3pt);}
}, postaction={decorate}, edge label'={\(\kvec\)}] (o),
      (origin) -- [plain] (i1),
      (origin) -- [plain] (i2),
      (origin) -- [plain] (i3),
    };
  \end{feynman}
\end{tikzpicture}
+
\begin{tikzpicture}[baseline = (o.base),scale = 1.2]
\begin{feynman}
\vertex (o) at (0,0) {\(a\)};
\vertex (a) at (0.8,0);
\vertex (i3) at ($(a) + (0.75,0)$) {\(d\)};
\vertex (b) at ($(a) + (0.33,1)$);     
\vertex (c) at ($(a) + (1,0.4)$);    
\vertex (i1) at ($(b) + (0.5,0.6)$) {\(b\)};  
\vertex (i2) at ($(b) + (0.5,0.3)$) {\(c\)};  
\diagram* {
(o) -- [plain, edge label = \(\kvec\), decoration={markings, 
  mark=at position 0.75 with {\draw[-] (0,-3pt) -- (0,3pt);},
  mark=at position 0.80 with {\draw[-] (0,-3pt) -- (0,3pt);}
}, postaction={decorate},] (a),
(a) -- [plain] (i3),
(a) -- [anti fermion, decoration={markings, 
  mark=at position 0.80 with {\draw[-] (0,-3pt) -- (0,3pt);},
  mark=at position 0.75 with {\draw[-] (0,-3pt) -- (0,3pt);}
} ] (b),
(b) -- [anti fermion, decoration={markings, mark=at position 0.65 with {\draw[-] (0,-3pt) -- (0,3pt);}},] (c),
(a) -- [anti fermion, decoration={markings, mark=at position 0.65 with {\draw[-] (0,-3pt) -- (0,3pt);}},] (c),
(b) -- [plain] (i1),
(b) -- [plain] (i2),
};
\end{feynman}
\end{tikzpicture}
~~~ + \sym 
~+~
\begin{tikzpicture}[baseline = (o.base), scale = 1.2]
\begin{feynman}
\vertex (o) at (0,0) {\(a\)};
\vertex (a) at (0.8,0);
\vertex (i3) at ($(a) + (0.75,0)$) {\(d\)};
\vertex (b) at ($(a) + (0.33,1)$);     
\vertex (c) at ($(a) + (1,0.4)$);    
\vertex (i1) at ($(b) + (0.5,0.6)$) {\(b\)};  
\vertex (i2) at ($(b) + (0.5,0.3)$) {\(c\)};  
\diagram* {
(o) -- [plain, edge label = \(\kvec\), decoration={markings, 
  mark=at position 0.75 with {\draw[-] (0,-3pt) -- (0,3pt);},
  mark=at position 0.8 with {\draw[-] (0,-3pt) -- (0,3pt);}
}, postaction={decorate},] (a),
(a) -- [plain] (i3),
(a) -- [anti charged scalar, decoration={markings, 
  mark=at position 0.80 with {\draw[-] (0,-3pt) -- (0,3pt);},
  mark=at position 0.75 with {\draw[-] (0,-3pt) -- (0,3pt);}
} ] (b),
(b) -- [anti charged scalar, decoration={markings, mark=at position 0.65 with {\draw[-] (0,-3pt) -- (0,3pt);}}, postaction={decorate}] (c),
(a) -- [anti charged scalar, decoration={markings, mark=at position 0.65 with {\draw[-] (0,-3pt) -- (0,3pt);}}, postaction={decorate}] (c),
(b) -- [plain] (i1),
(b) -- [plain] (i2),
};
\end{feynman}
\end{tikzpicture}
~~~ + \sym 
\end{equation}
where $\sym$ refers to the diagrams generated by permutations of the incoming legs, whose components are labelled by $b$, $c$ and $d$. As can be seen by the double dashes on the outgoing (amputated) legs, every diagram is multiplied by the outgoing $\kvec^2$. We are interested only in such terms proportional to $\kvec^2$, which allows us safely set $\kvec_i=\zerovec$ inside the diagrams, as the additional contributions produced by $\kvec_i\ne\zerovec$ are UV-convergent. We thus arrive at 
\begin{equation} \label{eq:vertexuone_oneloop}
    \vertexuone{}{\kvec} = -\kvecsquared \frac{\Qtensor{}}{3}\left\{u_1 - u_1^2\frac{n+8}{3} \diagram_{\vertexrenormoneone[0.4]}(\zerovec) - g_{12}g_{21}3n\diagram_{\vertexrenormtwotwo[0.4]}(\zerovec) \right\} + \mathcal{O}(\kvec^4)
\end{equation}
with the loops labelled as $\diagram_{\vertexrenormoneone[0.2]}$ and $\diagram_{\vertexrenormtwotwo[0.2]}$, respectively, 
\begin{subequations} \label{eq:diagrams_vertexu1}
    \begin{align}
        \diagram_{\vertexrenormoneone[0.3]} &\defequal  \kint_1 \omegaint_1 \frac{2\Gamma_1 \kvec_1^4}{(-\imag\omega_1 + \Delta_1(\kvec_1))(\imag\omega_1 +  \Delta_1(\kvec_1))^2}   \nonumber\\
        &= \frac{\Gamma_1}{\Done^2} \kint_1 \frac{1}{(\kvecsquared_1 + \tau_1/\Done)^2}  = \frac{\Gamma_1}{\Done^2} \loopfactor \left(\frac{\tau_1}{ \Done}\right)^{-\epsilon/2}   \\
        \diagram_{\vertexrenormtwotwo[0.3]} &\defequal  \kint_1 \omegaint_1 \frac{2 \Gamma_2 \kvec_1^4}{(-\imag\omega_1  + \Delta_2(\kvec_1))(i \omega_1 + \Delta_2(\kvec_1))^2} \nonumber\\
        &= \frac{\Gamma_2}{\Dtwo^2} \kint_1  \frac{1}{(\kvecsquared_1 + \tau_2/\Dtwo)^2}  = \frac{\Gamma_2}{\Dtwo^2} \loopfactor \left(\frac{\tau_2}{\Dtwo}\right)^{-\epsilon/2} \ ,
    \end{align}
\end{subequations}
using the renormalised inverse propagator $-\imag\omega+\deltaik=-\imag\omega+\kvecsquared (\tau_i + D_i \kvecsquared)$.
Substituting \Erefs{diagrams_vertexu1} into \Eref{vertexuone_oneloop}, to one loop, the vertex therefore takes the form
\begin{equation} \label{eq:vertex1300}
    \vertexuone{}{\kvec} = -\kvec^2\uone \frac{\Qtensor{}}{3}\left[ 1 - \uone \frac{\numcomponent + 8}{6} \frac{\Gamma_1}{\Done^2} \loopfactor \left(\frac{\tau_1}{ \Done}\right)^{-\epsilon/2}  - \frac{\gonetwo \gtwoone}{\uone} \frac{3\numcomponent}{2} \frac{\Gamma_2}{\Dtwo^2} \loopfactor \left(\frac{\tau_2}{ \Dtwo}\right)^{-\epsilon/2}  \right] + \mathcal{O}(\kvec^4)
\end{equation}
The vertex $\vertexutwo{}{}$ follows the same pattern, 
\begin{equation}
    \vertexutwo{}{\kvec} = 
        \begin{tikzpicture}[baseline = (o.base), scale = 1.2]
  \begin{feynman}
   \vertex (origin) at (0,0) ;
    \vertex (o) at (-0.6,0) {\(a\)};
    \vertex (i1) at (0.6,0) {\(c\)};
    \vertex (i2) at (0.6,0.6) {\(b\)};
    \vertex (i3) at (0.6,-0.6) {\(d\)};
    \diagram* {
      (origin) -- [scalar, decoration={markings, 
  mark=at position 0.25 with {\draw[-] (0,-3pt) -- (0,3pt);},
  mark=at position 0.30 with {\draw[-] (0,-3pt) -- (0,3pt);}
}, postaction={decorate}, edge label'={\(\kvec\)}] (o),
      (origin) -- [scalar] (i1),
      (origin) -- [scalar] (i2),
      (origin) -- [scalar] (i3),
    };
  \end{feynman}
\end{tikzpicture}
+ 
\begin{tikzpicture}[baseline = (o.base), scale = 1.2]
\begin{feynman}
\vertex (o) at (0,0) {\(a\)};
\vertex (a) at (0.8,0);
\vertex (i3) at ($(a) + (0.75,0)$) {\(d\)};
\vertex (b) at ($(a) + (0.33,1)$);     
\vertex (c) at ($(a) + (1,0.4)$);    
\vertex (i1) at ($(b) + (0.5,0.6)$) {\(b\)};  
\vertex (i2) at ($(b) + (0.5,0.3)$) {\(c\)};  
\diagram* {
(o) -- [scalar, edge label = \(\kvec\), decoration={markings, 
  mark=at position 0.80 with {\draw[-] (0,-3pt) -- (0,3pt);},
  mark=at position 0.75 with {\draw[-] (0,-3pt) -- (0,3pt);}
}, postaction={decorate},] (a),
(a) -- [scalar] (i3),
(a) -- [anti charged scalar, decoration={markings, 
  mark=at position 0.80 with {\draw[-] (0,-3pt) -- (0,3pt);},
  mark=at position 0.75 with {\draw[-] (0,-3pt) -- (0,3pt);}
} ] (b),
(b) -- [anti charged scalar, decoration={markings, mark=at position 0.65 with {\draw[-] (0,-3pt) -- (0,3pt);}},] (c),
(a) -- [anti charged scalar, decoration={markings, mark=at position 0.65 with {\draw[-] (0,-3pt) -- (0,3pt);}},] (c),
(b) -- [scalar] (i1),
(b) -- [scalar] (i2),
};
\end{feynman}
\end{tikzpicture}
~~~ + \sym 
~+~
\begin{tikzpicture}[baseline = (o.base), scale = 1.2]
\begin{feynman}
\vertex (o) at (0,0) {\(a\)};
\vertex (a) at (0.8,0);
\vertex (i3) at ($(a) + (0.75,0)$) {\(d\)};
\vertex (b) at ($(a) + (0.33,1)$);     
\vertex (c) at ($(a) + (1,0.4)$);    
\vertex (i1) at ($(b) + (0.5,0.6)$) {\(b\)};  
\vertex (i2) at ($(b) + (0.5,0.3)$) {\(c\)};  
\diagram* {
(o) -- [scalar, edge label = \(\kvec\), decoration={markings, 
  mark=at position 0.80 with {\draw[-] (0,-3pt) -- (0,3pt);},
  mark=at position 0.75 with {\draw[-] (0,-3pt) -- (0,3pt);}
}, postaction={decorate},] (a),
(a) -- [scalar] (i3),
(a) -- [anti fermion, decoration={markings, 
  mark=at position 0.80 with {\draw[-] (0,-3pt) -- (0,3pt);},
  mark=at position 0.75 with {\draw[-] (0,-3pt) -- (0,3pt);}
} ] (b),
(b) -- [anti fermion, decoration={markings, mark=at position 0.65 with {\draw[-] (0,-3pt) -- (0,3pt);}}, postaction={decorate}] (c),
(a) -- [anti fermion, decoration={markings, mark=at position 0.65 with {\draw[-] (0,-3pt) -- (0,3pt);}}, postaction={decorate}] (c),
(b) -- [scalar] (i1),
(b) -- [scalar] (i2),
};
\end{feynman}
\end{tikzpicture}
~~~ + \sym .
\end{equation}
These are the same diagrams as those that renormalise $\vertexuone{}{}$, \Eref{diagrams_vertexu1}, but with solid and dashed lines swapped. Therefore, the contribution to the vertex $\vertexutwo{}{}$ will be that of $\vertexuone{}{}$ with indices $1$ and $2$ swapped, \Eref{vertex1300}, 
\begin{equation} \label{eq:vertex0013}
    \vertexutwo{}{\kvec} = -\kvec^2\utwo \frac{\Qtensor{}}{3}\left[ 1 - \utwo \frac{\numcomponent + 8}{6} \frac{\Gamma_2}{\Dtwo^2} \loopfactor \left(\frac{\tau_2}{ \Dtwo}\right)^{-\epsilon/2} - \frac{\gonetwo \gtwoone}{\utwo} \frac{3\numcomponent}{2} \frac{\Gamma_1}{\Done^2} \loopfactor  \left(\frac{\tau_1}{ \Done}\right)^{-\epsilon/2} \right] + \mathcal{O}(\kvec^4)
\end{equation}

We next consider the quartic coupling between the two fields $\vertexgone{}{}$. The one loop diagrams that contribute to it are given by 
    \begin{align}
        \vertexgone{}{\kvec} &= 
        \begin{tikzpicture}[baseline = (o.base), scale = 1.2]
  \begin{feynman}
    \vertex (origin) at (0,0) ;
    \vertex (o) at (-0.6,0) {\(a\)};
    \vertex (i1) at (0.6,0) {\(c\)};
    \vertex (i2) at (0.6,0.6) {\(b\)};
    \vertex (i3) at (0.6,-0.6) {\(d\)};
    \diagram* {
      (origin) -- [plain, decoration={markings, 
  mark=at position 0.20 with {\draw[-] (0,-3pt) -- (0,3pt);},
  mark=at position 0.25 with {\draw[-] (0,-3pt) -- (0,3pt);}
}, postaction={decorate}, edge label'={\(\kvec\)}] (o),
      (origin) -- [scalar] (i1),
      (origin) -- [plain] (i2),
      (origin) -- [scalar] (i3),
    };
  \end{feynman}
\end{tikzpicture}
~+~
\begin{tikzpicture}[baseline = (o.base), scale = 1.2]
\begin{feynman}
\vertex (o) at (0,0) {\(a\)};
\vertex (a) at (0.8,0);
\vertex (i3) at ($(a) + (0.75,0)$) {\(b\)};
\vertex (b) at ($(a) + (0.33,1)$);     
\vertex (c) at ($(a) + (1,0.4)$);    
\vertex (i1) at ($(b) + (0.5,0.6)$) {\(c\)};  
\vertex (i2) at ($(b) + (0.5,0.3)$) {\(d\)};  
\diagram* {
(o) -- [plain, edge label = \(\kvec\), decoration={markings, 
  mark=at position 0.75 with {\draw[-] (0,-3pt) -- (0,3pt);},
  mark=at position 0.8 with {\draw[-] (0,-3pt) -- (0,3pt);}
}, postaction={decorate},] (a),
(a) -- [plain] (i3),
(a) -- [anti fermion, decoration={markings, 
  mark=at position 0.80 with {\draw[-] (0,-3pt) -- (0,3pt);},
  mark=at position 0.75 with {\draw[-] (0,-3pt) -- (0,3pt);}
} ] (b),
(b) -- [anti fermion, decoration={markings, mark=at position 0.65 with {\draw[-] (0,-3pt) -- (0,3pt);}}, postaction={decorate}] (c),
(a) -- [anti fermion, decoration={markings, mark=at position 0.65 with {\draw[-] (0,-3pt) -- (0,3pt);}}, postaction={decorate}] (c),
(b) -- [scalar] (i1),
(b) -- [scalar] (i2),
};
\end{feynman}
\end{tikzpicture}
+ \sym +
\begin{tikzpicture}[baseline = (o.base),scale = 1.2]
\begin{feynman}
\vertex (o) at (0,0) {\(a\)};
\vertex (a) at (0.8,0);
\vertex (i3) at ($(a) + (0.75,0)$) {\(b\)};
\vertex (b) at ($(a) + (0.33,1)$);     
\vertex (c) at ($(a) + (1,0.4)$);    
\vertex (i1) at ($(b) + (0.5,0.6)$) {\(c\)};  
\vertex (i2) at ($(b) + (0.5,0.3)$) {\(d\)};  
\diagram* {
(o) -- [plain, edge label = \(\kvec\), decoration={markings, 
  mark=at position 0.75 with {\draw[-] (0,-3pt) -- (0,3pt);},
  mark=at position 0.80 with {\draw[-] (0,-3pt) -- (0,3pt);}
}, postaction={decorate},] (a),
(a) -- [plain] (i3),
(a) -- [anti charged scalar, decoration={markings, 
  mark=at position 0.80 with {\draw[-] (0,-3pt) -- (0,3pt);},
  mark=at position 0.75 with {\draw[-] (0,-3pt) -- (0,3pt);}
} ] (b),
(b) -- [anti charged scalar, decoration={markings, mark=at position 0.65 with {\draw[-] (0,-3pt) -- (0,3pt);}},] (c),
(a) -- [anti charged scalar, decoration={markings, mark=at position 0.65 with {\draw[-] (0,-3pt) -- (0,3pt);}},] (c),
(b) -- [scalar] (i1),
(b) -- [scalar] (i2),
};
\end{feynman}
\end{tikzpicture}
~~~ + \sym 
~~~ + 
\begin{tikzpicture}[baseline = (o.base),scale = 1.2]
\begin{feynman}
\vertex (o) at (0,0) {\(a\)};
\vertex (a) at (0.8,0);
\vertex (i3) at ($(a) + (0.75,0)$) {\(d\)};
\vertex (b) at ($(a) + (0.33,1)$);     
\vertex (c) at ($(a) + (1,0.4)$);    
\vertex (i1) at ($(b) + (0.5,0.6)$) {\(b\)};  
\vertex (i2) at ($(b) + (0.5,0.3)$) {\(c\)};  
\diagram* {
(o) -- [plain, edge label = \(\kvec\), decoration={markings, 
  mark=at position 0.75 with {\draw[-] (0,-3pt) -- (0,3pt);},
  mark=at position 0.80 with {\draw[-] (0,-3pt) -- (0,3pt);}
}, postaction={decorate},] (a),
(a) -- [scalar] (i3),
(a) -- [anti fermion, decoration={markings, 
  mark=at position 0.80 with {\draw[-] (0,-3pt) -- (0,3pt);},
  mark=at position 0.75 with {\draw[-] (0,-3pt) -- (0,3pt);}
} ] (b),
(b) -- [anti charged scalar, decoration={markings, mark=at position 0.65 with {\draw[-] (0,-3pt) -- (0,3pt);}},] (c),
(a) -- [anti charged scalar, decoration={markings, mark=at position 0.65 with {\draw[-] (0,-3pt) -- (0,3pt);}},] (c),
(b) -- [plain] (i1),
(b) -- [scalar] (i2),
};
\end{feynman}
\end{tikzpicture} 
+ \sym \nonumber  \\
& + 
\begin{tikzpicture}[baseline = (o.base),scale = 1.2]
\begin{feynman}
\vertex (o) at (0,0) {\(a\)};
\vertex (a) at (0.8,0);
\vertex (i3) at ($(a) + (0.75,0)$) {\(d\)};
\vertex (b) at ($(a) + (0.33,1)$);     
\vertex (c) at ($(a) + (1,0.4)$);    
\vertex (i1) at ($(b) + (0.5,0.6)$) {\(b\)};  
\vertex (i2) at ($(b) + (0.5,0.3)$) {\(c\)};  
\diagram* {
(o) -- [plain, edge label = \(\kvec\), decoration={markings, 
  mark=at position 0.75 with {\draw[-] (0,-3pt) -- (0,3pt);},
  mark=at position 0.80 with {\draw[-] (0,-3pt) -- (0,3pt);}
}, postaction={decorate},] (a),
(a) -- [scalar] (i3),
(a) -- [anti charged scalar, decoration={markings, 
  mark=at position 0.80 with {\draw[-] (0,-3pt) -- (0,3pt);},
  mark=at position 0.75 with {\draw[-] (0,-3pt) -- (0,3pt);}
} ] (b),
(b) -- [anti fermion, decoration={markings, mark=at position 0.65 with {\draw[-] (0,-3pt) -- (0,3pt);}},] (c),
(a) -- [anti fermion, decoration={markings, mark=at position 0.65 with {\draw[-] (0,-3pt) -- (0,3pt);}},] (c),
(b) -- [plain] (i1),
(b) -- [scalar] (i2),
};
\end{feynman}
\end{tikzpicture} 
+ \sym  ,
\end{align}
where the first two loops shown have non-linear couplings $u_1\gonetwo$ and $\gonetwo u_2$ and the final two loops shown have couplings $\gonetwo \gonetwo$ and $\gonetwo \gtwoone$ respectively. Once again, we label the diagrams as $\diagram_{\vertexrenormoneone[0.2]}$ and $\diagram_{\vertexrenormtwotwo[0.2]}$, \Erefs{diagrams_vertexu1}, as well as $\diagram_{\vertexrenormonetwo[0.2]}$ and $\diagram_{\vertexrenormtwoone[0.2]}$, to be calculated below,  and set the internal wave-vector of the diagrams to be $\zerovec$, whereby the vertex becomes
\begin{equation}
    \vertexgone{}{} = -\kvecsquared \kroneckerdelta{}\kroneckerdelta{cd}\left\{ g_{12} - g_{12}u_1 \frac{n+2}{3} \diagram_{\vertexrenormoneone[0.2]} - g_{12} u_2 \frac{n+2}{3} \diagram_{\vertexrenormtwotwo[0.2]} - 2g_{12}^2\diagram_{\vertexrenormonetwo} - 2g_{12}g_{21}\diagram_{\vertexrenormtwoone}\right\} \ .
\end{equation}
The only two-loop diagrams to be calculated are
\begin{subequations} \label{eq:final-integrals}
    \begin{align}
        \diagram_{\vertexrenormonetwo} &\defequal  \kint_1 \omegaint_1 \frac{2 \Gamma_2 \kvec_1^4}{(-\imag\omega_1 + \Delta_2(\kvec_1))(\imag\omega_1 + \Delta_2(\kvec_1))(\imag\omega_1 + \Delta_1(\kvec_1))}  \nonumber\\ 
        &= \frac{\Gamma_2}{\Dtwo(\Done + \Dtwo)} \kint_1 \frac{1}{(\kvec_1^2 + \tau_2/\Dtwo)(\kvecsquared_1 + (\tau_1 + \tau_2)/(\Done + \Dtwo))}  \nonumber\\
        & =  \frac{ \Gamma_2 }{\Dtwo(\Done + \Dtwo)} \loopfactor \left(\frac{\tau_2}{\Dtwo} \right)^{-\epsilon/2}\int_0^{1} dx\left[ x + (1-x) \frac{D_2}{D_1 + D_2} \frac{\tau_1 + \tau_2}{\tau_2}\right]^{-\epsilon/2} \label{eq:final_integrals_one}\\
        \diagram_{\vertexrenormtwoone} &\defequal  \kint_1 \omegaint_1 \frac{2 \Gamma_1 \kvec_1^4}{(-\imag\omega_1 + \Delta_1(\kvec_1))(\imag\omega_1 + \Delta_1(\kvec_1))(\imag\omega_1 + \Delta_2(\kvec_1))} + \mathcal{O}(\kvec^4) \nonumber\\ 
        & =  \frac{\Gamma_1}{\Done(\Done + \Dtwo)} \kint_1 \frac{1}{(\kvec_1^2 + \tau_1/\Done)(\kvecsquared_1 + (\tau_1 + \tau_2)/(\Done + \Dtwo))} \nonumber\\
        & =  \frac{ \Gamma_1 }{\Done(\Done + \Dtwo)} \loopfactor \left(\frac{\tau_1}{\Done} \right)^{-\epsilon/2}\int_0^{1} dx\left[ x + (1-x) \frac{D_1}{D_1 + D_2} \frac{\tau_1 + \tau_2}{\tau_1}\right]^{-\epsilon/2} \label{eq:final_integrals_two}
    \end{align}
\end{subequations}
where the remaining integrals in \Erefs{final_integrals_one} and \eref{final_integrals_two} are easily evaluated to be unity to leading order in $\epsilon$. In summary, we find
\begin{multline} \label{eq:vertex1102}
     \vertexgone{}{\kvec} = -\kvec^2 \gonetwo \kroneckerdelta{} \kroneckerdelta{cd} \left[ 1 - \uone \frac{\numcomponent + 2}{6}\frac{\Gamma_1}{\Done^2} \loopfactor \left(\frac{\tau_1}{ \Done}\right)^{-\epsilon/2}  - \utwo \frac{\numcomponent + 2}{6}\frac{\Gamma_2}{\Dtwo^2} \loopfactor \left(\frac{\tau_2}{ \Dtwo}\right)^{-\epsilon/2}  \right.  
     \\
    \left. - \frac{2 \gonetwo \Gamma_2}{\Dtwo(\Done + \Dtwo)} \loopfactor \left(\frac{\tau_2}{\Dtwo} \right)^{-\epsilon/2} 
    \Big(1+\OC(\epsilon)\Big)
    - \frac{2 \gtwoone \Gamma_1}{\Done(\Done + \Dtwo)} \loopfactor \left(\frac{\tau_1}{\Done} \right)^{-\epsilon/2}
    \Big(1+\OC(\epsilon)\Big)
    \right]
\end{multline}

We now consider the diagrams that renormalise $\vertexgtwo{}{}$,
\begin{align}
    \vertexgtwo{}{\kvec} &= 
    \begin{tikzpicture}[baseline = (o.base), scale = 1.2]
  \begin{feynman}
    \vertex (origin) at (0,0) ;
    \vertex (o) at (-0.6,0) {\(a\)};
    \vertex (i1) at (0.6,0) {\(c\)};
    \vertex (i2) at (0.6,0.6) {\(b\)};
    \vertex (i3) at (0.6,-0.6) {\(d\)};
    \diagram* {
      (origin) -- [scalar, decoration={markings, 
  mark=at position 0.25 with {\draw[-] (0,-3pt) -- (0,3pt);},
  mark=at position 0.30 with {\draw[-] (0,-3pt) -- (0,3pt);}
}, postaction={decorate}, edge label'={\(\kvec\)}] (o),
      (origin) -- [plain] (i1),
      (origin) -- [scalar] (i2),
      (origin) -- [plain] (i3),
    };
  \end{feynman}
\end{tikzpicture}
+ 
\begin{tikzpicture}[baseline = (o.base), scale = 1.2]
\begin{feynman}
\vertex (o) at (0,0) {\(a\)};
\vertex (a) at (0.8,0);
\vertex (i3) at ($(a) + (0.75,0)$) {\(b\)};
\vertex (b) at ($(a) + (0.33,1)$);     
\vertex (c) at ($(a) + (1,0.4)$);    
\vertex (i1) at ($(b) + (0.5,0.6)$) {\(c\)};  
\vertex (i2) at ($(b) + (0.5,0.3)$) {\(d\)};  
\diagram* {
(o) -- [scalar, edge label = \(\kvec\), decoration={markings, 
  mark=at position 0.80 with {\draw[-] (0,-3pt) -- (0,3pt);},
  mark=at position 0.75 with {\draw[-] (0,-3pt) -- (0,3pt);}
}, postaction={decorate},] (a),
(a) -- [scalar] (i3),
(a) -- [anti fermion, decoration={markings, 
  mark=at position 0.80 with {\draw[-] (0,-3pt) -- (0,3pt);},
  mark=at position 0.75 with {\draw[-] (0,-3pt) -- (0,3pt);}
} ] (b),
(b) -- [anti fermion, decoration={markings, mark=at position 0.65 with {\draw[-] (0,-3pt) -- (0,3pt);}},] (c),
(a) -- [anti fermion, decoration={markings, mark=at position 0.65 with {\draw[-] (0,-3pt) -- (0,3pt);}},] (c),
(b) -- [plain] (i1),
(b) -- [plain] (i2),
};
\end{feynman}
\end{tikzpicture}
~~~ + \sym 
~+~
\begin{tikzpicture}[baseline = (o.base), scale = 1.2]
\begin{feynman}
\vertex (o) at (0,0) {\(a\)};
\vertex (a) at (0.8,0);
\vertex (i3) at ($(a) + (0.75,0)$) {\(b\)};
\vertex (b) at ($(a) + (0.33,1)$);     
\vertex (c) at ($(a) + (1,0.4)$);    
\vertex (i1) at ($(b) + (0.5,0.6)$) {\(c\)};  
\vertex (i2) at ($(b) + (0.5,0.3)$) {\(d\)};  
\diagram* {
(o) -- [scalar, edge label = \(\kvec\), decoration={markings, 
  mark=at position 0.80 with {\draw[-] (0,-3pt) -- (0,3pt);},
  mark=at position 0.75 with {\draw[-] (0,-3pt) -- (0,3pt);}
}, postaction={decorate},] (a),
(a) -- [scalar] (i3),
(a) -- [anti charged scalar, decoration={markings, 
  mark=at position 0.80 with {\draw[-] (0,-3pt) -- (0,3pt);},
  mark=at position 0.75 with {\draw[-] (0,-3pt) -- (0,3pt);}
} ] (b),
(b) -- [anti charged scalar, decoration={markings, mark=at position 0.65 with {\draw[-] (0,-3pt) -- (0,3pt);}}, postaction={decorate}] (c),
(a) -- [anti charged scalar, decoration={markings, mark=at position 0.65 with {\draw[-] (0,-3pt) -- (0,3pt);}}, postaction={decorate}] (c),
(b) -- [plain] (i1),
(b) -- [plain] (i2),
};
\end{feynman}
\end{tikzpicture}
~~~ + \sym +
\begin{tikzpicture}[baseline = (o.base),scale = 1.2]
\begin{feynman}
\vertex (o) at (0,0) {\(a\)};
\vertex (a) at (0.8,0);
\vertex (i3) at ($(a) + (0.75,0)$) {\(c\)};
\vertex (b) at ($(a) + (0.33,1)$);     
\vertex (c) at ($(a) + (1,0.4)$);    
\vertex (i1) at ($(b) + (0.5,0.6)$) {\(d\)};  
\vertex (i2) at ($(b) + (0.5,0.3)$) {\(b\)};  
\diagram* {
(o) -- [scalar, edge label = \(\kvec\), decoration={markings, 
  mark=at position 0.75 with {\draw[-] (0,-3pt) -- (0,3pt);},
  mark=at position 0.80 with {\draw[-] (0,-3pt) -- (0,3pt);}
}, postaction={decorate},] (a),
(a) -- [plain] (i3),
(a) -- [anti fermion, decoration={markings, 
  mark=at position 0.80 with {\draw[-] (0,-3pt) -- (0,3pt);},
  mark=at position 0.75 with {\draw[-] (0,-3pt) -- (0,3pt);}
} ] (b),
(b) -- [anti charged scalar, decoration={markings, mark=at position 0.65 with {\draw[-] (0,-3pt) -- (0,3pt);}},] (c),
(a) -- [anti charged scalar, decoration={markings, mark=at position 0.65 with {\draw[-] (0,-3pt) -- (0,3pt);}},] (c),
(b) -- [plain] (i1),
(b) -- [scalar] (i2),
};
\end{feynman}
\end{tikzpicture} 
+ \sym \nonumber  \\
& + 
\begin{tikzpicture}[baseline = (o.base),scale = 1.2]
\begin{feynman}
\vertex (o) at (0,0) {\(a\)};
\vertex (a) at (0.8,0);
\vertex (i3) at ($(a) + (0.75,0)$) {\(c\)};
\vertex (b) at ($(a) + (0.33,1)$);     
\vertex (c) at ($(a) + (1,0.4)$);    
\vertex (i1) at ($(b) + (0.5,0.6)$) {\(d\)};  
\vertex (i2) at ($(b) + (0.5,0.3)$) {\(b\)};  
\diagram* {
(o) -- [scalar, edge label = \(\kvec\), decoration={markings, 
  mark=at position 0.75 with {\draw[-] (0,-3pt) -- (0,3pt);},
  mark=at position 0.80 with {\draw[-] (0,-3pt) -- (0,3pt);}
}, postaction={decorate},] (a),
(a) -- [plain] (i3),
(a) -- [anti charged scalar, decoration={markings, 
  mark=at position 0.80 with {\draw[-] (0,-3pt) -- (0,3pt);},
  mark=at position 0.75 with {\draw[-] (0,-3pt) -- (0,3pt);}
} ] (b),
(b) -- [anti fermion, decoration={markings, mark=at position 0.65 with {\draw[-] (0,-3pt) -- (0,3pt);}},] (c),
(a) -- [anti fermion, decoration={markings, mark=at position 0.65 with {\draw[-] (0,-3pt) -- (0,3pt);}},] (c),
(b) -- [plain] (i1),
(b) -- [scalar] (i2),
};
\end{feynman}
\end{tikzpicture} 
+ \sym
\end{align}
These are the same diagrams as those that renormalise $\vertexgone{}{}$ but with solid and dashed lines swapped. This means that the contribution to the vertex $\vertexgtwo{}{}$ will be that of $\vertexgone{}{}$, \Eref{vertex1102}, with indices $1$ and $2$ swapped:
\begin{multline} \label{eq:vertex0211}
    \vertexgtwo{}{\kvec} = -\kvec^2 \gtwoone \kroneckerdelta{} \kroneckerdelta{cd} \left[ 1 - \uone \frac{\numcomponent + 2}{6}\frac{\Gamma_1}{\Done^2} \loopfactor \left(\frac{\tau_1}{ \Done}\right)^{-\epsilon/2}  - \utwo \frac{\numcomponent + 2}{6}\frac{\Gamma_2}{\Dtwo^2} \loopfactor \left(\frac{\tau_2}{ \Dtwo}\right)^{-\epsilon/2}  \right.   \\
      \left.
      -\frac{2 \gonetwo \Gamma_2}{\Dtwo(\Done + \Dtwo)} \loopfactor \left( \frac{\tau_2}{\Dtwo}\right)^{-\epsilon/2}  
    \Big(1+\OC(\epsilon)\Big)
    -\frac{2 \gtwoone \Gamma_1}{\Done(\Done + \Dtwo)} \loopfactor \left( \frac{\tau_1}{\Done}\right)^{-\epsilon/2}  
    \Big(1+\OC(\epsilon)\Big)
    \right]
\end{multline}
This completes the derivations in the present appendix. Calculating the $Z$-factors, \Erefs{Zfactors}, and the Wilson $\gamma$-functions, \eg \Erefs{defnwilson}, from it, is a matter of collecting terms.
4
\section{Analysis of the one-loop flow functions} \label{app:AnalysisFlow}
In this section, we perform a detailed analysis of the flow functions given in \Erefs{Flowred},
\begin{subequations} \label{eq:Flowred_AppB}
    \begin{align}
            \betautildeone &= -\epsilon \utildeoneR + \frac{n + 8}{3} \utildeoneR^2 + 3 \sigma n \gtildeonetwoR^2 \label{eq:betautildeone_AppB}\\
    \betautildetwo &= -\epsilon \utildetwoR + \frac{n + 8}{3} \utildetwoR^2 + 3 \sigma n \gtildeonetwoR^2 \label{eq:betautildetwo_AppB}\\
    \betagtildeonetwo &= \gtildeonetwoR \left( -\epsilon + 4\frac{w  + \sigma}{w + 1} \gtildeonetwoR + \frac{n + 2}{3} \utildeoneR + \frac{n + 2}{3} \utildetwoR \right) \elabel{betagtildeonetwo_AppB}
\ ,
    \end{align}
\end{subequations}
and derive the results quoted in the main text, in particular the precise values of the two stable fixed points \Erefs{first_fixed_points} and \eref{second_fixed_point}.

We start our analysis by classifying the possible solutions. As a first criterion, we note that $\gtildeonetwostar = 0$ is always a solution of $\beta_{\gtildeonetwo}=0$, \Eref{betagtildeonetwo_AppB}, whereas the solution of $\beta_{\utildeone}=0$ depends on both $u_1$ and $g_12$, and similarly for $\beta_{\utildetwo}=0$. We therefore categorise the solution by triviality of $\gtildeonetwostar$. As a second criterion, subtracting $\beta_{\utildetwo}$ from $\beta_{\utildeone}$, we find that
\begin{equation}\elabel{beta_u_difference}
    \left( \frac{n+8}{3}(\utildeonestar + \utildetwostar) - \epsilon\right)(\utildeonestar - \utildetwostar) = 0
\end{equation}
from which we infer that either $\utildeonestar = \utildetwostar$ or $\utildeonestar + \utildetwostar = 3\epsilon/(n+8)$. These criteria, firstly $\gtildeonetwostar$ vanishing or not and secondly $\utildeonestar=\utildetwostar$ or $\utildeonestar + \utildetwostar = 3\epsilon/(n+8)$ (or both), fully map out the solutions, which we analyse in the following.

Starting with $\gtildeonetwostar=0$, the two remaining \Erefs{betautildeone_AppB} and \eref{betautildetwo_AppB} decouple and are thus easily solved by any $\utildeonestar,\utildetwostar\in\{0,3\epsilon/(n+8)\}$, in line with \Eref{beta_u_difference}:
\begin{equation} \elabel{fp_case_1}
    \left( \utildeonestar, \utildetwostar, \gtildeonetwostar\right) = \left\{(0,0,0); ~ \left(\frac{3\epsilon}{n+8},0,0 \right); ~ \left(0, \frac{3\epsilon}{n+8},0 \right); ~\left(\frac{3\epsilon}{n+8},\frac{3\epsilon}{n+8},0 \right)\right\}
\end{equation}
Using these solutions in the stability matrix in \Eref{stabilitymat}, we find that, for $\epsilon >0$, it is positive definite (all eigenvalues are positive) only in the last case in \Eref{fp_case_1} and only when $n>4$, with eigenvalues $\epsilon$, $\epsilon$ and $\epsilon (n-4)/(n+8)$. This gives the first stable fixed point quoted in \Eref{first_fixed_points}. 

We now analyse the case where $\gtildeonetwostar \neq 0$, considering first the case $\utildeonestar + \utildetwostar = 3\epsilon/(n+8)$. In this case, $\beta_{\gtildeonetwo}/\gtildeonetwo = 0$ can easily be solved to yield
\begin{equation} \label{eq:solution_two_g}
    \gtildeonetwostar  = \frac{3}{2(n+8)} \frac{w+1}{w+\sigma} \ .
\end{equation}
Using this expression in $\beta_{\utildeone} = 0 = \beta_{\utildetwo}$ and solving the quadratic for $\utildeonestar$ and $\utildetwostar$ yields
\begin{equation} \label{eq:solution_two_u}
    (\utildeonestar, \utildetwostar) = \left( \left[1\pm \sqrt{1-\sigma \frac{9n}{n+8} \left(\frac{w+1}{w+\sigma} \right)^2} \right] \frac{3\epsilon}{2(n+8)}, \left[1\mp \sqrt{1-\sigma \frac{9n}{n+8} \left(\frac{w+1}{w+\sigma} \right)^2} \right] \frac{3\epsilon}{2(n+8)}\right) \ ,
\end{equation}
to be read as \emph{two} solutions, as they need to fulfil the constraint $\utildeonestar + \utildetwostar = 3\epsilon/(n+8)$. However, it turns out that these solutions \Erefs{solution_two_g} with \eref{solution_two_u}, are \emph{always} unstable, as we discuss in the following. The determinant of the stability matrix \Eref{stabilitymat} is
\begin{equation}\label{eq:det_stabilitymat}
    \det(\Lambda) = -\frac{2\epsilon}{3(n+8)} (2(n+8) \utildeonestar - 3\epsilon)^2,
\end{equation}
which is negative, implying instability, unless $\utildeonestar = \frac{3\epsilon}{2(n+8)}$, in which case the determinant vanishes. For this edge case, $\utildeonestar = \frac{3\epsilon}{2(n+8)}$, the bare parameters $\sigma$ and $w$ have to be fine-tuned such that in \Eref{solution_two_u}, the argument of the square root needs to vanish. 

However, in this edge case, one eigenvalue of the stability matrix \Eref{stabilitymat} still turns out to be negative. In summary, $\gtildeonetwostar \neq 0$ with $\utildeonestar + \utildetwostar = 3\epsilon/(n+8)$ produces no stable fixed point.

We finally analyse the case $\gtildeonetwostar\ne0$ and $\utildeonestar =  \utildetwostar=\utildestar$. Using $\beta_{\gtildeonetwo} = 0$ to express $\gtildeonetwostar$ in terms of $\utildestar$,
\begin{equation} \label{eq:g_and_u}
    \gtildeonetwostar = \frac{w+1}{4(w+\sigma)}\left( \epsilon- \frac{2(n+2)}{3}\utildestar\right),
\end{equation}
and using this to determine the eigenvalues of the stability matrix, we find that any stable solution needs to satisfy
\begin{equation} \label{eq:stability_cond}
    \utildestar > \frac{\epsilon}{4} \ ,
\end{equation}
with an additional constraint of $\utildestar  > 3\epsilon/(2(n+8))$ which is satisfied for any $n\ge1$. Using \Eref{g_and_u} to express $\beta_{\utildeone}$ solely in terms of $\utildestar$, we can find the explicit solution for $\utildestar$ as the root of $\beta_{\utildeone}$,
\begin{equation} \label{eq:utilde_soln}
    \utildestar = \frac{3\epsilon}{2(n+8)} \frac{1 + \sigma \frac{n(n+2) (w+1)^2}{4(w+\sigma)^2} \pm \sqrt{1 - \sigma \frac{n(4-n) (w+1)^2}{4(w+\sigma)^2}}}{1 + \sigma \frac{n(n+2)^2 (w+1)^2}{4(n+8) (w+\sigma)^2}} \ .
\end{equation}
Adding and subtracting $\epsilon/4$ from \Eref{utilde_soln}, after some algebra we can rearrange the expression for $\utildestar$ into
\begin{equation} \label{eq:utilde_soln_simpler}
    \utildestar_{\pm} = \frac{\epsilon}{4} \left( 1 + \frac{(4-n) h(\sigma,w,n)}{(n+2)h(\sigma,w,n) \pm 6} \right)
\end{equation}
with
\begin{equation} \label{eq:hfactor_app}
    h(\sigma,w,n) \defequal \sqrt{1 - \sigma \frac{n(4-n)(w+1)^2}{4(w+\sigma)^2}} \ .
\end{equation}
Physical fixed points require $h(\sigma, w, n)$ to be real, in which case it is understood to take the positive root, as suggested in \Eref{utilde_soln}. Using \Eref{utilde_soln_simpler} in \Eref{g_and_u}, we obtain an explicit expression for
\begin{equation} \label{eq:gtilde_soln}
    \gtildeonetwostar = \tilde{g}^{*}_{\pm} = \frac{4-n}{4(n+2)} \frac{w+1}{w+\sigma} \frac{\epsilon}{\frac{6}{n+2} \pm h(\sigma, w,n)}.
\end{equation}

To determine the stability of the solution, we return to \Eref{stability_cond}, requiring 
\begin{equation}\elabel{h_condition}
    \frac{(4-n) h(\sigma,w,n)}{(n+2)h(\sigma,w,n) \pm 6} \ge 0 \ ,
\end{equation}
given \Eref{utilde_soln_simpler}. If there is a physical solution, $h(\sigma,w,n)\in\Rset^{+}$, then \Eref{h_condition} is fulfilled for $\utildestar_{+}$ and $n<4$. For $n>4$, the fixed point $\utildestar_{+}$ is unstable, whereas for $n=4$ it is marginal, coinciding with the rightmost fixed point in \Eref{fp_case_1}.

As for $\utildestar_{-}$, for $n<4$ the denominator in \Eref{h_condition} needs to be positive, requiring $h(\sigma,w,n)>6/(n+2)$, 
\begin{equation}\elabel{uminus_sigma_condition}
    0> \sigma^2 + 2\left(w + \frac{n(n+2)^2}{8(n+8)}(w+1)^2 \right)\sigma + w^2
    \ ,
\end{equation}
which in turn implies
\begin{equation} \label{eq:sigma_condition}
    0 > (\sigma + \sigma_{+})(\sigma + \sigma_{-})
\end{equation}
with 
\begin{equation} \label{eq:def_sigma_pm}
    \sigma_{\pm} \defequal \left( w + \frac{n(n+2)^2}{8(n+8)} (w+1)^2\right)\left( 1 \pm \sqrt{1 - \left(\frac{w}{ w + \frac{n(n+2)^2}{8(n+8)} (w+1)^2} \right)^2}\right) \ .
\end{equation}
Given that $w$ is non-negative, both $\sigma_{+}$ and $\sigma_{-}$ are bound to be non-negative and real, so that $\sigma\in(-\sigma_+,-\sigma_-)$ for \Eref{sigma_condition} to hold for $n<4$. Again, for $n=4$, the fixed point value $\utildestar_{-}$ in \Eref{utilde_soln_simpler} reduces to the right-most fixed point in \Eref{fp_case_1}. Finally, when $n>4$, we require $h(\sigma, w, n)<6/(n+2)$ in \Eref{h_condition}. The ensuing analysis in fact reproduces condition \Eref{sigma_condition} with \Eref{def_sigma_pm}, with the factor $(n-4)$ in the definition of $h(\sigma, w, n)$, \Eref{hfactor_app}, cancelling as $1-6/(n+2)=(n-4)/(n+2)$. 

To conclude this case, the fixed point $\utildestar_{-}$ we started to analyse with \Eref{uminus_sigma_condition}
\begin{equation} \label{eq:second_fixed_point_app}
    (\utildeonestar \scalebox{1.5}{,} \utildetwostar \scalebox{1.5}{,} \gtildeonetwostar) =
        \left(\frac{\epsilon}{4} \left( 1 + \frac{(4-n) h(\sigma,w,n)}{(n+2)h(\sigma,w,n) - 6} \right)\scalebox{1.5}{,} \frac{\epsilon}{4} \left( 1 + \frac{(4-n) h(\sigma,w,n)}{(n+2)h(\sigma,w,n) - 6} \right)\scalebox{1.5}{,} \frac{4-n}{4(n+2)} \frac{w+1}{w+\sigma} \frac{\epsilon}{\frac{6}{n+2} - h(\sigma,w,n)}\right)
\end{equation}
is stable for all $n\ne4$, provided $\sigma \in (-\sigma_{+}, -\sigma_{-})$.

In summary, we find the following stable fixed points: Firstly, the right-most of \Eref{fp_case_1}, \Eref{first_fixed_points} in the main text, $\utildeonestar = \utildetwostar = 3\epsilon/(n+8)$ and $\gtildeonetwostar=0$, for $n>4$, secondly \Eref{utilde_soln_simpler}, \Eref{first_fixed_points}, $u^*_1=u^*_2=\utildestar_{+}$ and $\gtildeonetwostar = \tilde{g}^{*}_{+}$ given in \Eref{gtilde_soln}, \Eref{second_fixed_point}, for n<4, and finally \Eref{second_fixed_point_app}, \Eref{second_fixed_point}, $u^*_1=u^*_2=\utildestar_{-}$ and $\gtildeonetwostar = \tilde{g}^{*}_{-}$, for $n<4$ and $\sigma \in (-\sigma_{+}, -\sigma_{-})$.

\end{document}

%% file: diagrams_small.tex
\newcommand{\massrenormone}[1][0.3]{%
  \scalebox{#1}{\begin{tikzpicture}[baseline = (a.base)]
   \begin{feynman}
    \vertex (a) at (0.8,0);
    \vertex (b) at (1.7,1);
    \diagram* {
      (a) -- [anti fermion, line width=2pt, decoration={markings, mark=at position 0.65 with {\draw[-] (0,-3pt) -- (0,3pt);}}, postaction={decorate}, bend left=40] (b),
      (a) -- [anti fermion, line width=2pt, decoration={markings, mark=at position 0.65 with {\draw[-] (0,-3pt) -- (0,3pt);}}, postaction={decorate}, bend right=40] (b),
    };
  \end{feynman}
\end{tikzpicture}%
}
}

\newcommand{\massrenormtwo}[1][0.3]{%
  \scalebox{#1}{\begin{tikzpicture}[baseline = (a.base)]
   \begin{feynman}
    \vertex (a) at (0.8,0);
    \vertex (b) at (1.7,1);
    \diagram* {
      (a) -- [anti charged scalar, line width=2pt, double distance=1pt, decoration={markings, mark=at position 0.65 with {\draw[-] (0,-3pt) -- (0,3pt);}}, postaction={decorate}, bend left=40] (b),
      (a) -- [anti charged scalar, line width=2pt, double distance=1pt, decoration={markings, mark=at position 0.65 with {\draw[-] (0,-3pt) -- (0,3pt);}}, postaction={decorate}, bend right=40] (b),
    };
  \end{feynman}
\end{tikzpicture}}%
}

\newcommand{\vertexrenormoneone}[1][0.3]{
\scalebox{#1}{
\begin{tikzpicture}[baseline = (a.base),scale = 1.2]
\begin{feynman}
\vertex (a) at (0.8,0);
\vertex (b) at ($(a) + (0.33,1)$);     
\vertex (c) at ($(a) + (1,0.4)$);    
\diagram* {
(a) -- [anti fermion, line width=2pt, decoration={markings, 
  mark=at position 0.80 with {\draw[-] (0,-3pt) -- (0,3pt);},
  mark=at position 0.75 with {\draw[-] (0,-3pt) -- (0,3pt);}
} ] (b),
(b) -- [anti fermion, line width=2pt, decoration={markings, mark=at position 0.65 with {\draw[-] (0,-3pt) -- (0,3pt);}},] (c),
(a) -- [anti fermion, line width=2pt, decoration={markings, mark=at position 0.65 with {\draw[-] (0,-3pt) -- (0,3pt);}},] (c),
};
\end{feynman}
\end{tikzpicture}
}
}

\newcommand{\vertexrenormonetwo}[1][0.3]{
\scalebox{#1}{
\begin{tikzpicture}[baseline = (a.base),scale = 1.2]
\begin{feynman}
\vertex (a) at (0.8,0);
\vertex (b) at ($(a) + (0.33,1)$);     
\vertex (c) at ($(a) + (1,0.4)$);    
\diagram* {
(a) -- [anti fermion, line width=2pt, decoration={markings, 
  mark=at position 0.80 with {\draw[-] (0,-3pt) -- (0,3pt);},
  mark=at position 0.75 with {\draw[-] (0,-3pt) -- (0,3pt);}
} ] (b),
(b) -- [anti charged scalar, line width=2pt, decoration={markings, mark=at position 0.65 with {\draw[-] (0,-3pt) -- (0,3pt);}},] (c),
(a) -- [anti charged scalar, line width=2pt, decoration={markings, mark=at position 0.65 with {\draw[-] (0,-3pt) -- (0,3pt);}},] (c),
};
\end{feynman}
\end{tikzpicture}
}
}

\newcommand{\vertexrenormtwoone}[1][0.3]{
\scalebox{#1}{
\begin{tikzpicture}[baseline = (a.base),scale = 1.2]
\begin{feynman}
\vertex (a) at (0.8,0);
\vertex (b) at ($(a) + (0.33,1)$);     
\vertex (c) at ($(a) + (1,0.4)$);    
\diagram* {
(a) -- [anti charged scalar, line width=2pt, decoration={markings, 
  mark=at position 0.80 with {\draw[-] (0,-3pt) -- (0,3pt);},
  mark=at position 0.75 with {\draw[-] (0,-3pt) -- (0,3pt);}
} ] (b),
(b) -- [anti fermion, line width=2pt, decoration={markings, mark=at position 0.65 with {\draw[-] (0,-3pt) -- (0,3pt);}},] (c),
(a) -- [anti fermion, line width=2pt, decoration={markings, mark=at position 0.65 with {\draw[-] (0,-3pt) -- (0,3pt);}},] (c),
};
\end{feynman}
\end{tikzpicture}
}
}

\newcommand{\vertexrenormtwotwo}[1][0.3]{
\scalebox{#1}{
\begin{tikzpicture}[baseline = (a.base),scale = 1.2]
\begin{feynman}
\vertex (a) at (0.8,0);
\vertex (b) at ($(a) + (0.33,1)$);     
\vertex (c) at ($(a) + (1,0.4)$);    
\diagram* {
(a) -- [anti charged scalar, line width=2pt, decoration={markings, 
  mark=at position 0.80 with {\draw[-] (0,-3pt) -- (0,3pt);},
  mark=at position 0.75 with {\draw[-] (0,-3pt) -- (0,3pt);}
} ] (b),
(b) -- [anti charged scalar, line width=2pt, decoration={markings, mark=at position 0.65 with {\draw[-] (0,-3pt) -- (0,3pt);}},] (c),
(a) -- [anti charged scalar, line width=2pt, decoration={markings, mark=at position 0.65 with {\draw[-] (0,-3pt) -- (0,3pt);}},] (c),
};
\end{feynman}
\end{tikzpicture}
}
}